\documentclass[a4paper,fleqn,usenatbib]{mnras}

\usepackage[T1]{fontenc}
\usepackage{ae,aecompl}

\usepackage{subfigure}
\usepackage{graphicx}	
\usepackage{amsmath}	
\usepackage{amssymb}	
\usepackage{color}
\usepackage{mathptmx}
\usepackage{epsfig}
\usepackage{wasysym}
\usepackage{xfrac}
\usepackage{pifont}
\usepackage[table,xcdraw]{xcolor}
\usepackage{booktabs}
\usepackage{multirow}
\usepackage{comment}
\usepackage{soul}
\usepackage{tikz}
\usepackage{booktabs}

\definecolor{grey}{rgb}{0.75,0.75,0.75}
\definecolor{Orange}{rgb}{1.0,0.5,0.15}
\definecolor{brown}{rgb}{0.7,0.25,0.0}
\definecolor{pink}{rgb}{1.0,0.5,0.5}
\definecolor{darkerred}{rgb}{0,0.5,0.5}
\definecolor{darkerblue}{rgb}{0,0,0.8}
\definecolor{lightblue}{rgb}{0.12, 0.56, 1.0}
\definecolor{Blue}{rgb}{0,0.08,0.65}
\definecolor{Red}{rgb}{0.65,0.08,0.05}
\definecolor{Green}{rgb}{0.15,0.45,0.25}

\definecolor{purple}{rgb}{0.5,0,0.87}

\usetikzlibrary{positioning,decorations.pathreplacing,shapes}

\title[Characterising tidal features with the Rubin Obs.]{Preparing for low surface brightness science with the Vera C. Rubin Observatory: characterisation of tidal features from mock images}

\author[G. Martin et al.]{G. Martin,$^{1,2}$\thanks{E-mail: garrethmartin@kasi.re.kr, garrethmartin@arizona.edu}
A. E. Bazkiaei,$^{3,4,5}$\thanks{E-mail: amir.ebadati-bazkiaei@mq.edu.au}
M. Spavone,$^{6}$ 
E. Iodice,$^{6}$ 
J. C. Mihos,$^{7}$ 
M. Montes,$^{8}$\thanks{STScI Prize Fellow} \newauthor
J. A. Benavides,$^{9,10}$ 
S. Brough,$^{11}$ 
J. L. Carlin,$^{12}$ 
C. A. Collins,$^{13}$ 
P. A. Duc,$^{14}$ \newauthor
F. A. G\'{o}mez,$^{15,16}$ 
G. Galaz,$^{17}$ 
H. M. Hern\'{a}ndez-Toledo,$^{18}$ 
R. A. Jackson,$^{19}$ 
S. Kaviraj,$^{20}$ \newauthor
J. H. Knapen,$^{21,22}$ 
C. Mart\'inez-Lombilla,$^{11}$ 
S. McGee,$^{23}$ 
D. O'Ryan,$^{24}$ 
D. J. Prole,$^{3,4}$ \newauthor
R. M. Rich,$^{25}$ 
J. Rom\'an,$^{21,22}$ 
E. A. Shah,$^{26}$\thanks{LSSTC DSFP Fellow}
T. K. Starkenburg,$^{27}$ 
A. E. Watkins,$^{20}$ \newauthor
D. Zaritsky,$^{2}$ 
C. Pichon,$^{28,29,30}$ 
L. Armus,$^{31}$ 
M. Bianconi,$^{23}$ 
F. Buitrago,$^{32,33}$ 
I. Bus\'a,$^{34}$ \newauthor
F. Davis,$^{20}$ 
R. Demarco,$^{35}$ 
A. Desmons,$^{11}$ 
P. Garc\'{i}a,$^{36}$ 
A. W. Graham,$^{37}$ 
B. Holwerda,$^{38}$ \newauthor
D. S. -H. Hon,$^{37}$ 
A. Khalid,$^{11}$ 
J. Klehammer,$^{14}$ 
D. Y. Klutse,$^{39}$ 
I. Lazar,$^{20}$ 
P. Nair,$^{40}$ \newauthor
E. A. Noakes-Kettel,$^{20}$ 
M. Rutkowski,$^{41}$ 
K. Saha,$^{42}$ 
N. Sahu,$^{43,37}$ 
E. Sola,$^{14}$ \newauthor
J. A. V\'{a}zquez-Mata,$^{18,44}$ 
A. Vera-Casanova,$^{16}$ 
I. Yoon$^{45}$ \newauthor
\large (Affiliations can be found after the list of references)
}

\pubyear{2021}
\begin{document}
\label{firstpage}
\pagerange{\pageref{firstpage}--\pageref{lastpage}}
\maketitle


\begin{abstract}
Tidal features in the outskirts of galaxies yield unique information about their past interactions and are a key prediction of the hierarchical structure formation paradigm. The Vera C. Rubin Observatory is poised to deliver deep observations for potentially of millions of objects with visible tidal features, but the inference of galaxy interaction histories from such features is not straightforward. Utilising automated techniques and human visual classification in conjunction with realistic mock images produced using the \textsc{NewHorizon} cosmological simulation, we investigate the nature, frequency and visibility of tidal features and debris across a range of environments and stellar masses. In our simulated sample, around 80 per cent of the flux in the tidal features around Milky Way or greater mass galaxies is detected at the 10-year depth of the Legacy Survey of Space and Time ($30-31$ mag arcsec$^{-2}$), falling to 60 per cent assuming a shallower final depth of 29.5 mag arcsec$^{-2}$. The fraction of total flux found in tidal features increases towards higher masses, rising to 10 per cent for the most massive objects in our sample ($M_{\star}\sim10^{11.5}~{\rm M_{\odot}}$). When observed at sufficient depth, such objects frequently exhibit  many distinct tidal features with complex shapes. The interpretation and characterisation of such features varies significantly with image depth and object orientation, introducing significant biases in their classification. Assuming the data reduction pipeline is properly optimised, we expect the Rubin Observatory to be capable of recovering much of the flux found in the outskirts of Milky Way mass galaxies, even at intermediate redshifts ($z < 0.2$).
\end{abstract}



\section{Introduction}

Hierarchical structure formation scenarios \citep[e.g.][]{Fall1980,Bosch2002,Agertz2011} predict that massive galaxies acquire much of their stellar mass through a combination of continuous cold gas accretion and mergers with smaller objects \citep[e.g.][]{Press1974,Moster2013,Kaviraj2015a,Rodriguez2016,Martin2018_progenitors,Davison2020,Martin2021}. As a consequence, mergers are also expected to play a significant role in driving the evolution of galaxy properties, for example, by triggering \citep[][]{Schweizer1982,Mihos1996,Duc1997,Elbaz2003,Kaviraj2011,Lofthouse2017,Martin2017} or quenching \citep[][]{Schawinski2014, Barro2017, Pontzen2017, Kawinwanichakij2017} star formation in the host galaxy or by driving its morphological evolution \citep[e.g.][]{Toomre1977,Dekel2009,Conselice2009,Taranu2013,Naab2014,Fiacconi2015,Graham2015,Gomez2017,Deeley2017,Welker2017,martin2018_sph,Jackson2019}. Signatures of past mergers take the form of faint extended tidal features such as tails \citep[e.g.][]{Pfleiderer1963, Toomre1972,Peirani2010,Kaviraj2014b,Kaviraj2019} or plumes \citep[e.g.][]{Lauer1988} -- which are typically produced by major mergers -- and streams \citep[e.g.][]{Johnston1999,Shipp2018,Martinez-Delgado2021} or shells \citep[e.g.][]{Malin1983,Quinn1984} -- which mainly arise from minor interactions -- as well as in the structure of the surrounding diffuse light \citep[e.g.,][]{Johnston2002,Choi2002,Graham2002,Seigar2007,Kaviraj2012,Montes2019,Iodice2019,Monachesi2016,Monachesi2019}. These features, which arise from many different types of encounter, hold a fossil record of the host galaxy's past interactions and mergers which can be used to reconstruct its assembly history and dynamical history \citep[][]{Johnston2008,Martinez-Delgado2009,Belokurov2017,Ren2020,Spavone2020,Montes2020,VeraCasanova2021}. However, the majority of tidal features are expected to have surface brightnesses fainter than 30 mag arcsec$^{-2}$ in the $r$-band \citep[][]{Johnston2008}. Although pushing towards these kinds of limiting surface brightnesses remains extremely challenging, it is nevertheless desirable to do so, being necessary to uncover a more detailed history of local Universe. This is not only vital for our understanding of hierarchical galaxy assembly \citep[e.g.][]{Johnston2001,Wang2012}, but also serves as a novel galactic scale probe of more fundamental physics such as theories of gravity \citep[e.g.][]{Gentile2007,Renaud2016} and dark matter \citep[][]{Dubinski1996,Kesden2006,Dumas2015,vanDokkum2018,Montes2020}. In particular, tidal structure is a powerful tracer of the underlying galactic halo potential \citep[e.g.][]{Dubinski1999,Varghese2011,Bovy2016,Ibata2020,Malhan2021}.

Over the last few decades, advances in the sensitivity and field of view of modern instruments \citep[e.g.][]{Miyazaki2002,Kuijken2002,Mihos2005,Miyazaki2012,Diehl2012,Abraham2014,Torrealba2018} and increasing sophistication of observational and data-analysis techniques \citep[e.g.][]{Mihos2005,Akhlaghi2015,Pawlik2016,Prole2018,Morales2018,Rich2019,Tanoglidis2021,Zaritsky2021} have permitted relatively large studies that concentrate on the low surface brightness (LSB) regime which tidal features inhabit. This has made possible the detailed characterisation of the LSB components of galaxies \citep[e.g.][]{Fong2018,Bilek2020} and allowed studies of their prevalence \citep[e.g.][]{Hood2018}.

The 10-year Legacy Survey of Space and Time (LSST), which will take place at the Vera C. Rubin Observatory \citep[][]{Olivier2008,Ivezic2019}, will lead to a step change in the depth and detail that can be achieved by wide area surveys. Data from the 10-year survey will vastly increase the number of known objects with tidal features. While deep observations tracing low surface brightness structures and galaxies have been possible previously \citep[e.g.][]{Martinez-Delgado2009,Kim2012,Beaton2014,Duc2015,Mihos2015,Fong2018,Zaritsky2019,Iodice2019,Trujillo2021}, LSST will offer a distinct advantage as these studies have generally been limited to small fields or targeted observations of individual galaxies (typically emphasising cluster environments) or else do not have the requisite depth to detect a significant fraction of prominent tidal features. LSST uniquely combines very deep imaging \citep[$r$-band depth better than 30.5~mag~arcsec$^{-2}$ with $10^{\prime\prime}\times10^{\prime\prime}$ binning;][]{Laine2018,Brough2020} with a wide area covering the whole Southern sky (18,000 square degrees). This will enable detailed statistical studies of tidal features within a representative volume of the Universe for the first time.

It is expected that the raw data produced by the Rubin Observatory will be of sufficient quality to study low surface brightness features \citep[][]{Robertson2017,Kaviraj2020,Trujillo2021}. However, a number of other obstacles still remain if the available data are to be exploited to their full potential. The characterisation of tidal features requires not only sufficiently deep imaging but also bespoke data reduction suitable for LSB science and a thorough understanding of biases and uncertainties present in the data.

Follow-up observations for the full population of galaxies with LSB features that will be revealed by LSST will be intractable, especially as tidal features and disturbed morphologies are expected to be ubiquitous in massive galaxies \citep[e.g.][]{Tal2009,Cibinel2019} and likely remain at least somewhat common in lower mass galaxies \citep[e.g.][]{MartinezDelgado2012,Martin2021}. Analysis of the majority of galaxies will therefore be based primarily on available 2-d photometric information. This means that additional information such as spectroscopy and multi-wavelength data, which can reveal important information about the distances, 3-d distribution, kinematics, environments, baryonic content and stellar populations of galaxies \citep[e.g.][]{Bournaud2004,Kadowaki2017,Junais2020,Karunakaran2020}, will be unavailable for a majority of objects. Analysis of the majority of galaxies will therefore be limited to Rubin Observatory data.

With regards to the characterisation of the tidal features themselves, automated methods \citep[e.g.][]{Grillmair1995,Conselice2000,Rockosi2002,Lotz2004,Hendel2019,Pearson2021} can help to define the structure of galaxies and identify merging systems and their tidal features, but a full characterisation of these galaxies and their tidal features befitting the quality of the available photometric data will require detailed visual inspection by human classifiers \citep[e.g.][]{Darg2010,Bilek2020}. Visual inspection relies on a significant level of domain knowledge and physical intuition for interpretation. This inevitably introduces some level of subjectivity, especially in the absence of precise redshifts, kinematics or other 3-d information. While machine learning and machine vision techniques can help alleviate reliance on human classifiers \citep[e.g.][]{Beck2018,Hendel2019,Walmsley2019}, continuous human intervention will likely still be required. Training sets, will still need to be constructed and labelled by human classifiers, and as the coadded LSST images become deeper they will need to be routinely updated \citep[][]{Martin2020}. Some level of bias is therefore unavoidable and its nature may evolve with a number of factors including limiting surface brightness, galaxy mass and orientation \citep[e.g.][]{Mantha2019,Muller2019,Blumenthal2020,Lambrides2021}.

Some sources of bias, such as the effect of projection, are intrinsic to observations, while others like image depth, can be improved with longer exposure times. For example, depending on the angle at which a given LSB structure is observed, the efficiency with which they are detected can be impacted \citep[][]{Mancillas2019,VeraCasanova2021} or their nature can change so that the same structures appear either stream-like or shell-like from different angles \citep[][]{Hendel2015,Greco2018b}. Additionally different classes of tidal features may become more or less detectable over time, can persist over differing timescales \citep[][]{Johnston1999,Bullock2005,Mancillas2019,VeraCasanova2021} or else may transform into different classes of tidal structures \citep[][]{Foster2014,Hendel2015}. Other unrelated structures like galactic cirrus \citep[][]{Deschenes2016,Roman2020} or instrumental artefacts \citep[][]{Tanoglidis2021,Chang2021} can be misclassified or otherwise inhibit the detection of tidal features. At higher redshifts, it can also become increasingly difficult to interpret images as the angular scale of objects decreases and they become more poorly resolved.

The purpose of this paper is to evaluate the expected performance of the LSST at recovering all forms of tidal features and diffuse light occurring in the outskirts of galaxies based on realistic mock images produced using the \textsc{NewHorizon} simulation \citep[][]{Dubois2021}. We make predictions as a function of a galaxy's physical properties, redshift and imaging depth. We then explore how stellar mass, ex-situ mass fraction, redshift, limiting surface brightness and orientation may affect or bias the visual characterisation of galaxies by expert human classifiers across different types of tidal features.

\begin{itemize}
    \item In Section \ref{sec:method} we present an overview of the \textsc{NewHorizon} simulation along with the relevant physics and the method for producing mock images and merger trees as well as outline our visual classification scheme.
    \item In Section \ref{sec:properties} we explore the properties of the extended light around galaxies using automated techniques to separate the LSB components. We study the spatial and surface brightness distributions as well as the fraction of tidal flux that we expect to detect at various limiting surface brightnesses and redshifts.  We additionally consider how tidal flux fraction evolves with galaxy mass and accretion history.
    \item In Section \ref{sec:visual_characterisation} we present the results of visual classifications of our mock images by human classifiers. We discuss the frequency of different classes of tidal feature as a function of galaxy mass and limiting surface brightness and look at how limiting surface brightness, redshift and projection can introduce biases.
    \item In Section \ref{sec:summary} we summarise our results.
\end{itemize}

Throughout this paper we adopt a $\Lambda$CDM cosmology consistent with \citet[][]{Komatsu2011} ($\Omega_{\rm m}=0.272$, $\Omega_\Lambda=0.728$, $\Omega_{\rm b}=0.045$, $H_0=70.4 \ \rm km\,s^{-1}\, Mpc^{-1}$) and we primarily assume a \citet[][]{Salpeter1955} initial mass function (IMF).

\section{Method}

We employ the state-of-the-art cosmological hydrodynamical simulation, \textsc{NewHorizon}, in order to produce realistic mock observations of galaxies and their outskirts, companions and satellites within a self-consistent cosmological context. These objects have known properties and interaction histories which can be used to test the efficacy of observational assumptions and techniques.

\label{sec:method}

\subsection{The \textsc{NewHorizon} simulation}

\label{sec:NH}

The \textsc{NewHorizon} simulation\footnote{\href{http://new.horizon-simulation.org}{http://new.horizon-simulation.org}} \citep[][]{Dubois2021} is a zoom-in of the (142~Mpc)$^{3}$ parent Horizon-AGN simulation \citep{Dubois2014,Kaviraj2017}. Initial conditions are generated using cosmological parameters that are compatible with \emph{WMAP7} $\Lambda$CDM cosmology \citep{Komatsu2011} ($\Omega_{\rm m}=0.272$, $F\Omega_\Lambda=0.728$, $\sigma_8=0.81$, $\Omega_{\rm b}=0.045$, $H_0=70.4 \ \rm km\,s^{-1}\, Mpc^{-1}$, and $n_s=0.967$). Within the original Horizon-AGN volume, a spherical volume with a diameter of 20~Mpc and an effective resolution of $4096^3$ is defined, corresponding to a dark matter (DM) mass resolution and \textit{initial} gas mass resolution of $m_{DM}=1.2\times 10^6 \ \rm M_\odot$ and $m_{gas}=2\times 10^5 \ \rm M_\odot$. \textsc{NewHorizon} uses the adaptive mesh refinement (AMR) code \textsc{Ramses} \citep[][]{Teyssier2002} and gas is evolved with a second-order Godunov scheme and the approximate Harten-Lax-Van Leer-Contact \citep{Toro1999} Riemann solver with linear interpolation of the cell-centred quantities at cell interfaces.

\textsc{NewHorizon} combines high stellar mass ($1.3\times10^{4}{\rm M_{\odot}}$) and spatial resolution ($\sim34$~pc), with a contiguous volume of $(16~{\rm Mpc})^{3}$. The volume probes field and group environments, but does not extend to dense clusters (the maximum halo mass is $M_{h}\sim10^{13}~{\rm M_{\odot}}$). Given the diffuse nature of galaxy stellar haloes, the trade off between resolution and volume is an important consideration. We find \textsc{NewHorizon} to be a better compromise than similar simulations like Illustris TNG50 \citep[][]{Nelson2019} or \textsc{Romulus25} \citep[][]{Tremmel2017}, both of which trade larger volumes for lower mass resolution. The closest similar simulation in terms of mass resolution is TNG50, with a volume of $(50~{\rm Mpc})^{3}$ and a stellar mass resolution ($8.5\times10^{4}{\rm M_{\odot}}$). For comparison, most observed tidal features individually account for $\sim0.1$ per cent to a few per cent of the total stellar mass of a system \citep[e.g.][]{vanDokkum2019,Fensch2020}, meaning these tidal features would resolved with only $\sim100 - 1000$ particles and $\sim800 - 8000$ particles for a galaxy of $M_{\star}=10^{10}~{\rm M_{\odot}}$ by TNG50 and \textsc{NewHorizon} respectively.

The additional resolution of \textsc{NewHorizon} is therefore important in order to sample as much as possible the LSB outskirts of galaxies at surface brightness limits that contemporary or forthcoming instrument will be capable of targeting (see Section \ref{sec:resolution}). \textsc{NewHorizon} also has sufficient volume to yield a reasonable sample of massive galaxies ($M_{\star}>10^{10}~{\rm M_{\odot}}$) and provides a realistic distribution of galaxies, as well as fully simulating the cosmological context required to produce galaxies with \textit{ab initio} realistic interaction and formation histories (as opposed to zoom-in simulations of individual haloes, where the zoom region must be carefully selected to avoid bias).

\textsc{NewHorizon} reproduces key galaxy properties with good agreement to observed quantities. The galaxy stellar mass function, galaxy size-mass relation,  halo mass-stellar mass relation as well as the evolution of galaxy morphology and cosmic star formation rate densities show fair agreement with observationally derived relations. There is however significant uncertainty from cosmic variance owing to size of the simulated region. Relevant to this study, \textsc{NewHorizon} appears to deviate from observations at the high mass end or the galaxy size-mass relation and at the low mass end of the halo-mass stellar-mass relation. Galaxies appear somewhat more compact than expected at $M_{\star} \gtrsim 10^{11}~{\rm M_{\odot}}$ and have stellar masses that are relatively too massive for halo masses of $M_{h} \lesssim 10^{11}~{\rm M_{\odot}}$. We refer readers to Sections 3.2., 3.6., 3.7. and 3.9  of \citet[][]{Dubois2021} for a more detailed description of the galaxy stellar mass function, halo mass- stellar mass relation, size-mass relation and kinematics respectively.

\subsubsection{Numerical resolution limit for detecting tidal features}
\label{sec:resolution}

Because the stellar particle mass resolution of a simulation places limits on its ability to resolve structures, we first attempt to estimate the numerical limits that the resolution of \textsc{NewHorizon} places on our ability to resolve tidal features. We restrict our analysis to shells, which we are typically fainter than tails (see Figure \ref{fig:classifications_hist}) and therefore more susceptible to resolution effects. In Appendix \ref{sec:res_limit} we describe an analytical method using analytical shell profiles \citep[][]{Sanderson2013} and \citet{Sersic1968} profiles. We find that  this is dependent on a number of factors including the galactocentric radius of the tidal feature and the shape and brightness of the galaxy profile.

Even with its relatively high stellar mass resolution, for the most massive galaxies in our sample, we do not expect \textsc{NewHorizon} to resolve shells with surface brightnesses comparable to LSST 10-year depth ($\mu_{r}^{\rm lim}(3\sigma,10^{\prime\prime}\times10^{\prime\prime})\approx30.5$~mag arcsec$^{-2}$) close to the central regions of galaxies ($r<4.5~R_{\rm eff}$). For less massive galaxy models, it is possible to detect faint shells at significantly smaller radii but, for the full sample, we would likely require significantly better mass resolution to enable us to detect all tidal features. As the underlying radial and surface brightness distribution of shells and other tidal features is not known, it is difficult to estimate how significantly this affects our results, but shells are typically resolved down to sufficiently small radii so as to have negligible impact on the number of detected tidal features for depths realistically achievable by LSST. At significantly higher limiting surface brightnesses more care is needed in interpreting results. Shells which would be observationally detectable may not be sufficiently resolved at radii significantly larger than $10~R_{\rm eff}$, meaning the frequency of tidal features at very faint limiting surface brightens is likely underestimated, particularly around more massive galaxies. We refer to Appendix \ref{sec:res_limit} for a more detailed discussion.

\subsection{Galaxy Sample}
\label{sec:sample}

We use the structure finder \textsc{AdaptaHOP} \citep{Aubert2004} to separately detect both galaxies, haloes along with their respective sub-structures based on the distribution of dark matter and star particles in the simulation box respectively. The centre of each galaxy or halo is recursively determined by seeking the centre of mass in a shrinking sphere,  while decreasing its radius by 10 per cent recurrently down  to a minimum radius of 0.5 kpc~\citep{Power2003}. We impose a minimum structure size of 100 dark matter particles and 50 star particles as well as requiring and an average overdensity of 80 times the critical density for dark matter haloes and 160 times the critical density for galaxies \citep[see][for details]{Aubert2004}. Halo virial masses and radii are obtained by computing the kinetic and gravitational energy within ellipsoids, stopping once virial equilibrium is sufficiently well verified \citep[][]{Dubois2021}.

\begin{figure}
    \centering
    \includegraphics[width=0.45\textwidth]{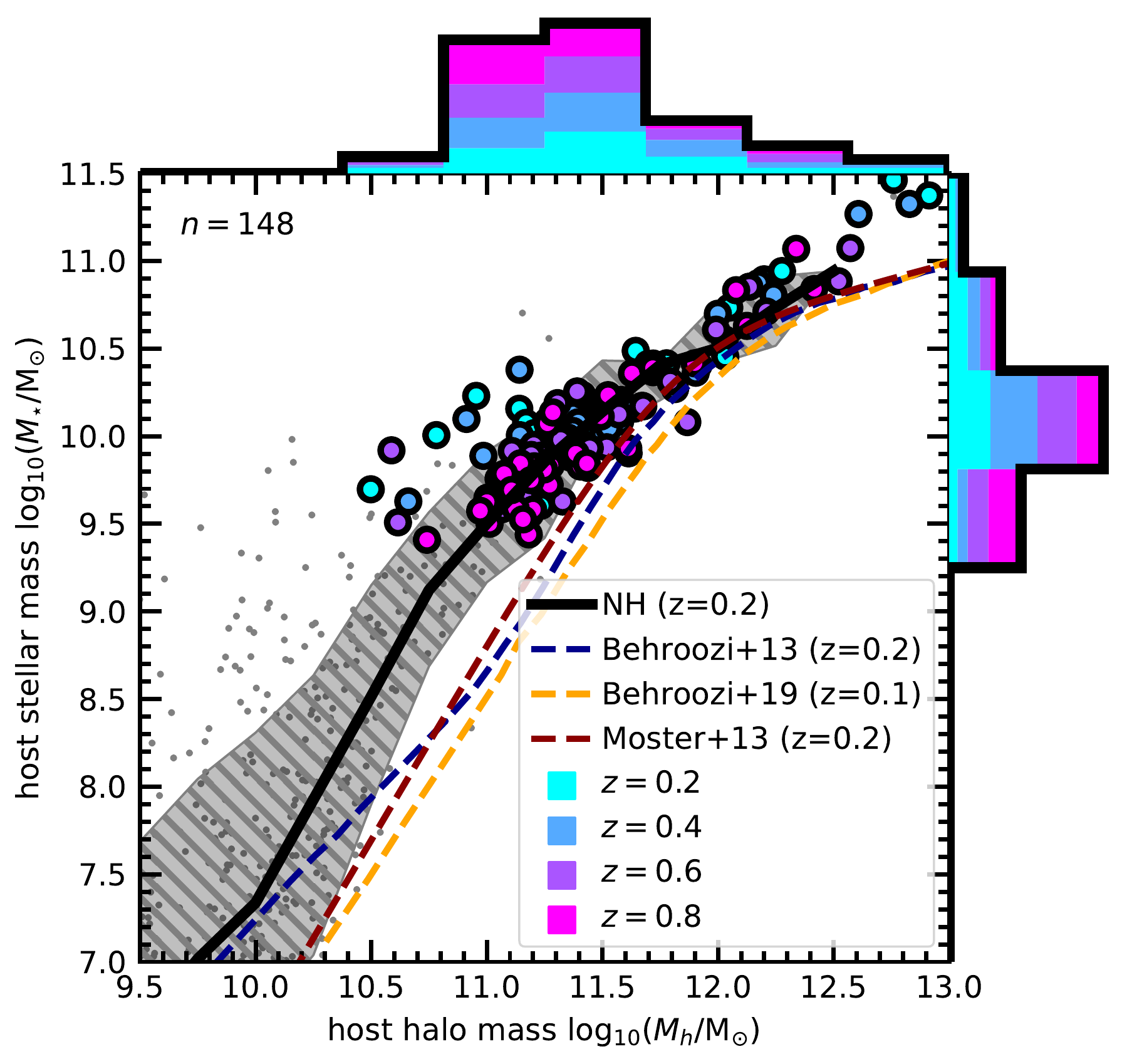}
    \caption{Scatter plot and stacked histograms showing the distribution of host galaxy halo masses and  stellar masses selected at $z=0.2$ along with their progenitors at $z=0.4, 0.6$ and 0.8. There are a total of 148 objects (37 unique objects at 4 different snapshots). Objects in our sample are indicated by coloured points while all other objects are indicated by smaller grey points. The thick black line and shaded region indicate the median halo mass--stellar mass relation and its $1\sigma$ scatter. Also indicated by coloured dashed lines show relations from the literature \citep[][]{Behroozi2013, Moster2013, Behroozi2019}.}
    \label{fig:mass_distribution}
\end{figure}

We select 30 host galaxies with stellar masses greater than $10^{10}$~M$_{\odot}$ with a supplementary sample of 7 host galaxies with stellar masses of $10^{9.5}$~M$_{\odot}<M_{\star}<10^{10}$~M$_{\odot}$, which were selected to better probe trends for lower mass galaxies. We do not include any galaxies whose haloes are contaminated by low-resolution dark matter particles from outside of the high-resolution zoom region. In total, this sample consists of 37 objects at $z=0.2$\footnote{The lowest redshift to which the simulation had been run at the time of analysis.} and their progenitors at $z=0.4, 0.6$ and $0.8$ giving a total of 148 objects across 4 different redshifts. Figure \ref{fig:mass_distribution} shows the stellar mass ($M_{\star}$) and halo mass ($M_{h}$) distribution of host galaxies in our sample presented as a scatter plot and stacked histograms for each redshift. All galaxies in our sample are resolved with a minimum of $\sim250,000$ star particles and an average of $\sim10^{6}$ star particles. We select galaxies only based on the criteria above, making no attempt to preferentially select galaxies with prominent tidal features. Apart from environmental bias due to the size of the simulated volume (see Section \ref{sec:NH}), the sample presented in this paper is therefore unbiased with respect to accretion history and representative of the intermediate and high mass populations found in the simulation at low-to-intermediate redshift as a whole.

The thick black line and filled region indicate the median halo mass--stellar mass relation and its $1\sigma$ scatter at $z=0.2$.  While there is good qualitative agreement for more massive central haloes compared to best fit semi-empirical relations from \citet[][]{Behroozi2013} and \citet[][]{Moster2013} and compared to the empirical model of \citet[][]{Behroozi2019}, below the knee of the relation there is significant overestimation in baryon conversion efficiency. This is at least partially a consequence of the limited volume of \textsc{NewHorizon}, which lacks clusters, rich group haloes and very rarefied environments. Note that in denser environments the median halo mass--stellar mass relation tends towards lower star formation efficiencies -- possibly a result of earlier formation times driving more efficient feedback and more self-regulation or by environmental quenching \citep[e.g.][]{GarrisonKimmel2019,Samuel2022}. Better agreement with observations (an improvement of 0.3 to 0.6 dex at halo masses smaller than $10^{11}$~M$_{\odot}$) is achieved when we weight our sample to account for the fact that under-dense environments are over-represented in \textsc{NewHorizon} compared with the parent Horizon-AGN simulation. As a result of the discrepancy, the total accreted stellar mass in central haloes is likely overestimated compared with a more representative sample. For a given halo mass, this may result in elevated tidal feature strength or greater quantities of diffuse light around \textsc{NewHorizon} galaxies compared with their observed counterparts.

Another important consideration, which we do not investigate here, is how resolution effects and implementation of sub-grid physics impact the orbital sub-structures that are produced in our synthetic galaxies and their haloes. One example is the over or under production of bars, explored in \citet[][]{Reddish2021}, which could potentially inhibit the detection of tidal features or otherwise result in misclassification. Perhaps more important to this study are the orbits and phase-space correlations of the satellite galaxies that are responsible for producing tidal features \citep[e.g.][]{Pawlowski2021}. We defer a full discussion of agreement with observed quantities and phase-space analysis to an upcoming paper (Uzeirbegovic, in preparation).

\begin{figure*}
    \centering
    \includegraphics[width=0.95\textwidth]{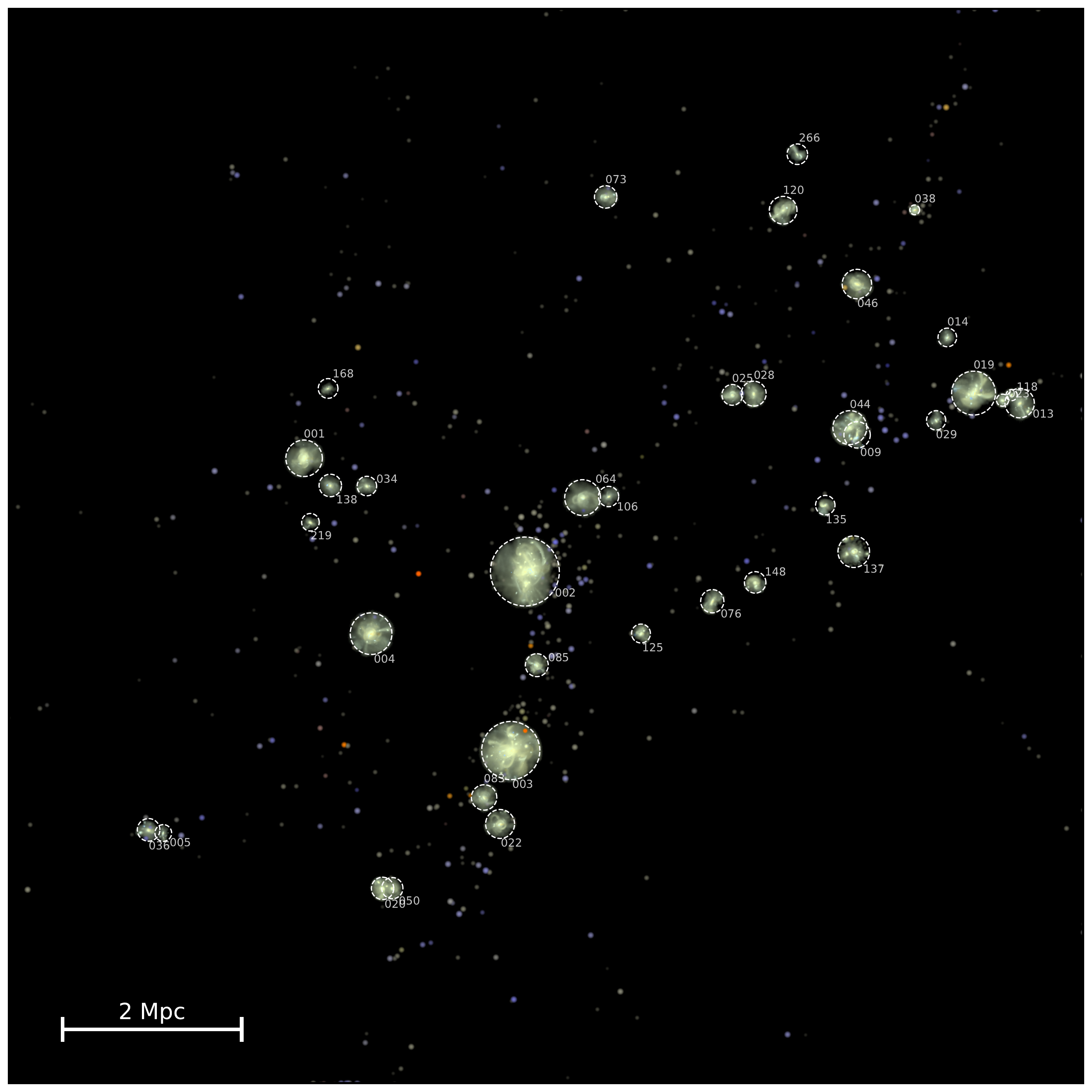}
    \caption{$g$, $r$, $i$ false colour mock image showing the distribution of light within $1~R_{\rm vir}$ of each of the massive galaxies in our sample in the context of the $\sim(16 \,\rm Mpc)^3$ simulation volume at a single simulation snapshot corresponding to $z=0.4$. Dashed circles enclose the virial radius of each object and the numbers indicate their object ID. Less massive objects which we do not use in our sample ($10^{7.5} < M_{\star}/{\rm M_{\odot}}<10^{9.5}$) are indicated by coloured points with the brightness and colour corresponding to mass (brighter objects are more massive) and specific star formation rate (bluer objects are more star forming) respectively.}
    \label{fig:collage}
\end{figure*}

In Figure \ref{fig:collage}, we present $g$,$r$,$i$ false colour images of each object in our sample for the snapshot corresponding to a redshift of $z=0.4$ in the context of the larger cosmic structure and with the same scale. The distribution of LSB structure is shown out to $1~R_{\rm vir}$ for each galaxy, with every other galaxy in the simulation with $10^{7.5} < M_{\star}/{\rm M_{\odot}} < 10^{9.5}$ shown as a point source whose brightness and colour correspond to their mass and specific star formation rate respectively. The images are stretched so that black corresponds to 35 mag arcsec$^{-2}$. The process of producing these images is described in the next section.

\subsection{Mock images}
\label{sec:mock_images}

The analysis of mock observations \citep[e.g.][]{Jonsson2006,Naab2014,Choi2018,Baes2020,Olsen2021} is the most direct method of comparing models and making predictions based on theoretical or synthetic data. In the following section we describe how we produce Rubin-like mock images for each of the galaxies in our sample.

\begin{figure*}
    \centering
    \includegraphics[width=0.95\textwidth]{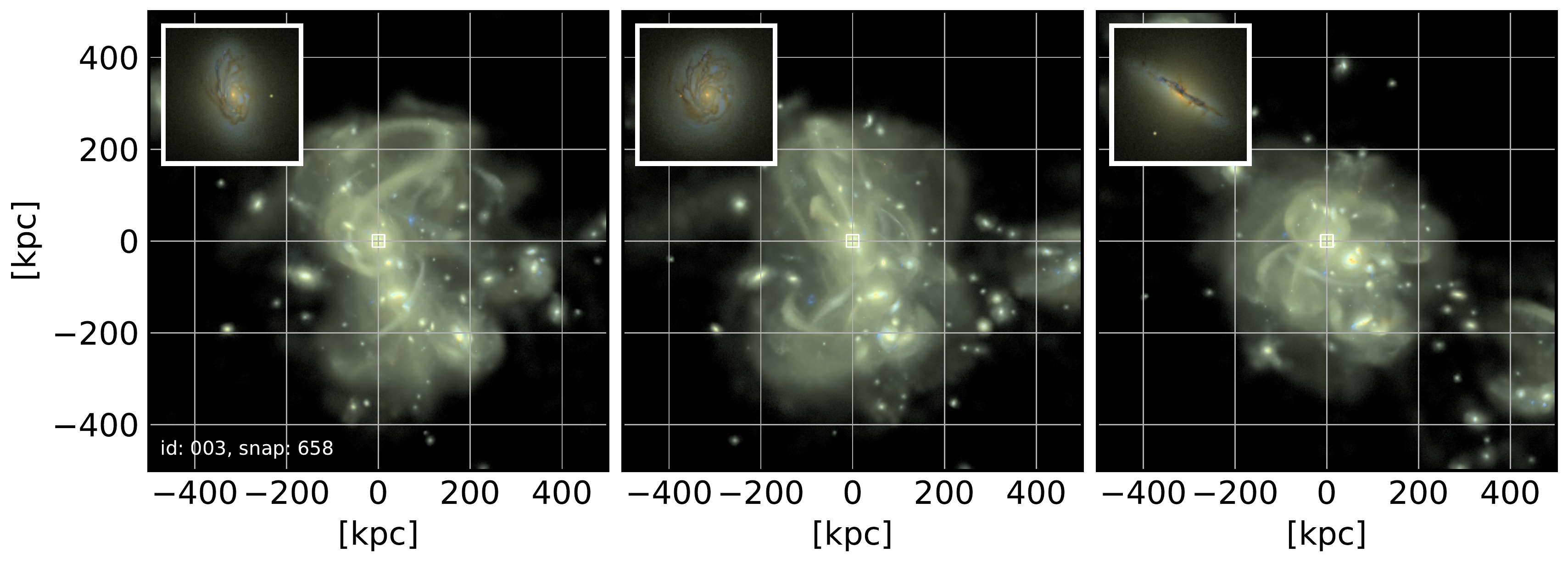}
    \caption{$g, r, i$ false colour mock images of the same 1~Mpc field. From left to right, the panels show projections in $xy$, $xz$ and $yz$. The total stellar mass enclosed in the image is $3.2\times10^{11}\,$M$_{\odot}$ with the host accounting for $2.9\times10^{11}\,$M$_{\odot}$ (90 per cent). Images are produced using the arcsin stretch scheme from \citet{Lupton2004} with the $g, r$ and $i$ bands rescaled by a factor of 1, 0.7 and 0.5 respectively, black corresponds to surface brightnesses fainter than $\sim 35$~mag\,arcsec$^{-2}$. Inset panels show an enlarged image of the host galaxy produced using only particles detected by \textsc{AdaptaHOP} as part of the primary structure. The corresponding location and scale of the inset plot ($\sim 50$ kpc) is indicated by the small white box at the centre of each panel. An interactive version of this plot showing multiple examples can be found at \href{https://garrethmartin.github.io/files/example_images.html}{garrethmartin.github.io/files/example\_images.html}. A video showing this object rotated through multiple projections and with different limiting surface brightnesses can be found at \href{https://www.youtube.com/watch?v=hZg_5FbnnyE}{youtube.com/watch?v=hZg$\_$5FbnnyE}. At such high limiting surface brightnesses, the morphology of the extended light can appear radically different depending on projection.}
    \label{fig:mock_image}
\end{figure*}

\subsubsection{Star particle fluxes}
\label{sec:raw_mock_images}

We produce mock images by first extracting star particles in a (1~Mpc)$^{3}$ cube centred around each host galaxy. Spectral energy distributions (SEDs) for each star particle are calculated from a grid of \citet[][BC03 hereafter]{Bruzual2003} simple stellar population (SSP) models interpolated to the age and metallicity of each star particle. We assume a single \citet[][]{Salpeter1955} IMF for all objects\footnote{Note that, for the purposes of calculating stellar feedback and mass loss, the \textsc{NewHorizon} simulation assumes a \citet[][]{Chabrier2003} IMF \citep[see Section 2.4. of][]{Dubois2021}.}. If we instead consider a \citet[][]{Chabrier2003} IMF, this changes the brightness of the central galaxy and its tidal features roughly equally so that they are both become slightly brighter overall. There is not, therefore, any qualitative impact on our results other than to increase surface brightnesses by roughly 0.6 mag arcsec$^{-2}$ (or equivalently reducing the limiting surface brightness by the same amount) with negligible scatter introduced. Changing the IMF from \citet[][]{Salpeter1955} to \citet[][]{Chabrier2003} confers a less than a 2 per cent change in the quantities presented in Section \ref{sec:tidal_flux}.

We account for the effects of dust via a dust screen model in front of each star particle, so that the dust column density in each AMR gas cell is given by:
\begin{equation}
N_{\rm cell} = \rho\, Z\,  \Delta r \times {\rm GDR},
\end{equation}
where $\rho$ is the gas density of the cell, $Z$ is the metallicity, $\Delta r$ is the length of the cell along a given line of sight and ${\rm GDR}$ is the gas-to-dust ratio, for which we assume a value of 0.4 \citep[e.g.][]{Draine2007}. The total column density in front of each star particle, $N$, is calculated by summing along the line-of-sight. By calculating dust attenuation separately for each particle, we ensure that the geometry of the spatial distribution of dust within and around the galaxy is taken into account. Note that, since we focus on the outskirts of galaxies where there is very little gas and dust, modelling dust attenuation is only relevant for observational predictions for the flux of the host galaxy.

Using the $R = 3.1$ Milky Way dust grain model of \citet[][]{Weingartner2001}, we then produce the dust attenuated SED:
\begin{equation}
I({\lambda})_{\rm attenuated} = I(\lambda) e^{-\kappa(\lambda) N},
\end{equation}
where $I(\lambda)$ is the SED's luminosity density as a function of wavelength and $\kappa(\lambda)$ is the dust opacity as a function of wavelength from \citet[][]{Weingartner2001}. The luminosity of each star particle is calculated by first summing the resultant luminosity of the attenuated SEDs once they have been redshifted and convolved with the LSST $u$, $g$ ,$r$ ,$i$, $z$ and $y$ bandpass transmission functions \citep{Olivier2008}. The apparent magnitude of each star particle is calculated taking into account mass loss from stellar winds and the distance modulus.  

\subsubsection{Smoothing}
Where the density of star particles falls below a few particles per $0.2^{\prime\prime}$ pixel of the Rubin Observatory LSSTCam, it is necessary to apply smoothing in order to better represent the distribution of stellar mass in phase space and remove unrealistic variation between adjacent pixels (usually only an issue in the extreme outskirts of galaxies). To achieve this, we use an adaptive smoothing scheme\footnote{The adaptive smoothing code used in this paper is available from \href{https://github.com/garrethmartin/smooth3d}{github.com/garrethmartin/smooth3d}} following a similar procedure to the \textsc{adaptiveBox} method employed by \citet[][]{Merritt2020}.

We first create a super-sample from the original star particles by splitting them into a large number of smaller particles and then distribute them according to the local density as follows:
\begin{enumerate} 
    \item Calculate the distance to the 5th nearest neighbour for each star particle, $d_{k=5}$.
    \item Split each star particle into 500 equal flux particles whose positions are drawn from a Gaussian distribution about the centre of the original particle and with a standard deviation equal to $d_{k=5}$ such that $P(x,y,z)\sim\mathcal{N}([x_{0},y_{0},z_{0}], \sigma=d_{k=5})$.
    \item Create a 2D image by collapsing the particles along one of the axes and summing the flux across a 2D grid with elements of $0.2^{\prime\prime}\times0.2^{\prime\prime}$.
\end{enumerate}

Figure \ref{fig:mock_image} shows an example of a false colour smoothed mock image for one of our simulated galaxies in 3 different projections. In these images, black corresponds to a surface brightness fainter than  $\sim 35$~mag\,arcsec$^{-2}$. At very low surface brightnesses (significantly in excess of those currently accessible), almost all objects in our sample display multiple distinct tidal features, often with complex morphologies. Viewed at different angles, the shape and number of visible tidal features can change radically. Examples of additional objects in different projections can be seen in a supplementary interactive version of Figure \ref{fig:mock_image}, found at \href{https://garrethmartin.github.io/files/example_images.html}{garrethmartin.github.io/files/example\_images.html}. We return to the issue of how robustly tidal features are classified in multiple projections later in Section \ref{sec:visual}.

\subsubsection{Mock observations}
\label{sec:making_mocks_class}
For every object we produce smoothed mock images in 3 projections ($xy$, $xz$ and $yz$) and at distances corresponding to a range of redshifts ($z = 0.05$ to $z=0.8$) as described above. Each image is then convolved with a point spread function (PSF)\footnote{The PSF FWHM is always larger than the smoothing length in regions of interest (i.e. for the galaxy and dense tidal features as defined in Section \ref{sec:identification}).}. We use the $g$-band Hyper Suprime-Cam \citep[HSC;][]{Miyazaki2012} 1D PSF measured by \citet[][]{Montes2021}\footnote{Measured to $289^{\prime\prime}$ and extrapolated to $420^{\prime\prime}$ based on a power law fit.} as we find that a Gaussian or Moffat distribution do not adequately describe the shape of the PSF at large radii (see Appendix \ref{sec:PSF_visibility} for discussion of the suitability of various PSF models and the possible effect of the PSF on our ability to detect tidal features). We note that the full width at half maximum (FWHM) of the PSF measured by \citet[][]{Montes2021} is slightly broader than the  expected median FWHM of the Rubin Observatory PSF ($0.7^{\prime\prime}$ in the $r$-band \citep[][]{Ivezic2019} compared to a FWHM $1.07^{\prime\prime}$ in the $g$-band obtained by \citet[][]{Montes2021} for HSC) and therefore slightly overestimates the likely effect of the PSF.

Finally, we add random Gaussian noise, $\mathcal{N}(0, \sigma_{\rm noise})$, where $\sigma_{\rm noise}$ is the standard deviation of the noise per pixel. We assume that the background is perfectly subtracted so that there is no variation in the noise level across the image. In reality, this is not a realistic assumption as the detection of LSB sources is sensitive to a host of factors. These include sky estimation \citep[e.g. see Section 4.1 and Figure 5 of][]{Aihara2019} and observing techniques, how CCD artefacts and biases \citep[e.g.][]{Baumer2017} are accounted for, as well as source extraction and masking methods. We also choose to neglect other instrumental and astrophysical contaminants (e.g. foreground and background objects, Galactic cirrus, scattered light, ghosts and diffraction spikes) which may be present in real imaging. Although it is possible to mitigate some of this contamination \citep[e.g.][]{Slater2009,Fliri2016,Roman2020,Tanoglidis2021b}, visibility will always be reduced under realistic conditions.\footnote{See \citet[][]{Mihos2019} for a review of recent advances and challenges in deep imaging.} Our results therefore represent a best case estimate. Predictions for the LSST final limiting surface brightness vary fairly significantly between $\sim$29.5 mag~arcsec$^{-2}$ (P. Yoachim, private communication) and $\sim$31 mag~arcsec$^{-2}$ \citep[e.g.][]{Laine2018,Brough2020} and up to 32 mag~arcsec$^{-2}$ \citep[][]{Brough2020} in the deep drilling fields and these limits may differ between objects dependent on the severity of sky subtraction bias.

Of course, the final depth achieved by LSST will depend strongly on how well the data are reduced and optimised for LSB science. Given a typical sky brightness in the $r$-band \citep[21.2 mag arcsec$^{-2}$][]{Ivezic2019} and a limiting surface brightness 31 mag arcsec$^{-2}$ requires that the sky background is characterised with an uncertainty greater than 1/10000. The current best
practices \citep[e.g.][]{Ji2018} allow characterisation of the sky background down to 4 parts in one
million (which, in theory, would enable tidal features to be analysed down to $\sim34.5$ mag arcsec$^{-2}$), meaning limiting surface brightnesses greater than 31 mag arcsec$^{-2}$ are achievable if LSST operates, at least in theory.

For comparison, \citet[][]{Kniazev2004} measure an SDSS $r$-band limiting surface brightness\footnote{Note that estimates of the limiting surface brightness are not always directly comparable as they can vary depending on the exact methodology used.} of 26.2 mag arcsec$^{-2}$, and the IAC Stripe82 Legacy Project are able to achieve an $r$-band limiting surface brightness of 28.7 mag arcsec$^{-2}$ in the SDSS Stripe82 calibration area \citep[][]{Roman2017}. Additionally, many contemporary wide-area surveys \citep[e.g.][]{Aihara2018,Dey2019,Gwyn2012} now reach limiting surface brightnesses in excess of 28 mag arcsec$^{-2}$. Some of the deepest imaging currently available corresponds to targeted ground based campaigns \citep[e.g.][]{Fry1999,Cappellari2011a,Abraham2014,Ferreras2014,Duc2015,Trujillo2016,Iodice2016,Merritt2016,Spavone2017,Iodice2019,Mihos2017,Montes2021,Ragusa2021} which can achieve limiting surface brightness of around 30 mag arcsec$^{-2}$ (typically requiring much longer integration times and/or with relatively limited spatial resolution and field of view compared to the Rubin Observatory).

We consider multiple noise levels, which are calculated assuming a range of limiting surface brightnesses ($3\sigma$ in a $10^{\prime\prime}\times10^{\prime\prime}$ box). These are converted to a per-pixel $1\sigma$ variance by rearranging the equation found in \citet[][Appendix A]{Roman2020} (for simplicity, we neglect the zero point):
\begin{equation}
    \sigma_{\rm noise} = \frac{10^{-0.4\, \mu_{r}^{\rm lim}(n\sigma,\Omega\times\Omega)}\,pix\,\Omega}{n},
\end{equation}
\noindent where $\Omega$ is the size of one of the sides of the box in arcseconds, $pix$ is the pixel scale in arcseconds/pixel, $n$ is the number of Gaussian standard deviations and $\mu_{r}^{\rm lim}$ is the $r$-band limiting surface brightness, in this case $\mu_{r}^{\rm lim}(3\sigma,10^{\prime\prime}\times10^{\prime\prime})$. Finally, we perform a $5\times5$ re-binning of the images to an angular scale of $1^{\prime \prime}$. As the FWHM of the PSF is also around $1^{\prime \prime}$, this represents the maximum binning we can perform before the images start to lose fidelity.

For the reasons outlined above, we do not target any specific prediction for image depth but instead opt to explore a range of values for $\mu_{r}^{\rm lim}(3\sigma,10^{\prime\prime}\times10^{\prime\prime})$. We choose values between 28 and 31 mag arcsec$^{-2}$, which roughly encompass expected depths from a single exposure and close to the upper end of predictions of the depth of a full 10-year stacked exposure (825 visits).

For each limiting surface brightness, we produce images assuming different redshifts. While the largest telescopes are capable of producing sufficiently deep imaging of local galaxies ($z<0.01$) with minutes to a few hours of integration time \citep[e.g.][]{Sand2009,Trujillo2021}, the Rubin Observatory will be capable of collecting much larger samples at higher redshifts ($0.05<z<0.1$) thanks to the large survey area of LSST. Therefore, we do not consider the very local Universe, instead picking a range of redshifts starting at $z=0.05$ and extending to high redshift ($z=0.8$), with cosmological redshift and surface brightness dimming taken into account, in order to match the capabilities of the Rubin Observatory and test its ability to resolve tidal features in the more distant Universe.

\begin{figure}
    \centering
    \includegraphics[width=0.45\textwidth]{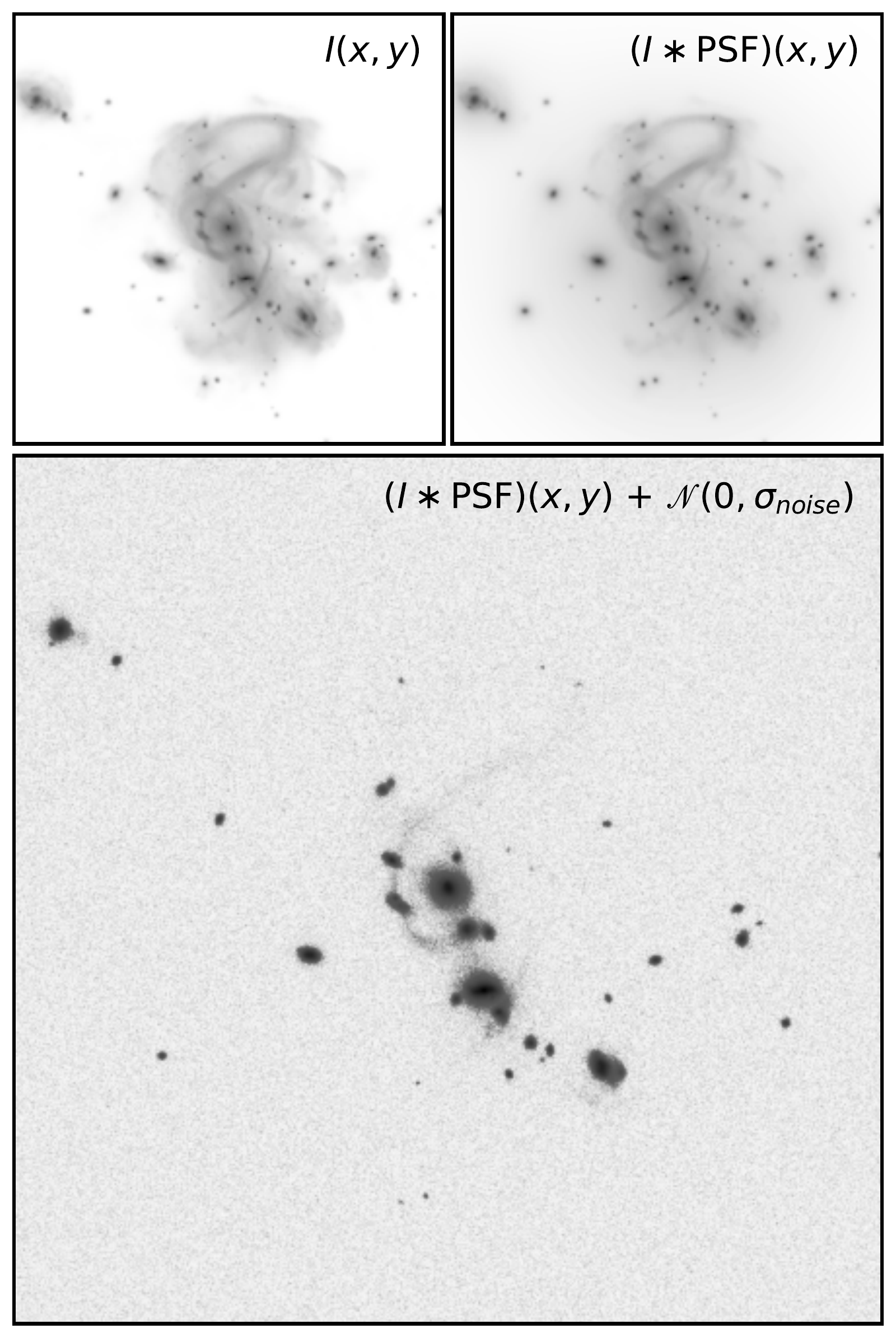}
    \caption{Plot demonstrating the steps taken to produce the mock images used for classification. As in Figure \ref{fig:mock_image}, white corresponds to surface brightnesses fainter than $\sim 35$~mag\,arcsec$^{-2}$. \textbf{Top left}: $r$-band mock image of a galaxy at $z=0.2$ \textbf{Top right}: the image is convolved with the HSC PSF derived by \citet[][]{Montes2021} \textbf{Bottom}: Gaussian random noise is added to the convolved image.}
    \label{fig:making_mocks}
\end{figure}

Figure \ref{fig:making_mocks} illustrates the process of producing a single mock observation in the $r$-band. Moving clockwise from the top left panel we show the original $r$-band mock image created as described in Section \ref{sec:raw_mock_images}, the same image convolved with the PSF and finally, with Gaussian random noise added.

Table \ref{table:parameters} shows the full range of parameters used to produce mock observations, which add up to a total of 60 different variations per object. Values for the limiting surface brightness in brackets indicate the equivalent if we choose a \citet[][]{Chabrier2003} IMF\footnote{A \citet[][]{Kroupa2002} IMF also yields very similar results} instead of a \citet[][]{Salpeter1955} IMF. We also produce an extra set of mock images for $\mu_{r}^{\rm lim}$ = 35~mag\,arcsec$^{-2}$ and $z=0.05$ which we use as a `ground truth' for the other mock observations. This value is informed by the stellar mass resolution of the simulation, since we do not expect tidal features to contain enough particles to produce sufficient signal-to-noise at stellar mass surface densities equivalent to $\sim35$~mag\,arcsec$^{-2}$. Visual inspection confirms that we do not visually detect additional structures in the diffuse light beyond $35$~mag\,arcsec$^{-2}$. Given the finite resolution of the simulation and the hierarchical nature of galaxy assembly, it is likely that additional tidal features would become visible with finer resolution (see also Appendix \ref{sec:res_limit}).

\begin{table}
\caption{Range of parameters used to generate mock observations: \textit{a}, limiting $r$-band surface brightness for a $10^{\prime\prime}\times10^{\prime\prime}$ box with bracketed values showing the equivalent for a \citet[][]{Chabrier2003} IMF; \textit{b}, redshift corresponding to viewing distance; \textit{c}, axis of projection; \textit{d}, point spread function}
\begin{tabular}{@{}llllll@{}}
\toprule
Parameter & Values \\
\midrule
$\mu_{r}^{\rm lim}(3\sigma,10^{\prime\prime}\times10^{\prime\prime})\,^{a}$ & $[28 (27.43), 29 (28.43), 30 (29.43), 31 (30.43)]$ \\
$z\,^{b}$                    & $[0.05,0.1,0.2,0.4,0.8]$ \\
$\pi\,^{c}$                  & $[xy, xz, yz]$   \\
PSF$\,^{d}$ & \citet[][]{Montes2021}  \\ \bottomrule
\end{tabular}
\label{table:parameters}
\end{table}

\subsection{Measuring tidal features}
\label{sec:identification}

We perform a separate measurement of the galaxy tidal features based on the distribution of particles in the simulation. In order to do this, we separately define tidal features or tidal debris as any star particles within the $(1~{\rm Mpc})^{3}$ volume that are not detected as part of an object or sub-structure by \textsc{AdaptaHOP} (see Section \ref{sec:sample}). We define \textit{tidal features}, as opposed to more diffuse \textit{tidal debris} based on a $k=5$ nearest neighbour stellar mass weighted density\footnote{$\rho_{\star} = \sum_{i=1}^{k}m_{\star,i} \frac{3}{4 \pi d_{k}^{3}}$} threshold.

In order to determine the optimum density threshold we first assume that tidal features have a higher spatial frequency on average than the diffuse component, so that increasing the density threshold until the high frequency component is minimised should allow us effectively segregate features above and below a certain spatial frequency threshold. Based on the measurements of \citet[][]{Sola2022}, we choose a threshold value of $50~$kpc, which is a little larger than the average width of the largest tidal tail measured. To determine the density threshold corresponding to this spatial frequency, we calculate the 2-D Fourier transform of images produced at decreasing density thresholds until the power at frequencies smaller than $50~$kpc approaches a minimum value \citep[essentially applying a high-pass filter as in e.g.][]{Popesso2012}, arriving at an optimum density of $\rho_{\star} > 500~{\rm M_{\odot}\, kpc^{-3}}$.

Using these definitions we identify the following regions:

\begin{itemize}
    \item \textit{structure} $(\mathbf{S})$ -- members of any structure or sub-structure found by \textsc{AdaptaHOP}
    \item \textit{host} $(\mathbf{H})$ -- members of the structure identified as the host galaxy
    \item \textit{tidal debris} $(\mathbf{T})$ -- not members of any structure or sub-structure and where $\rho_{\star} < 500~{\rm M_{\odot}\, kpc^{-3}}$ 
    \item \textit{dense tidal features} $(\mathbf{T_{d}})$ -- not members of any structure or sub-structure and where $\rho_{\star} > 500~{\rm M_{\odot}\, kpc^{-3}}$ 
\end{itemize}

\begin{figure*}
    \centering
    \includegraphics[width=0.95\textwidth]{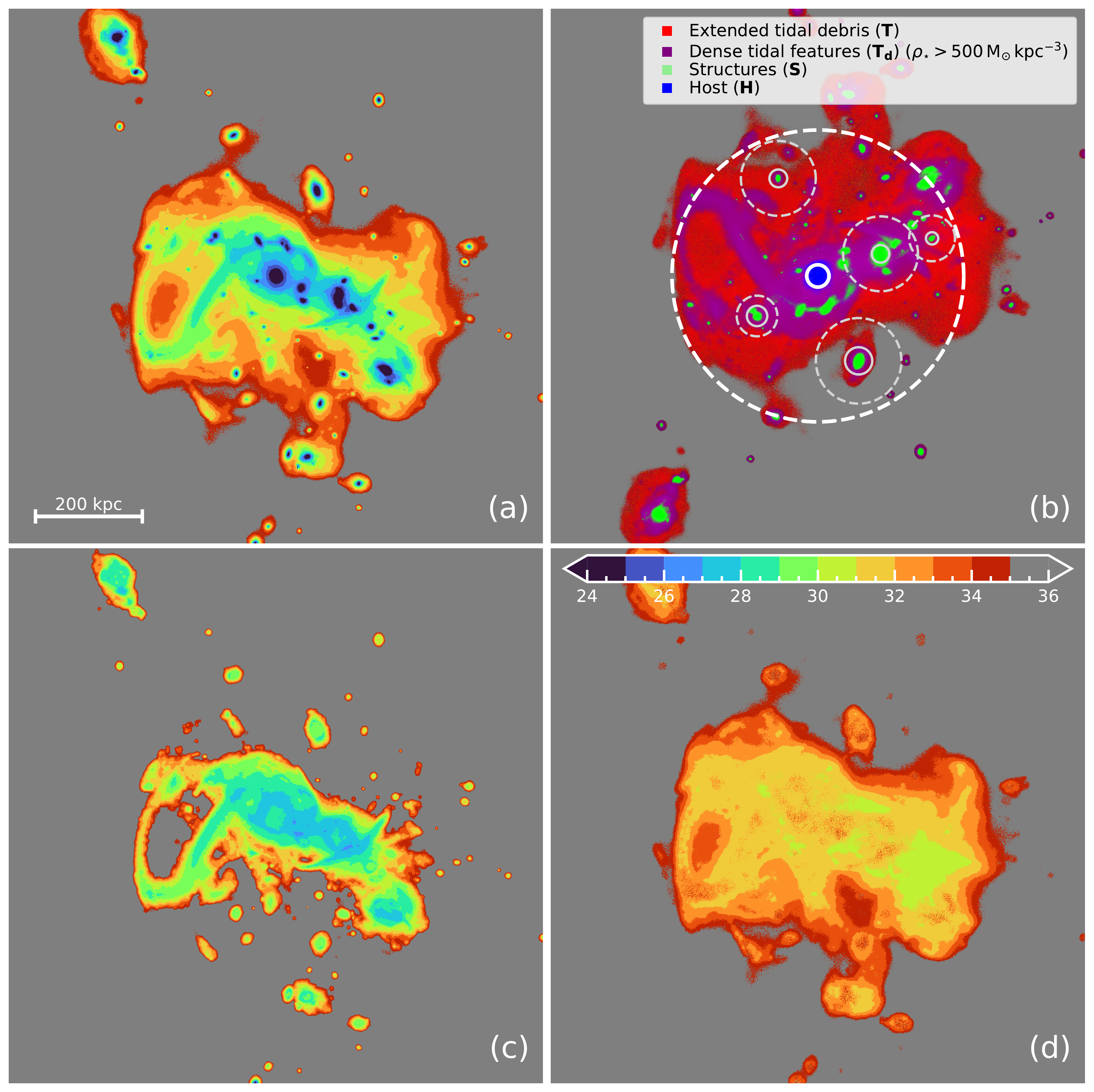}
    \caption{\textbf{(a)} $r$-band surface brightness map of one of the same mock fields as shown in Figure \ref{fig:mock_image}.
    \textbf{(b)} A plot showing the same field with colours corresponding to tidal debris (\textbf{T}, red), dense tidal features ($\mathbf{T_{d}}$, purple), structures identified by \textsc{AdaptaHOP} (\textbf{S}, green) and the central host galaxy (\textbf{H}, blue) with an equivalent stretch to the first image. The solid and dashed white circles indicate $7~R_{\rm eff}$ and $R_{\rm vir}$ of the host respectively. Grey circles indicate the same for the most massive satellites.
    \textbf{(c)} $r$-band surface brightness map created from particles identified as dense tidal features, where $\rho_{\star} > 500$~M$_{\odot}\,$kpc$^{-3}$ (typical surface brightnesses of $\mu_{r} \sim 28-32$~mag$\,$arcsec$^{-2}$).
    \textbf{(d)}: surface brightness map created from particles identified as tidal debris where, $\rho_{\star} < 500$~M$_{\odot}\,$~kpc$^{-3}$ (typical surface brightnesses of $\mu_{r} \gtrsim 32$~mag$\,$arcsec$^{-2}$).
Image scale in proper kpc is indicated by the scale bar in the bottom left of panel \textbf{(a)} and the colour bar at the top of panel \textbf{(d)} indicates the scale of the surface brightness maps in mag arcsec$^{-2}$. Pixels fainter than 35 mag$\,$arcsec$^{-2}$ or brighter than 24~mag$\,$arcsec$^{-2}$ are coloured grey and dark blue respectively.}
    \label{fig:residual_image}
\end{figure*}

Using these definitions, we can again calculate star particle fluxes as described in Section \ref{sec:raw_mock_images} and create smoothed images using only the star particles identified as making up tidal features or tidal debris. Figure \ref{fig:residual_image} shows an example of this process: the surface brightness map created using all star particles is shown in the top left and the tidal feature and tidal debris surface brightness maps (panels \textbf{c} and \textbf{d}) are produced from particles that are identified in panel \textbf{b} in purple (dense tidal features; \textbf{c}) and red (extended tidal debris; \textbf{d}).

Our definition of tidal debris includes the diffuse material and debris in the outskirts of satellite galaxies. Their contribution to the total tidal flux is typically small -- for the most massive 20th percentile of haloes, we find that, of the total tidal flux found within $1~R_{\rm vir}$ of the host galaxy, a median / mean of 7 / 8 cent and 24 / 25 per cent is found within $5~R_{\rm eff}$ and $10~R_{\rm eff}$ of its satellite galaxies respectively. For any individual halo this value never exceeds 15 per cent or 50 per cent for $5~R_{\rm eff}$ and $10~R_{\rm eff}$ respectively (larger values come from systems with an ongoing major or minor merger where the host and one of its minor companions are close). The proportion of tidal flux contained in satellites decreases further for less massive haloes.

\subsection{Merger trees}

We construct merger trees for each galaxy according to the method of \citet[][]{Tweed2009} based on the stellar particles of galaxies identified using \textsc{AdaptaHOP}. The time resolution of the merger trees is $\sim$15~Myr, enabling us to track in detail the main progenitors (the object in the chain of the most massive progenitors at each snapshot) and mass assembly of each galaxy. We follow the stellar mass evolution and stellar mass accretion history of the host galaxy, which in turn allows us to determine the merger history and ex-situ mass fraction of the galaxy.

We identify stars as ex-situ by iterating along the main branch of the merger tree. For each snapshot, star particles which were formed after the previous snapshot and which are identified by \textsc{AdaptaHOP} as members of the host galaxy in the current snapshot are marked as in-situ, then at the final snapshot $t_{\rm max}$, any star particles identified by \textsc{AdaptaHOP} as members of the host galaxy which are not marked as in-situ are considered to have formed ex-situ. The ex-situ mass fraction, $f_{\rm \rm exsitu}$, between the current time, $t_{\rm max}$, and some previous time, $t_{\rm min}$, can then be defined for for each host galaxy as follows:

\begin{equation}
    f_{\rm exsitu}(t_{\rm max}, t_{\rm min}) = \frac{\sum\{m_{\star}\, |\, t_{\rm min} < t_{\rm birth} < t_{\rm max}\ \land\ {\rm ex-situ} \}}{\sum\{m_{\star}\, |\, t_{\rm min} < t_{\rm birth} < t_{\rm max}\}},
    \label{eqn:exsitu}
\end{equation}

\noindent where $m_{\star}$ is the set of particles all identified by \textsc{AdaptaHOP} at $t_{\rm max}$, $\{m_{\star}\, |\, t_{\rm min} < t_{\rm birth} < t_{\rm max}\}$ is the subset of these particles with formation times, $t_{\rm birth}$, between $t_{\rm min}$ and $t_{\rm max}$ and similarly, $\{m_{\star}\, |\, t_{\rm min} < t_{\rm birth} < t_{\rm max}\ \land\ {\rm ex-situ} \}$ is the sub-set of all these particles formed between $t_{\rm min}$ and $t_{\rm max}$ which were formed ex-situ. Using this parameter, we are able to quantify how the visibility of tidal features correlates with the accretion history of each system. Throughout the rest of this paper we adopt a value of $t_{\rm min}$ equal to the earliest time that the main progenitor can be traced so that $f_{\rm exsitu}$ encompasses the entire lifetime of the object. As we discuss in Appendix \ref{sec:exsitu_interval}, increasing or reducing $t_{\rm min}$ does not have a statistically significant effect on either the correlation between $f_{\rm exsitu}$ and halo mass or the correlation between $f_{\rm exsitu}$ and tidal flux fraction, $f_{\rm tidal}$ (see Section \ref{sec:tidal_flux}).


Our definition of $f_{\rm exsitu}$ includes only stars that at are identified as part of the host galaxy by \textsc{AdaptaHOP} (i.e. parts of panel (b) of Figure \ref{fig:residual_image} colour coded in blue), so the set of particles used to calculate the ex-situ mass and tidal flux are mostly mutually exclusive. $f_{\rm exsitu}$ should be considered as a measure of the ex-situ mass fraction within the central galaxy itself rather than of the entire system including the extended envelope.

\subsection{Visual classification}
\label{sec:visual_classificatuion}

\begin{figure*}
    \centering
    \includegraphics[width=0.95\textwidth]{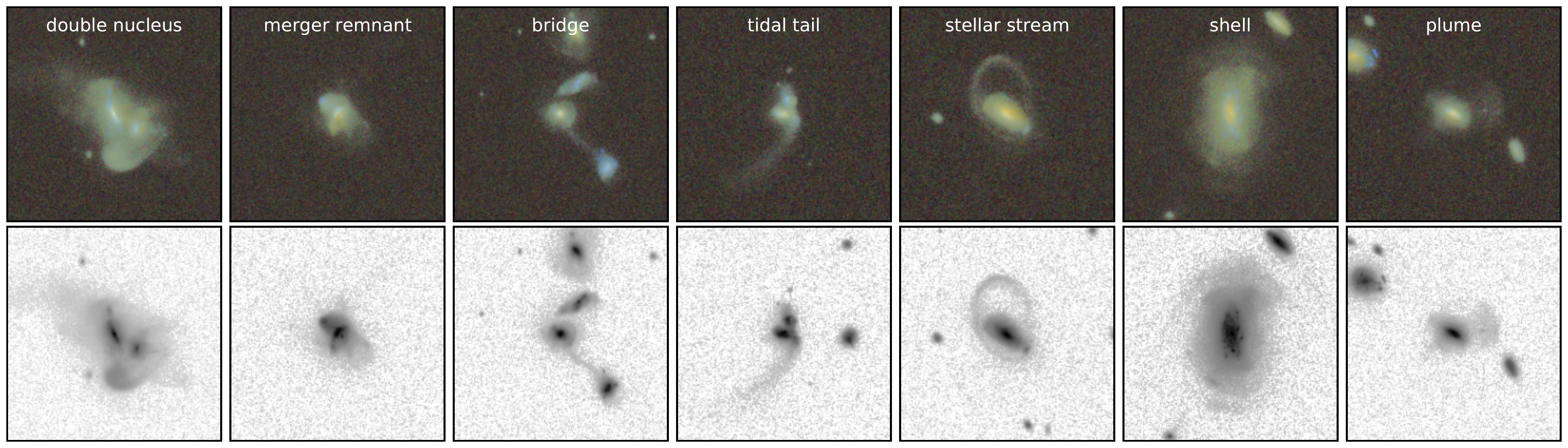}
    \caption{Example 100 Mpc $\times$100 Mpc thumbnail images for each category. For each object, we show the arcsinh stretched $g,r,i$ false colour image (top row) and greyscale surface brightness map (bottom row).}
    \label{fig:categories}
\end{figure*}

Except for images deemed too noisy to effectively classify\footnote{All objects had sufficient signal-to-noise to make classification possible except at $z=0.8$, where only half of objects at $\mu_{r}^{\rm lim}(3\sigma,10^{\prime\prime}\times10^{\prime\prime}) = 30-31$~mag\,arcsec$^{-2}$ were classifiable and almost no objects at $\mu_{r}^{\rm lim}(3\sigma,10^{\prime\prime}\times10^{\prime\prime}) = 28-29$~mag\,arcsec$^{-2}$ were classifiable.}, which were rejected based on visual inspection, we perform visual classifications for all of the permutations of each object (totalling $\sim 8000$ unique images). Images were shared among 45 expert classifiers so that each permutation was classified separately by at least two people. A subset of $\sim 600$ images were classified 5 times each in order to more robustly measure variation between classifiers. To maximise reproducibility, all classifiers were asked to follow a set of detailed instructions that included prototypical examples of each class of tidal feature and asked to classify around 300 individual images. For each permutation, classifiers viewed a set of 3 greyscale surface brightness maps with a fixed noise level but with a maximum stretch set to 27, 29 and 33 mag arcsec$^{-2}$ and an arcsinh stretched $g,r,i$ false colour image which were cropped to a physical size of 100 kpc $\times$ 100 kpc. They were asked to count the number of instances that they identified certain categories of tidal feature with various signatures of mergers and interactions considered -- interacting galaxies with double nuclei, merger remnants, bridges, tidal tails, stellar streams, shells and plumes \citep[e.g.][]{Lofthouse2017, Bilek2020}.  Classifiers were asked to make their determinations according to the following criteria, taking into account the surrounding context and their physical interpretation of the image:
\begin{itemize}
    \item \textit{Stellar streams} -- A stream of stars that can appear to trace an ellipse, spiral or straight line depending on the angle at which they are viewed.
    \item \textit{Tidal tails} -- Differing from stellar streams in that a tidal tail must originate from material ejected from the host galaxy or a massive companion. They are therefore likely to be associated with interactions between similar mass galaxies and consist of material that has been unbound from a disrupted galaxy rather than gradually stripped.
    \item \textit{Asymmetric stellar haloes} -- Diffuse, low surface brightness features in the outskirts of the galaxy that do not have a well defined structure like stellar streams or objects where the overall structure of the diffuse stellar halo is clearly not symmetric.
    \item \textit{Shells} -- radial structures consisting of concentric arcs or ring-like structures that do not cross one another.
    \item \textit{Tidal bridges} -- a tidal tail that connects two interacting galaxies. Bridges are formed from the material that gets dragged out during high mass ratio mergers, rather than material has been gradually stripped away over many orbits.
    \item \textit{Merger remnants} -- A strongly morphologically disturbed galaxy where the merging objects have recently coalesced. May also exhibit tidal tails but there should be no indication of a second galaxy.
    \item \textit{Double nuclei} -- visibly merging / interacting with both objects still clearly separated. There must be some sign of an interaction taking place (i.e. not only close pairs).
\end{itemize}
Examples of each of these categories are shown in Figure \ref{fig:categories}.


In comparison to simpler categorisation schemes, such as separating galaxies into elliptical and spiral morphologies \citep[e.g.][]{Lintott2008,Uzeirbegovic2020}, a higher level of domain knowledge is required to perform detailed classifications of galaxy tidal features. This stems from the fact that correctly interpreting tidal features can depend on context and is often reliant on a physical understanding of the interactions taking place. For example, stellar streams can form shell-like morphologies \citep[][]{Foster2014} or tidal tails may appear similar to linear streams when observed edge-on. Even for expert classifiers, characterising tidal features can still be quite uncertain, especially at high redshifts or at low limiting surface brightness, which both significantly alter the appearance of tidal features. The reliability of visual classifications is discussed in Section \ref{sec:visual}.

\section{Properties and detectability of the tidal features and diffuse light around galaxies}
\label{sec:properties}
\subsection{Quantifying the distribution and strength of tidal features}

In this section we consider how light is distributed around galaxies and their extended envelopes. We explore the distribution of light as a function of surface brightness in the main body of galaxies and in their extended envelopes as well as the total fraction of light that makes up different regions of the galaxy. In Appendix \ref{sec:z_evo}, we show that the mass accretion histories of the galaxies in our sample are sufficiently stochastic that any average evolution between $z=0.2$ and $z=0.8$ is largely washed out. Based on this finding, we treat each instance of the same galaxy across the 4 snapshots considered as independent objects. If we restrict our analysis only to galaxies at a single snapshot rather than include their progenitors at different redshifts, we do not see a notable difference in our results.

\subsubsection{Distribution of light in extended structures}

\begin{figure}
    \centering
    \includegraphics[width=0.45\textwidth]{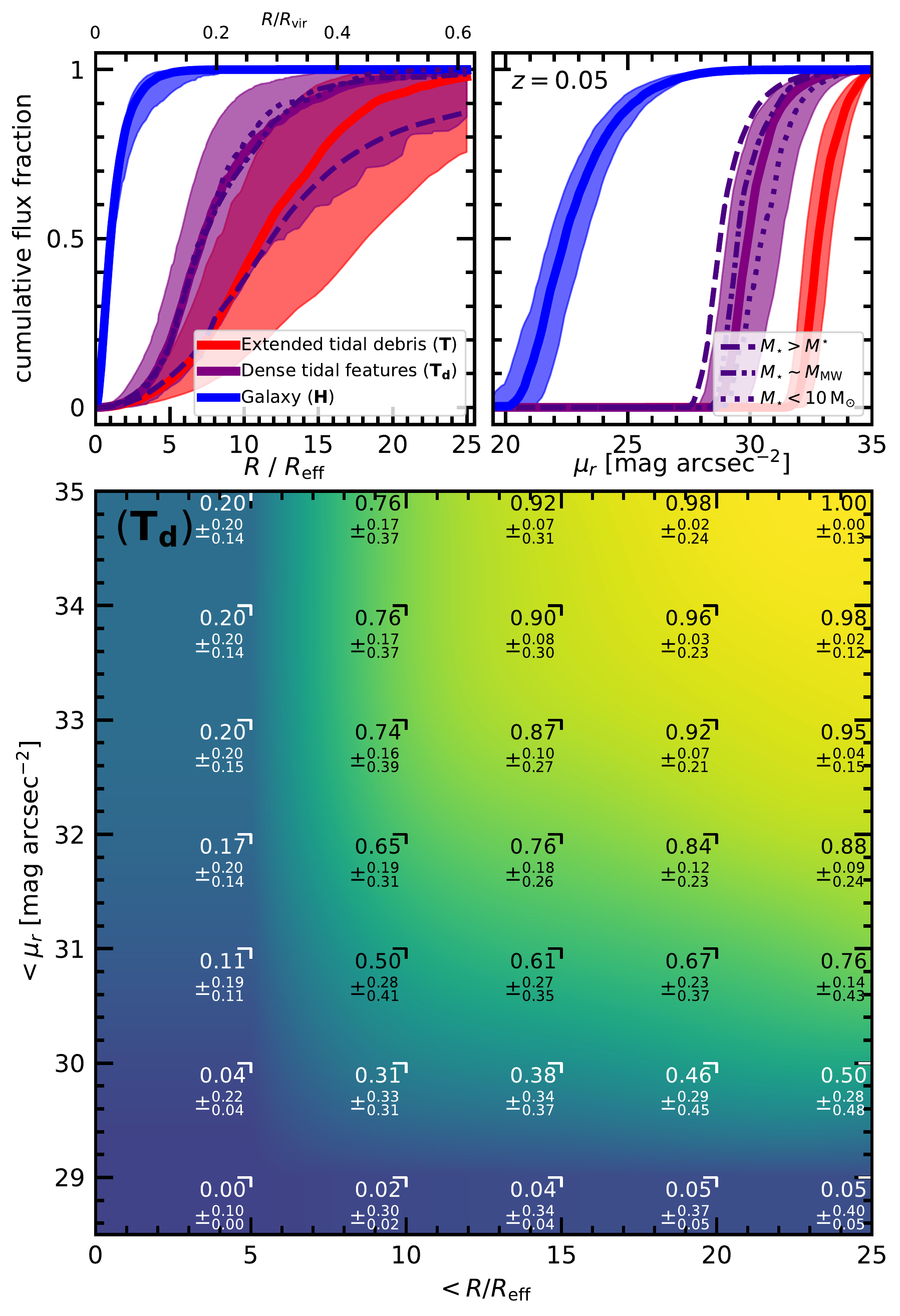}
    \caption{\textbf{Top left}: The median cumulative fraction of flux as a function of projected radius ($\mu_{r} < 35, R < R_{\rm vir}$) for the host galaxy ($\mathbf{H}$, blue), dense tidal features ($\mathbf{T_{d}}$, purple) and extended tidal debris ($\mathbf{T}$, red). Coloured filled regions indicate the central 68th percentile ($1\sigma$) of the distribution. The top $x$-axis gives an indication of the approximate equivalent value of $R/R_{\rm vir}$ corresponding to the value of $R/R_{\rm eff}$ (based on median measured ratio for $R_{\rm vir}$ to $R_{\rm eff}$ of $1:40$.) shown on the bottom $x$-axis. \textbf{Top right}: The median cumulative fraction of flux ($\mu_{r} < 35, R < R_{\rm vir}$) as a function of surface brightness. For both panels, median dense tidal feature profiles for 3 different mass ranges are shown with corresponding line styles indicated in the legend of the top right panel \textbf{Bottom}: colour plot showing the joint distribution of the cumulative flux fraction for dense tidal features as a function of projected radius and surface brightness. Values shown indicate the median and $1\sigma$ scatter of the cumulative flux fraction where brackets to the top right of each value indicate the maximum radius and surface brightness that the cumulative flux fraction is calculated within. Typically, flux from extended tidal debris is distributed towards larger radii and lower surface brightnesses than the dense tidal features, although there is significant scatter in the distribution of both components. In some cases, a non-negligible fraction of flux lies even beyond 25~$R_{\rm eff}$.}
    \label{fig:detection_grid}
\end{figure}

We first discuss how tidal feature flux is distributed, spatially and as a function of surface brightness.

The top left and right panels of Figure \ref{fig:detection_grid} show the distribution of flux as a function of the galaxy's 2-d effective radius and of surface brightness respectively. In both cases, galaxies are observed at a redshift of $z=0.05$ and an angular resolution of $0.2^{\prime\prime}$. The median fraction of cumulative flux contained within pixels brighter than 35~mag$\,$arcsec$^{-2}$ and $R < R_{\rm vir}$ is shown separately for the host galaxy $(\mathbf{H})$, dense tidal features $(\mathbf{T_{d}})$ and extended tidal debris $(\mathbf{T})$, indicated by thick blue, purple and red lines respectively. Within the host galaxy itself, the majority of pixels lie at low surface brightnesses (50 per cent fainter than 27~mag$\,$arcsec$^{-2}$) and small radii $(R<5~R_{\rm eff})$. Median profiles for the dense tidal features are plotted for 3 different mass ranges with $M_{\star}>M^{\star}$ $(M_{\star}>10^{10.8}\,{\rm M_{\odot}}$), Milky Way mass galaxies ($10^{10.25}{\rm M_{\odot}}<M_{\star}<10^{10.75}{\rm M_{\odot}}$) and low-mass galaxies ($M_{\star}<10^{10}{\rm M_{\odot}}$) indicated by dashed, dash-dotted and dotted purple lines respectively.

Typically, a majority of the flux from dense tidal features is found at smaller radii than extended tidal debris, with 50 per cent of flux contained within 7~$R_{\rm eff}$ compared with 10~$R_{\rm eff}$ for extended tidal debris. There is significant scatter in the cumulative flux distributions for both  dense tidal features and extended tidal debris, meaning that, in many cases, a majority of the tidal flux lies at a large separation from the central galaxy. Typically, close to 100 per cent of tidal flux is contained within 25~$R_{\rm eff}$ or $\sim0.6~R_{\rm vir}$ (where $R_{\rm eff}\sim 4$~kpc for $M^{\star}$ galaxies on average), but a fairly substantial proportion lies beyond this in some cases. In particular more massive galaxies tend to have tidal features whose flux extends further into the halo, with an average of 40 per cent of the total stellar halo flux lies beyond 25~$R_{\rm eff}$ in the most massive galaxies ($M_{\star} > 10^{11}{\rm M_{\odot}}$). This is significantly further into the stellar halo than many contemporary studies are typically able to probe. For example \citet[][]{Merritt2016} and \citet{Trujillo2021} measure galaxy surface brightness profiles out to 15 - 20~$R_{\rm eff}$. While this is partially limited by the depth of imaging available (LSST will be similarly limited), at increasingly large radii, measurements of the surface brightness profile are increasingly likely to be contaminated by nearby bright objects. Note that studies which do probe deeper into the stellar halo \citep[e.g.][]{Buitrago2017,Borlaff2019} do appear to detect a larger fraction of flux outside of the main galaxy.

Without very coarse binning, and for the majority of galaxies in our sample $(M_{\star}\sim M^{\star})$, surface brightness limits significantly fainter than 32 mag arcsec$^{-2}$ would be needed to recover a significant fraction of the flux found in extended diffuse light, which, in integrated light, is beyond the capabilities of any contemporary or forthcoming instrument including the Rubin Observatory. However, as we discuss in Section \ref{sec:recovered}, the brighter parts of denser tidal features (and therefore much of the total light) will be detectable in many cases.

Finally, the bottom panel of Figure \ref{fig:detection_grid} shows the joint distribution of the cumulative flux fraction for dense tidal features as a function of projected radius and surface brightness (i.e. the fraction of total flux in dense tidal features that is contained in pixels that are both within a given radius and brighter than a given surface brightness). Again, the majority of the total flux in dense tidal features resides within relatively small radii and at surface brightnesses that are in reach of LSST. For example, on average 50 per cent of flux lies within 10~$R_{\rm eff}$ and in pixels brighter than 31~mag arcsec$^{-2}$) with a $1\sigma$ scatter of $\pm^{\scriptscriptstyle 28}_{\scriptscriptstyle 41}$ percentage points, and 76 per cent lies within 15~$R_{\rm eff}$ and $\mu_{r}=32$ mag arcsec$^{-2}$ with a $1\sigma$ scatter of $\pm^{\scriptscriptstyle 18}_{\scriptscriptstyle 26}$ percentage points. As these values indicate, there is significant scatter in the flux fraction, which appears particularly large at the expected LSST surface brightness limits ($30-31$~mag arcsec$^{-2}$).

\subsubsection{Proportion of galaxy flux in tidal features}
\label{sec:tidal_flux}

In this section, we consider the amount of flux found in dense tidal features and extended tidal debris compared to the host galaxy -- the \textit{tidal flux fraction}, $f_{\rm tidal}$ -- defined as:

\begin{equation}
    \label{eqn:ftidal}
    f_{\rm tidal} = \frac{F_{\rm \mathbf{T}} + F_{\rm \mathbf{T_{d}}}}{F_{\rm tot}}
\end{equation}

\noindent where $F_{\mathbf{T_{d}}}$ and $F_{\mathbf{T}}$ are the total flux  within a 3-d radius greater than $5~R_{\rm eff}$ and smaller than 1~$R_{\rm vir}$ found in dense tidal features and extended tidal debris respectively and $F_{\rm tot}$ is the total flux from every particle within 1~$R_{\rm vir}$ of the center of the host galaxy. A minimum radius of $5~R_{\rm eff}$ is sufficient to avoid any components of the galactic disc that may have been missed by \textsc{AdaptaHOP} (typically recently formed resolved clusters of stars that are rejected for falling below the minimum particle number threshold) but not large enough to exclude a significant contribution from the stellar halo \citep[e.g.][]{Abadi2006, Pillepich2015} outside of the central few effective radii{(however in almost all cases all particles within 5~$R_{\rm eff}$ are associated to a structure or sub-structure of the host galaxy and the contribution by other particles not associated with any structure or sub-structure within $5~R_{\rm eff}$ is negligible).}

\begin{figure}
    \centering
    \includegraphics[width=0.45\textwidth]{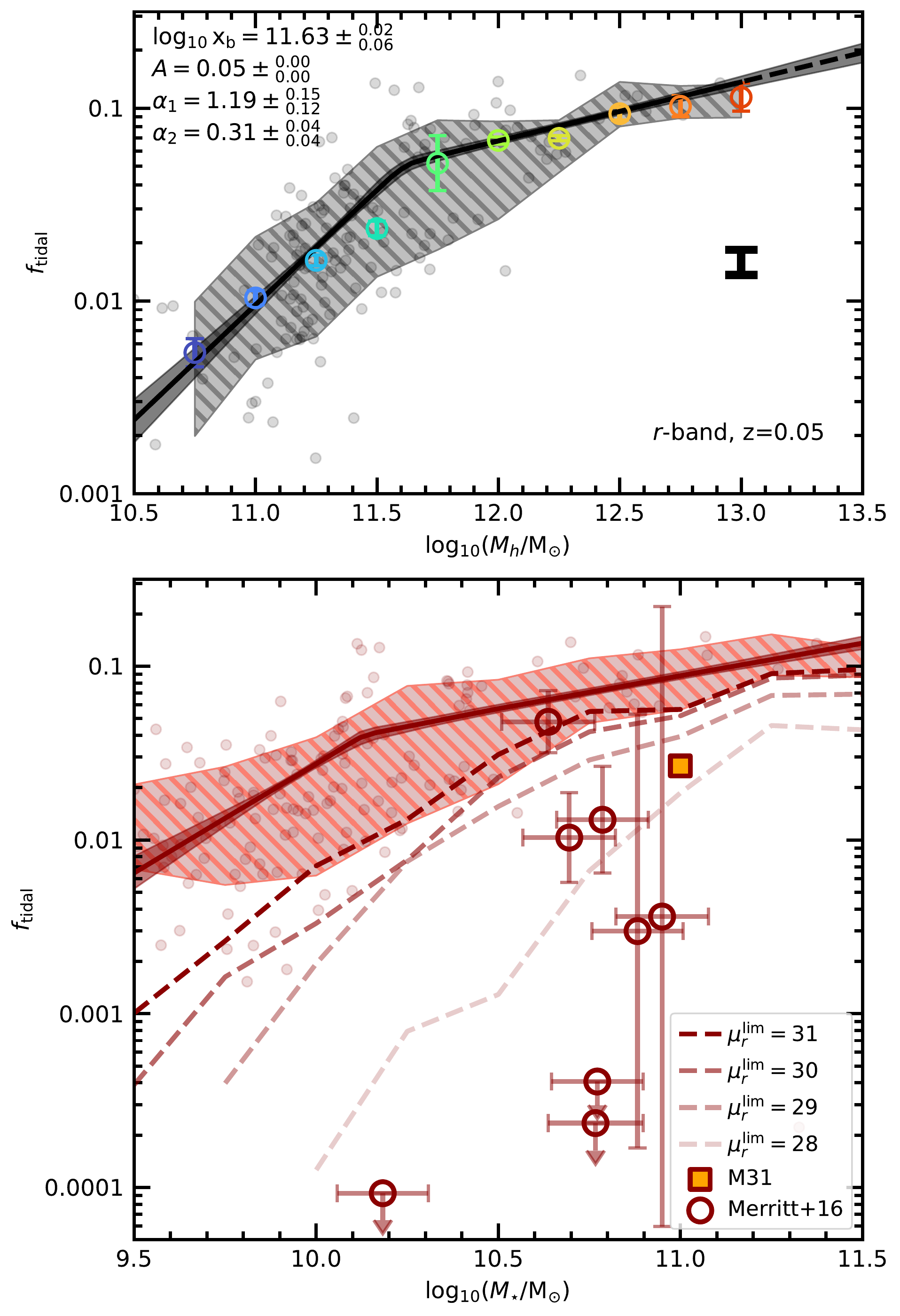}
    \caption{\textbf{Top:} Scatter plot showing the tidal flux fraction ($f_{\rm tidal}$, Equation \eqref{eqn:ftidal}) in the $r$-band as a function of halo mass. The black line and grey shaded region show a broken power law fit to the grey points representing individual \textsc{NewHorizon} galaxies and the associated $1\sigma$ uncertainty obtained from 100,000 bootstraps. The larger shaded and hatched region indicates the $1\sigma$ scatter of the grey points. The median $f_{\rm tidal}$ and associated errors in halo mass bins are indicated with coloured errorbars. The black error bar indicates the mean fractional variation in $f_{\rm tidal}$ due to orientation ($\sim 30$ per cent). \textbf{Bottom:} Similarly to the top panel, scatter points show $f_{\rm tidal}$ in the $r$-band as a function of stellar mass and a fit to $f_{\rm tidal}$ as a function of stellar mass is shown as a red line. Red open circles with error bars indicate the stellar halo mass fraction and associated $1\sigma$ uncertainties as a function of stellar mass from \citet[][]{Merritt2016}, with an additional data point for M31 \citep[][]{Courteau2011} plotted as a filled orange square. Dashed red lines indicate the median $f_{\rm tidal}$ that is recovered at different $r$-band limiting surface brightnesses ($3\sigma$, $10^{\prime\prime}\times10^{\prime\prime}$). As expected, $f_{\rm tidal}$ increases as a function of halo mass, however the normalisation and scatter of the relationship changes considerably with limiting surface brightness.}
    \label{fig:fraction_flux}
\end{figure}

We calculate $f_{\rm tidal}$ in each LSST band for each object, fitting a broken power law with the form

\begin{equation}
    f_{\rm tidal}(M) = \begin{cases}
    A[{\rm log}_{10}(M/M_{b})]^{\alpha_{1}}& \text{$M<M_{b}$}.\\
    A[{\rm log}_{10}(M/M_{b})]^{\alpha_{2}}& \text{$M>M_{b}$}.
  \end{cases}
\end{equation}

\noindent where $M$ is the galaxy mass (stellar mass or halo mass), $M_{b}$ is the mass at the break point, $A$ is the amplitude at $M_{b}$, and $\alpha_{1}$ and $\alpha_{2}$ are power law indices before and after $M_{b}$ respectively. 

Figure \ref{fig:fraction_flux} shows the distribution of $f_{\rm tidal}$ in the $r$-band as a function of halo mass (top panel) and stellar mass (bottom panel) as black or red coloured points. Solid black and red lines and smaller shaded regions show the broken power law fit to the grey points and the associated $1\sigma$ uncertainty obtained from 100,000 bootstraps. The larger hatched and shaded regions indicate the central 68th percentile ($1\sigma$) of the distribution of the grey and red points. In the top panel, we also show the median $f_{\rm tidal}$ and associated errors for a number of overlapping 0.5~dex wide mass bins as coloured errorbars. While there is considerable scatter as a result of the stochastic nature of galaxy accretion histories, the fraction of flux in tidal features increases towards higher masses on average for both stellar and halo mass. This is consistent with both observational and theoretical studies, which show that the merger rates and $f_{\rm exsitu}$ of galaxies are larger for more massive galaxies \citep[e.g.][]{Stewart2008,Rodriguez2015,Martin2019,Martin2021} (as we show later in Section \ref{sec:accreted_mass}, $f_{\rm exsitu}$ and $f_{\rm tidal}$ are correlated in our simulations). The fraction, $f_{\rm tidal}$, increases from less than 1 per cent for $M_{h} = 10^{10.5}{\rm M_{\odot}}$ up to around 10 per cent at $M_{h}\sim10^{13}{\rm M_{\odot}}$. We observe a break in the relation at a halo mass of $10^{11.6\pm^{\scriptscriptstyle 0.03}_{\scriptscriptstyle 0.06}}$~M$_{\odot}$ or a stellar mass of $10^{10.1\pm^{\scriptscriptstyle 0.01}_{\scriptscriptstyle 0.05}}$~${\rm M_{\odot}}$, which corresponds with the crossover mass at which elliptical galaxies begin to dominate and mergers become the dominant process driving the evolution of galaxies \citep[e.g.][]{Company2010,Robotham2014,Thanjavur2016}. One possible explanation is that the bulge itself is a remnant of past interactions \citep[e.g.][]{martin2018_sph,Park2019} \citep[but see also][]{Gargiulo2019}. Since more massive early-type galaxies typically formed their bulge at earlier times \citep[][]{Martin2018_progenitors} and typically exhibit fewer tidal features the earlier they formed \citep[][]{Yoon2020}, it is expected that the relationship between galaxy mass and $f_{\rm tidal}$ should weaken as spheroidal component of galaxies begin to dominate. It should be noted that while tidal features generally trace relatively recent events in a galaxy's accretion history, the extent that tidal features trace mass assembly or accretion history is complicated by the fact that the different types of tidal feature fade over different timescales and their flux may become more difficult to detect. Therefore, $f_{\rm tidal}$ can also be sensitive to a range of factors beyond bulk accretion history.

Based on an extrapolation of our fit (dashed line), we would expect to find over 20 per cent of flux in tidal features for the most massive galaxies $(M_{\star}>10^{12}{\rm M_{\odot}})$. We also note that, if we consider the fraction of stellar mass found in tidal features instead of flux, we obtain very similar results. It should be noted that we do not see any significant change in our results if we consider other LSST photometric bands ($u$, $g$, $i$, $z$ or $y$) except that the low mass slope becomes slightly shallower towards redder bands so that $\alpha_{1}=1.33\pm^{\scriptscriptstyle 0.22}_{\scriptscriptstyle 0.15}$ in the $u$-band falling to $\alpha_{1}=1.13\pm^{\scriptscriptstyle 0.14}_{\scriptscriptstyle 0.13}$ in the $y$-band.

The black error bar in Figure \ref{fig:fraction_flux} indicates the mean fractional variation in $f_{\rm tidal}$ due to orientation, which accounts for a variation of around 30 per cent. This variation is driven by changes in the effective gas geometry \citep[e.g.][]{Calzetti2001}, which act primarily to change the amount of dust attenuation and therefore the integrated flux of the galaxy. As dust column densities are much lower in the outskirts of galaxies, the integrated tidal flux in not similarly affected. As we discuss later, the fraction of \textit{observed} tidal debris is much more dependent on the geometry of the tidal features as they may move above or below the limiting surface brightness depending on the angle at which they are viewed.

Additionally, we compare our results with those of \citet[][]{Merritt2016}. Red open circles with error bars or limits in the bottom panel indicate the stellar halo mass fraction and associated $1\sigma$ uncertainties as a function of stellar mass (corresponding to the red top $x$-axis). The value for M31 \citep[][]{Courteau2011} is plotted as a filled orange square and re-scaled by \citet[][]{Merritt2016} to be consistent with their own definition. The quantity calculated by \citet[][]{Merritt2016} is derived in a similar way to $f_{\rm tidal}$. In both cases any flux within 5~$R_{\rm eff}$ is ignored and an attempt is made to remove any contaminating flux from the galaxy itself (see Section 3.2 of \citet[][]{Merritt2016} for a description of the method). \citet[][]{Merritt2016} only account for flux from the main galaxy out to 7~$R_{\rm eff}$, while we find that a large fraction of tidal flux lies beyond 7~$R_{\rm eff}$ (see Figure \ref{fig:detection_grid}) and that the galaxy itself does not account for a significant fraction of flux by 7~$R_{\rm eff}$. Discrepancies could arise if the light of the galaxy within 7~$R_{\rm eff}$ is not accurately subtracted. Therefore, we expect these two methods to be somewhat comparable, but systematic differences likely exist. Also note that local galaxy samples are dominated by galaxies with late-type morphology and there is a considerable difference in values measured in these galaxies and similar mass early-type galaxies at higher redshifts \citep[e.g.][]{Buitrago2017}.

While a na{\"i}ve comparison with the \citet[][]{Merritt2016} data points suggests a very significant level of disagreement with our simulated galaxies in both the normalisation and level of scatter in the data points, it is important to account for the fact that some fraction of the total flux will always be missed due to the finite limiting surface brightness of the observed data. \citet[][]{Merritt2016} achieve a limiting surface brightness of up to $\mu^{\rm lim}_{g}(3\sigma$, $60^{\prime\prime}\times60^{\prime\prime}) = 29.8$ mag arcsec$^{-2}$ which we convert to a limiting surface brightness of around $\mu^{\rm lim}_{r}(3\sigma$, $10^{\prime\prime}\times10^{\prime\prime})\sim28$ mag arcsec$^{-2}$ based on Appendix A of \citet[][]{Roman2020} and assuming a difference of 0.5 mag between the $g$ and $r$ bands. If we account only for detected flux (following the same method described later in Section \ref{sec:recovered}) we find that the normalisation of $f_{\rm tidal}$ falls as we move to lower limiting surface brightness (indicated by red dashed lines). Scatter also increases with shallower surface brightness limits. The standard deviation for approximately Milky Way mass galaxies ($10^{10.25}{\rm M_{\odot}}<M_{\star}<10^{10.75}{\rm M_{\odot}}$) increases from 0.4 dex at $\mu^{\rm lim}_{r}(3\sigma,10^{\prime\prime}\times10^{\prime\prime})=31$~mag~arcsec$^{-2}$ to 1.3 dex at $\mu^{\rm lim}_{r}(3\sigma,10^{\prime\prime}\times10^{\prime\prime})=28$~mag~arcsec$^{-2}$. This brings the data points from \citet[][]{Merritt2016} into closer agreement with the simulated data, however the increase in scatter is comparable to the change in normalisation (note that other studies \citep[e.g.][]{Monachesi2016,Harmsen2017} show a similar amount scatter and individual measurements \citep[e.g.][]{Carollo2010,Courteau2011,Deason2019,Smercina2020} span a similar range of values to \citet[][]{Merritt2016}). Again, we cannot be sure of the systematic differences between our methodology and that of \citet[][]{Merritt2016} (especially as we do not take azimuthal averages as they have done). However, this result shows that while the normalisation of $f_{\rm tidal}$ is expected to fall at brighter limiting surface brightness, the scatter is also expected to increase significantly, indicating that the uncertainty in the observations may currently be too large to make any valid comparison to theory.


\begin{figure}
    \centering
    \includegraphics[width=0.45\textwidth]{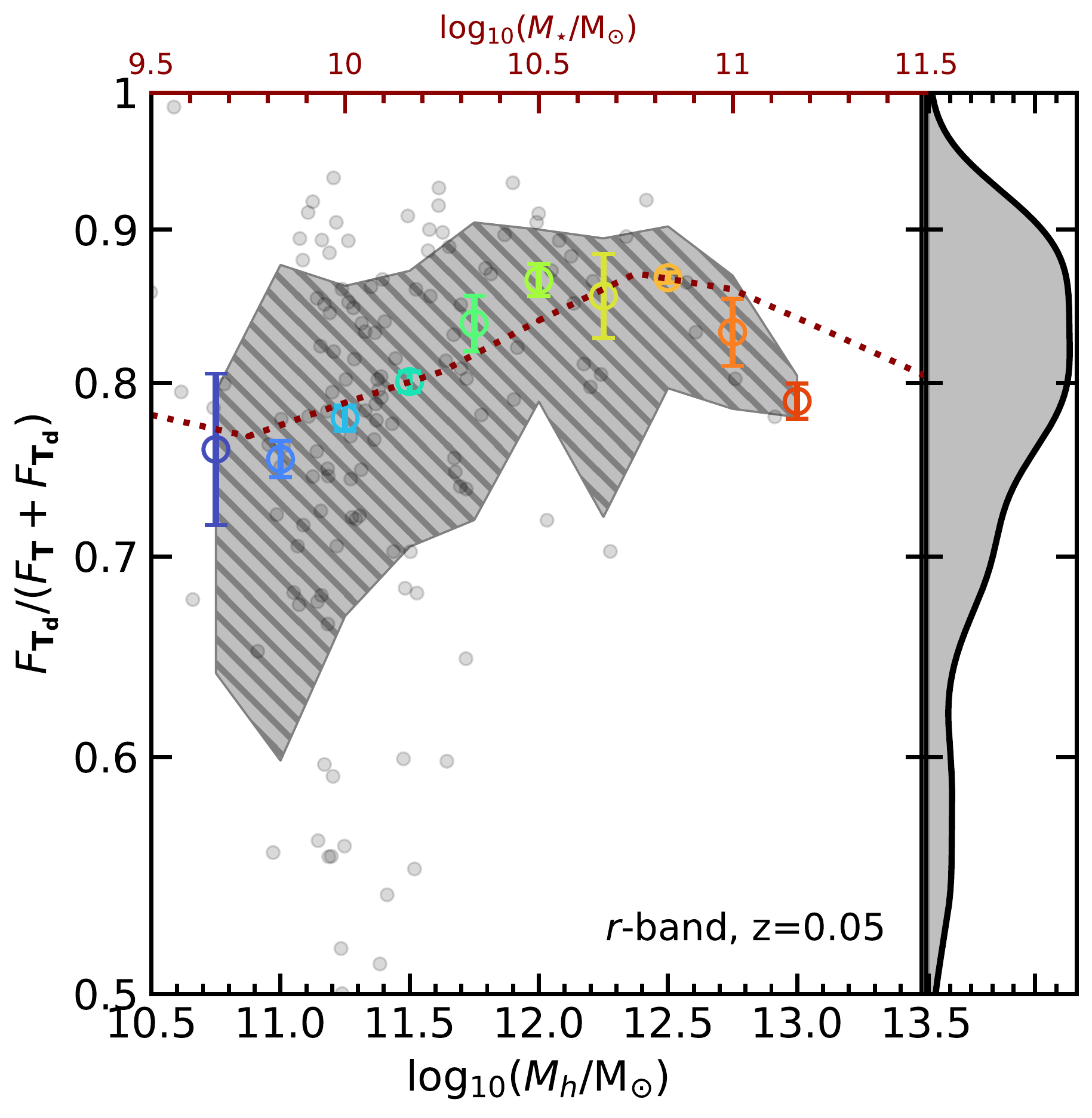}
    \caption{Scatter plot showing the fraction of tidal flux that is found in dense ($\rho_{\star}>500$~M$_{\odot}\,$kpc$^{-3}$) tidal features ($F_{{\mathbf T}_d} / (F_{{\mathbf T}_d}+ F_{{\mathbf T}})$) vs the total flux in all tidal features) as a function of halo mass. Coloured error bars show the median value in overlapping mass bins with its $1\sigma$ uncertainty determined from 100,000 bootstraps. The dotted red line indicates the same median trend with stellar mass (errors are similar). The hatched and shaded grey region indicates the $1\sigma$ scatter in the grey points. The histogram on the $y$-axis shows the distribution of the dense $f_{\rm tidal}$ marginalised over mass. The fraction of flux in dense tidal features appears to increase by a small amount with halo mass (with large scatter), but there is some indication that it begins to fall again around $L^{\star}$.}
    \label{fig:fraction_mass_tidal}
\end{figure}

In Figure \ref{fig:fraction_mass_tidal}, we consider the fraction of flux that is found in dense tidal features vs more extended tidal debris, $F_{{\mathbf T}_d} / (F_{{\mathbf T}_d}+ F_{{\mathbf T}})$, as a function of halo mass. Again, coloured error bars indicate the median and $1\sigma$ error for overlapping 0.5~dex wide mass bins and the grey filled region indicates the central 68th percentile ($1\sigma$) of the distribution of the grey points. The red dotted line shows the same relation as a function of stellar mass with the scale shown on the red top axes. For any given halo mass or stellar mass, around 80 per cent of the extended light of a galaxy is found in dense, generally higher surface brightness tidal features. The panel to the right of the main plot shows the probability density function (PDF) for $F_{{\mathbf T}_d} / (F_{{\mathbf T}_d}+ F_{{\mathbf T}})$ marginalised over halo mass. As the PDF shows, galaxies are rarely found with more than 30 per cent of their extended flux in extended tidal debris rather than dense tidal features ($<5$ per cent chance). Additionally, we do not find any galaxies in which the amount of flux in extended tidal debris outweighs that found in dense tidal features. This argues that observational studies which recover much of the coherent tidal features will likely not be missing significant amounts of the accreted mass. With this in mind, missing flux due to finite resolution may, in some cases, be a similarly important consideration in measuring halo flux \citep[in addition to the PSF and scattered light, as argued by e.g.][]{Abraham2014}, since smaller resolution elements allow tidal features to be analysed at higher surface brightness compared with the diffuse component of the halo \citep[see also the discussion in][]{Trujillo2021}.

There is evidence of only a small amount of evolution in the fraction of extended light found in dense tidal features, increasing towards intermediate halo masses and then declining (although it is difficult to say definitively, given the relatively small sample of galaxies at higher masses). However, over the full mass range, dense tidal features are responsible for almost all of the flux. A lack of any significant evolution as a function of halo mass suggests that the nature of the tidal features does not change over this mass range or perhaps that the timescales over which cohesive dense tidal features persist is not strongly affected by the host mass\footnote{Note that \citet[][]{Pillepich2018} show the ratio of ICL to all diffuse mass increases in more massive haloes and our own examination of mock images of massive central galaxies in the Horizon-AGN simulation \citep{Dubois2014} shows a relative dearth of distinct tidal features compared with lower mass galaxies in \textsc{NewHorizon}. However it is difficult to make any definitive statement due to the much more limited resolution of Illustris TNG300 and Horizon-AGN.}. We will return to this topic in Section \ref{sec:visual}.

\subsection{Tidal features and accreted mass}
\label{sec:accreted_mass}
\begin{figure}
    \centering
    \includegraphics[width=0.45\textwidth]{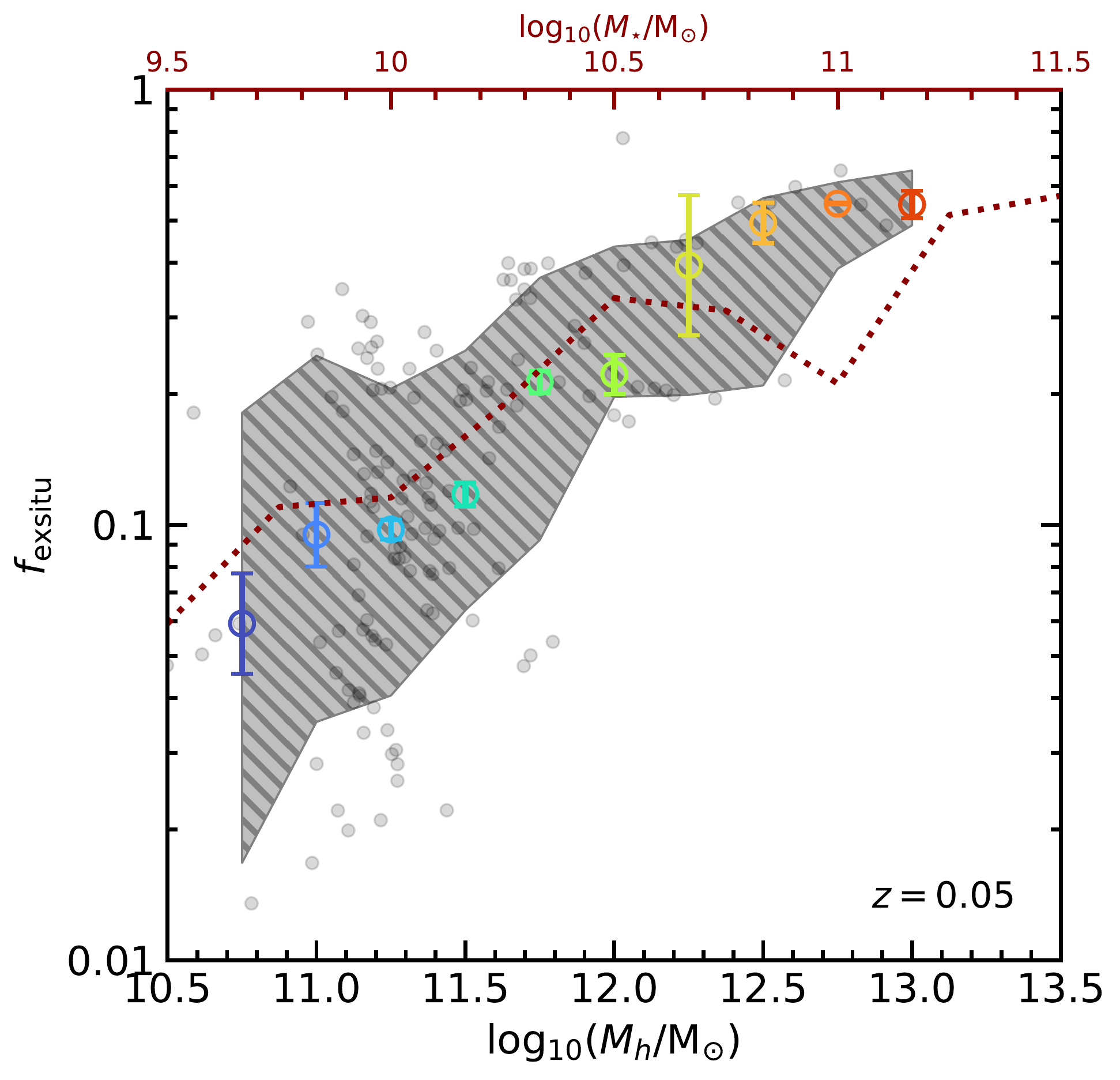}
    \caption{Scatter plot showing ex-situ mass fraction ($f_{\rm exsitu}$, Equation \eqref{eqn:exsitu}) as a function of halo mass. The shaded and hatched region indicates the $1\sigma$ scatter of the grey points. Coloured error bars show the median value in overlapping halo mass bins with $1\sigma$ error bars determined from 100,000 bootstraps. The dotted red line indicates the same median trend with stellar mass (errors are similar). We observe a clear and relatively strong correlation between galaxy mass and accreted mass.}
    \label{fig:fraction_mass}
\end{figure}

Figure \ref{fig:fraction_mass} shows the dependency of the ex-situ mass fraction on halo mass with respect to the total host stellar mass. The shaded region indicates the central 68th percentile ($1\sigma$) of the distribution of the grey points. We also show the median ex-situ mass fracton ($f_{\rm exsitu}$; Equation \eqref{eqn:exsitu}) and associated errors for a number of overlapping 0.5~dex wide mass bins as coloured errorbars. Finally the red dotted line shows the same relation as a function of stellar mass (scale shown on the red top axes). As many other theoretical and observational studies \citep[][]{Purcell2007,Oser2010,Dubois2013,Cooper2013,DSouza2014,Lee2015,Rodriguez2016,Harmsen2017,Pillepich2018,Spavone2018,Pillepich2018,Tacchella2019,Davison2020,Spavone2020,Martin2021} also predict, $f_{\rm exsitu}$ increases on average towards larger masses as a function of both halo and stellar mass. At lower masses ($M_{h} < 10^{11}{\rm M_{\odot}}$) less than 10 per cent of stellar mass is formed ex-situ on average, rising to around half at the highest masses shown. We see that $f_{\rm exsitu}$ follows a similar trend with halo and stellar mass as $f_{\rm tidal}$, with both increasing towards higher masses (Figure \ref{fig:fraction_flux}).

\begin{figure}
    \centering
    \includegraphics[width=0.45\textwidth]{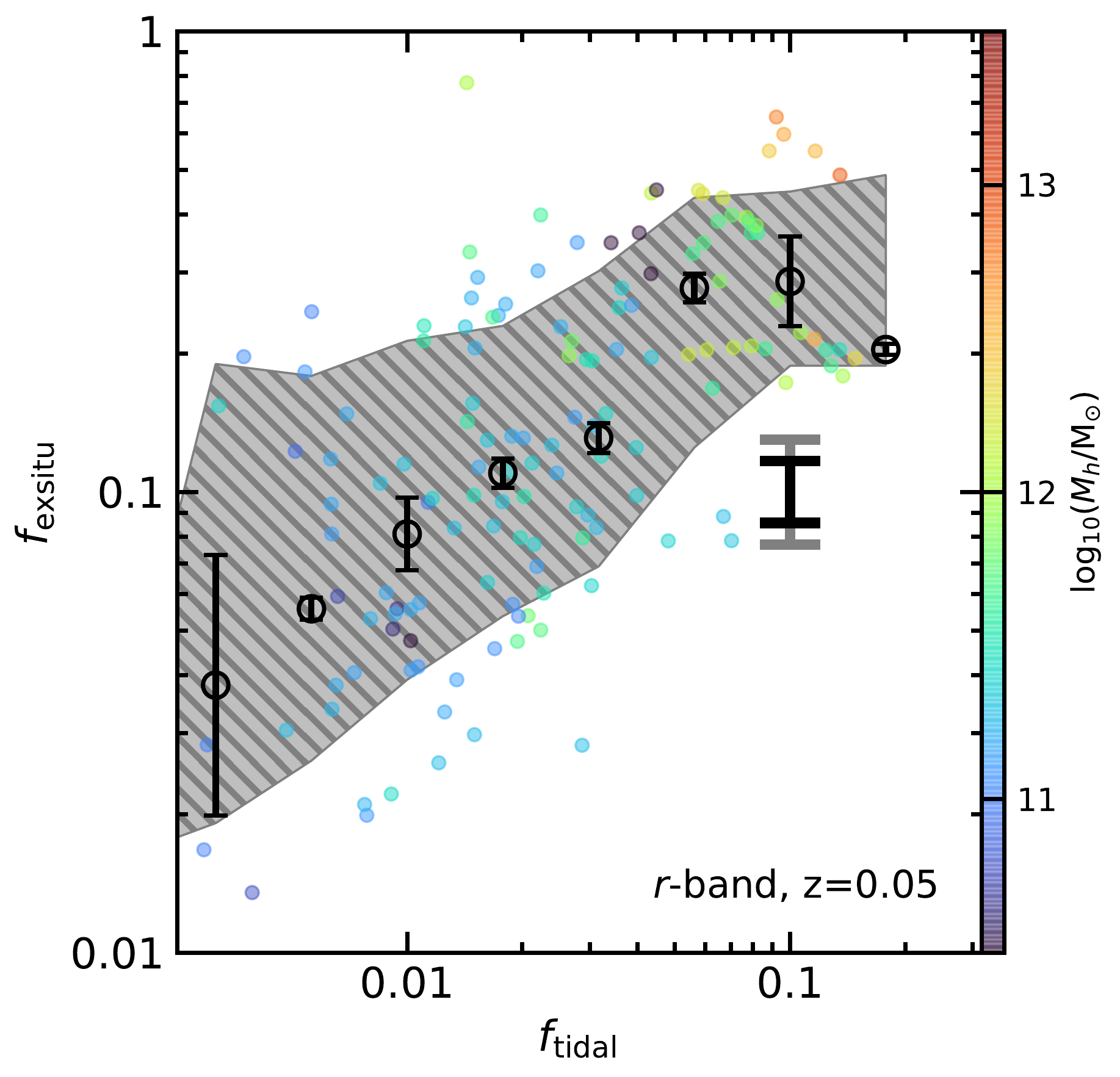}
    \caption{Scatter plot showing ex-situ mass fraction ($f_{\rm exsitu}$, Equation \eqref{eqn:exsitu}) as a function of tidal flux fraction ($f_{\rm tidal}$, Equation \eqref{eqn:ftidal}) in the $r$-band colour coded by halo mass. Black open circles with error bars indicate the median value in overlapping bins, with $1\sigma$ error bars determined from 100,000 bootstraps. The large overlapping black and grey error bars to the right indicate the mean fractional variation in ex-situ mass fraction over timescales of 2 Gyr and 4 Gyr (fractional errors of $\sim 30$ per cent and $\sim 50$ per cent respectively). We observe a similarly strong correlation between $f_{\rm exsitu}$ and $f_{\rm tidal}$ as we do with $f_{\rm exsitu}$ and galaxy mass. As the coloured points indicate, both $f_{\rm exsitu}$ and $f_{\rm tidal}$ are correlated with mass.}
    \label{fig:fraction_mass_exsitu}
\end{figure}

Finally, in Figure \ref{fig:fraction_mass_exsitu}, we plot $f_{\rm tidal}$ against $f_{\rm exsitu}$. The shaded region indicates the central 68th percentile ($1\sigma$) of the distribution of the grey points. We also show the median $f_{\rm exsitu}$ and associated errors for a number of overlapping bins in $f_{\rm tidal}$. The large overlapping error bars indicate the variation in $f_{\rm exsitu}$ over a 2 Gyr and 4 Gyr timescale respectively. While it is clear that $f_{\rm exsitu}$ correlates with $f_{\rm tidal}$ and halo mass, there appears to be a similar variance as a function of both variables.

In order to investigate whether accretion history ($f_{\rm exsitu}$) has a measurable influence in the strength of tidal features beyond the existing correlation of $f_{\rm exsitu}$ with halo mass, we calculate the partial distance correlation coefficient \citep[][]{Szekely2014} between $f_{\rm tidal}$ and $f_{\rm exsitu}$, controlling for the halo mass, $\mathfrak{R^{*}}({\rm log}_{10}\,f_{\rm tidal}, {\rm log}_{10}\,f_{\rm exsitu}; {\rm log}_{10}\,M_{\rm h})$, and between the halo mass and $f_{\rm exsitu}$, controlling for $f_{\rm tidal}$, $\mathfrak{R^{*}}({\rm log}_{10}\,M_{\rm h}, {\rm log}_{10}\,f_{\rm exsitu}; {\rm log}_{10}\,f_{\rm tidal}$). The partial correlation coefficients and associated $1\sigma$ uncertainties are $0.28\pm0.04$ and $0.16\pm0.04$ respectively (with full correlation coefficients of 0.61 and 0.54 respectively), indicating a stronger association between $f_{\rm exsitu}$ and $f_{\rm tidal}$. This suggests that the tidal mass fraction may be a better predictor of $f_{\rm exsitu}$. The statistical significance of this result (that there is a stronger correlation between $f_{\rm exsitu}$ and $f_{\rm tidal}$ than $f_{\rm exsitu}$ and halo mass) stands at $2.2\sigma$, giving a relatively weak indication that different accretion histories have a measurable impact on the amount of flux in the stellar halo. This result appears to be consistent with the idea that much of the tidal flux is contributed by recent mergers with a small number of relatively massive progenitors \citep[e.g.][]{Bullock2005, Purcell2007,Cooper2010}. However, the total fraction of accreted mass alone is a relatively weak predictor of the final structure of the stellar halo. For example \citet[][]{Rey2021} show that, even in galaxies with the same ex-situ mass fraction, the shape of the stellar halo is strongly sensitive to accretion history.

\subsection{Tidal feature detection}
\label{sec:recovered}

\begin{figure}
    \centering
    \includegraphics[width=0.45\textwidth]{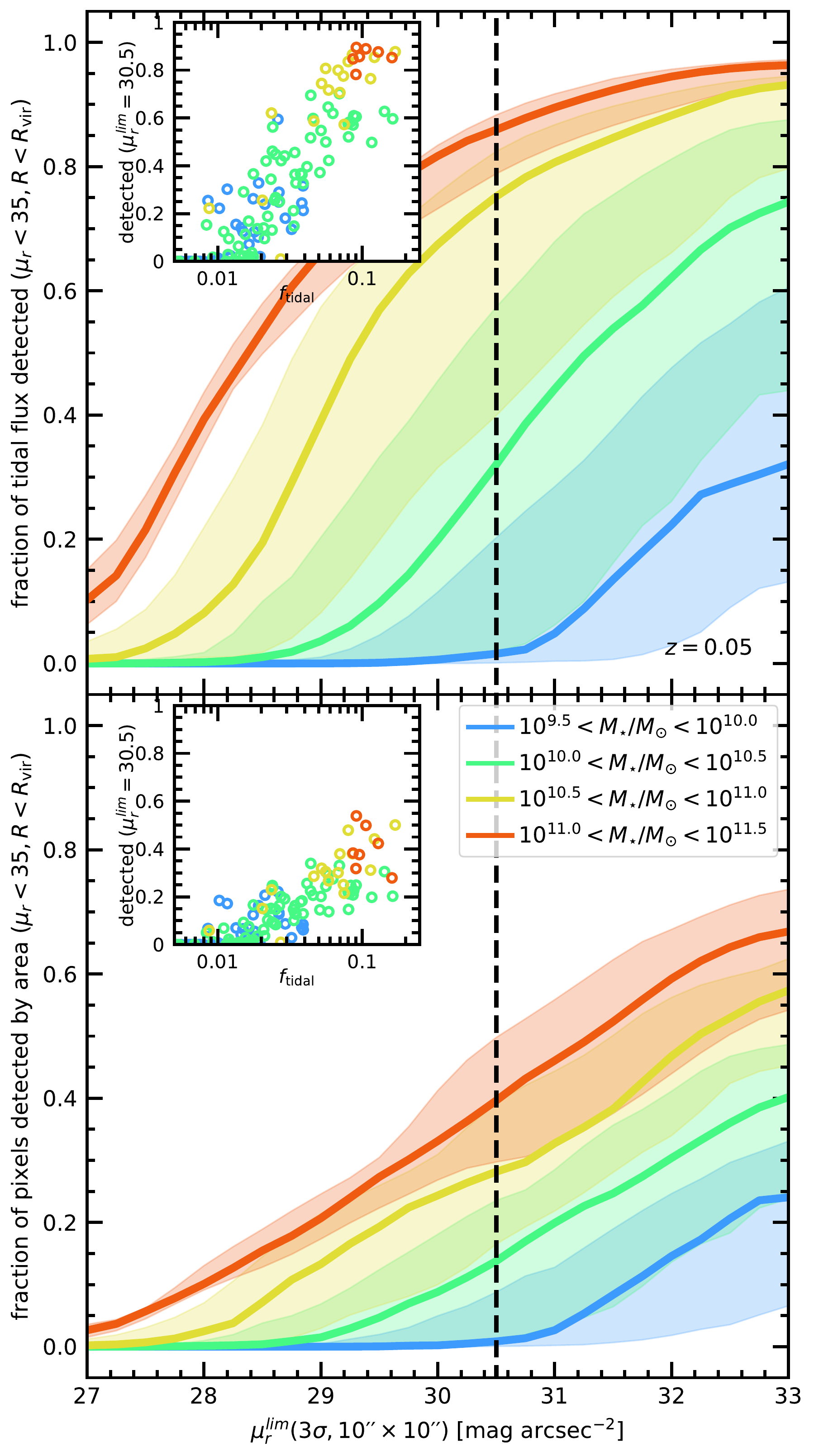}
    \caption{\textbf{Top}: the fraction of tidal flux found in detected structures as a function of $r$-band limiting surface brightness. Coloured lines indicate the median fraction of flux in detected structures for different stellar mass bins, and coloured regions indicate the central 68th percentile ($1\sigma$). The inset plot indicates the fraction of tidal flux detected per galaxy as a function of their tidal flux fraction for a limiting surface brightness of 30.5 mag arcsec$^{-2}$, and the colour of each point indicates which mass the galaxy is in. \textbf{Bottom}: the fraction of pixels detected as a function of $r$-band limiting surface brightness. Coloured lines indicate the median fraction of pixels detected in different stellar mass bins, and coloured regions indicate the central 68th percentile ($1\sigma$) of the distribution. The inset plot indicates the fraction of pixels detected per galaxy as a function of their tidal flux fraction for a limiting surface brightness of 30.5 mag arcsec$^{-2}$. The colour of each point indicates which mass bin the galaxy is in. An interactive version of the inset plots showing multiple surface brightness limits can be found at \href{https://garrethmartin.github.io/files/frac_recovered.html}{garrethmartin.github.io/files/frac\_recovered.html}.}
    \label{fig:frac_recovered}
\end{figure}

We also consider the fraction of tidal features that we expect to be detected at different limiting surface brightnesses. Because it may, in some cases, be possible to detect contiguous structures by eye, even if they are fainter than the surface brightness limit, we adopt a definition for \textit{detected structures} based on the connections between pixels that are $1\sigma$ above the noise level in images produced from particles that are part of dense tidal features only (so that contribution from well phase-mixed material is first removed i.e. Figure \ref{fig:residual_image}, panel \textbf{c}). We describe the procedure and present an example of the procedure performed on the same object for a range of limiting surface brightnesses in Appendix \ref{sec:binary_fill}. Qualitatively, our results do not change if we only consider pixels that are brighter than a given limiting surface brightness.

We define the total flux in detected structures within the original image (before noise is added) as a fraction of the total flux in pixels that are brighter than 35 mag arcsec$^{-2}$ and which are within a radius of 1~$R_{\rm vir}$. We define the detected area similarly by counting the total area of the detected mask as a fraction of the total area of pixels that are brighter than 35 mag arcsec$^{-2}$ and which are within a radius of 1~$R_{\rm vir}$.

In Figure \ref{fig:frac_recovered}, we show the fraction of flux (top) and the fraction of area (bottom) that is detected in different mass bins as a function of $r$-band limiting surface brightness ($3\sigma$, $10^{\prime\prime}\times10^{\prime\prime}$) at $z=0.05$. We consider only pixels with a surface brightness brighter than 35 mag arcsec$^{-2}$ and which are within 1~$R_{\rm vir}$ of the centre of the galaxy. More massive galaxies tend to have tidal features that are more easily detectable. For example, based on a predicted 10-year depth of $\mu_{r}^{\rm lim}(3\sigma,10^{\prime\prime}\times10^{\prime\prime})\approx30.5$~mag arcsec$^{-2}$ \citep[][]{Laine2018}, we can expect to detect a non-negligible fraction ($\gtrsim 25$ per cent) of the total area and most of the flux ($\gtrsim 80$ per cent) that makes up the dense tidal features of galaxies more massive than the Milky Way ($M_{\star} > 10^{10.5}~{\rm M_{\odot}}$). Even assuming a shallower final depth of 29.5 mag arcsec$^{-2}$, we still expect to detect more than 60 per cent of flux in tidal features for the same mass range.

A significant fraction of the mass associated with tidal features is detected at lower masses. The inset plots, which show the fraction of detected flux and area as a function of $f_{\rm tidal}$, indicate that this is largely a result of the fact that more massive galaxies tend to have stronger tidal features on average. As Figure \ref{fig:fraction_flux} shows, there is a significant spread in $f_{\rm tidal}$ such that lower mass galaxies can exhibit fairly strong tidal features. An interactive online supplement to this plot (\href{https://garrethmartin.github.io/files/frac_recovered.html}{garrethmartin.github.io/files/frac$\_$recovered.html}) shows a version of the two inset plots for different surface brightness limits.
We see that, for very weak tidal features (low values of $f_{\rm tidal}$), a majority of flux is still not detected even for very high limiting surface brightnesses (e.g. $\sim30$ per cent of flux detected for $f_{\rm tidal} = 0.01$ for a limiting surface brightness of 33 mag arcsec$^{-2}$). Therefore, in the nearby Universe ($z<0.05$), we can expect the LSST (or any similarly deep survey) to find a significant fraction of the mass associated with tidal features around intermediate and high mass galaxies ($M_{\star} > 10^{10.5}{\rm M_{\odot}}$). Most tidal features found around galaxies in the low mass regime, however, are likely to remain inaccessible at least in the near future. As we highlight later in Section \ref{sec:visual_z_mu}, this means that more massive galaxies exhibit tidal features that are both more frequent \textit{and} stronger.

\begin{figure}
    \centering
    \includegraphics[width=0.45\textwidth]{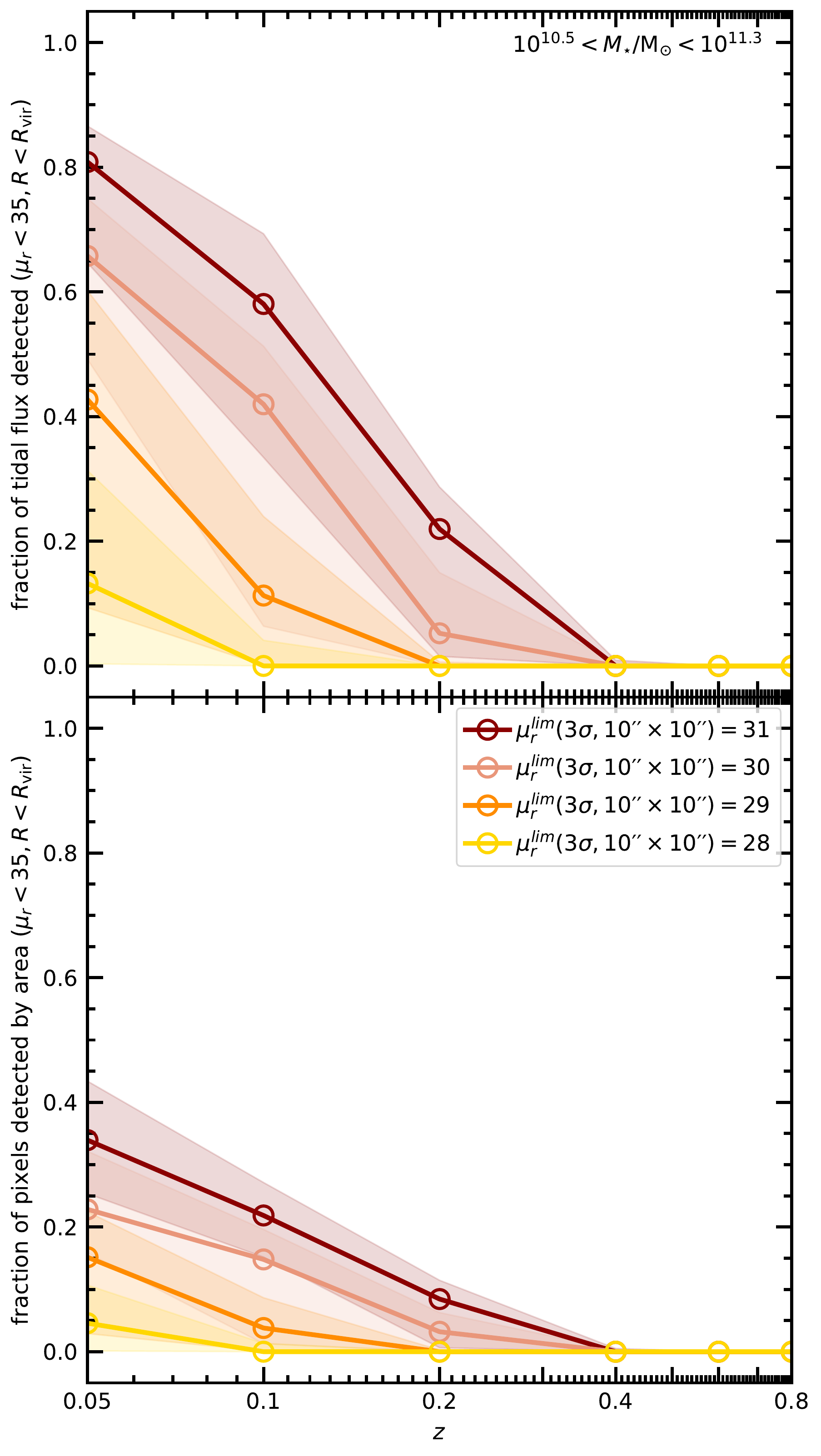}
    \caption{\textbf{Top} the fraction of tidal flux found in detected structures as a function of redshift in $\sim M^{\star}$ galaxies for different $r$-band limiting surface brightnesses. Coloured lines indicate the median fraction of flux in detected structures for a given redshift and coloured regions indicate the central 68th percentile ($1\sigma$) of the distribution. \textbf{Bottom}: the fraction of pixels detected as a function of redshift for different $r$-band limiting surface brightness. Coloured lines indicate the medial fraction of pixels detected in different stellar mass bins and coloured regions indicate the central 68th percentile ($1\sigma$) of the distribution. An interactive version of this plot showing multiple limiting surface brightnesses and stellar mass bins can be found at \href{https://garrethmartin.github.io/files/completeness.html}{garrethmartin.github.io/files/completeness.html}.}
    \label{fig:frac_recovered_2}
\end{figure}

Finally, we explore how our ability to detect tidal features around galaxies declines with redshift. Figure \ref{fig:frac_recovered_2} shows the same detected flux and detected area fractions as a function of the redshift at which each object is observed for approximately Milky Way mass galaxies and for a range of limiting surface brightnesses, which are indicated in the legend. Towards higher redshifts, the fraction of detected flux declines quite sharply such that we do not expect to be able to detect tidal features around any Milky Way mass galaxy after a redshift of $z=0.4$, even at 10-year LSST depth.

An interactive online supplement to this plot (\href{https://garrethmartin.github.io/files/completeness.html}{garrethmartin.github.io/files/completeness.html}) shows how the detected flux and detected area fraction evolve with redshift for different stellar mass bins and limiting surface brightnesses.

\section{Characterisation of tidal features by human classification}
\label{sec:visual_characterisation}
\subsection{Visual classification of mock images}
\label{sec:visual}

As discussed in Section \ref{sec:visual_classificatuion}, all images were classified independently by at least 2 people and a set of 600 images were independently classified by 5 people. Each of our classifiers has experience in the field of LSB science and therefore has a good level of domain knowledge and many are also experts in the morphological classification of galaxies.

\subsubsection{Collated feedback}

Classifiers were also asked to provide feedback commenting on their experience classifying the mock images. Classifiers were asked to evaluate their confidence in the reliability and reproducibility of their classifications, whether there were any particular categories that they found difficult to classify and whether they employed a particular strategy or methodology when performing the classifications. We list our conclusions based on classifier responses below:

\begin{itemize}

    \item Some classifiers were more conservative in their classifications than others, particularly in classifying mergers or fainter tidal features. Classifiers made what they felt was the most reasonable interpretations, but most felt that these choices were subjective and therefore liable to change between classifiers.

    \item Some classifiers were more confident in the reproducibility of their classification than others. Broadly, classifiers felt that they were able to identify the presence of tidal features very reliably, but felt that, in some cases, detailed characterisation and distinguishing between similar categories of tidal feature was difficult (e.g. differentiating tidal tails, stellar streams, plumes and shells from one another) and that precisely determining the frequency of these features in each image was also difficult.
    
    \item Classifiers generally felt that their classifications became more consistent as they classified more objects. In some cases, where classifiers went back through and repeated their classifications, they ended up revising some of their original classification and particularly the earliest images that they classified.
    
    \item In the case of faint or poorly resolved tidal features, it was often difficult to place these features into a specific category. Classifiers generally placed such features into miscellaneous category in these cases.
    
    \item  In the absence of 3-d kinematic information, ambiguities arise in the classification of certain features. Excluding tidal features within the field of view that were not associated with any interaction with the host galaxy added a degree of complexity to the classification as making this determination could be difficult, for example in the case where there is significant overlap between two objects but not clear tidal disturbance. For the deepest / highest spatial resolution images, classifications tended to become more difficult as the complexity of the morphology and environment of the tidal features became more apparent.

    \item Counting certain categories of tidal feature, like the number of shells, plumes or asymmetric features was not always simple. In a single halo, it is much easier to identify multiple distinct plumes than in a group environment, where it becomes much more difficult due to multiple overlapping features.
    
    \item For more distant (poorly resolved) objects there was a feeling among classifiers that they may have been susceptible to seeing asymmetries in the stellar halo that were not present.
    
    \item At low spatial resolution, distinguishing merging systems from close pairs is challenging due to a lack of resolved tidal features. In general, classifiers found the concept of double nuclei systems to be quite uncertain as there was a large degree of subjectivity in determining if two nuclei are close, share a common envelope, are just the result of projection, etc, which becomes more difficult for more distant objects.

\end{itemize}

Despite the high level of expertise of our classifiers, many of them still found the exercise challenging. Some of the difficulties raised by the classifiers could be alleviated by designing the study differently, but there are also factors that are more difficult or impossible to address. For example it is doubtful whether it is possible to produce entirely consistent classifications between classifiers due to the subjective nature of many of the decisions that classifiers are required to make. Additionally, classifiers likely have differing notions of what exactly constitutes a given class of tidal feature, a lack of any standard definitions likely compounds this in addition to making any comparisons between classifiers or between studies necessarily qualitative. Since it is impossible to completely standardise the classification process, this implies that there will always be differences between human classifiers.

\subsubsection{Census of tidal features by class}

\begin{figure*}
    \centering
    \includegraphics[width=0.95\textwidth]{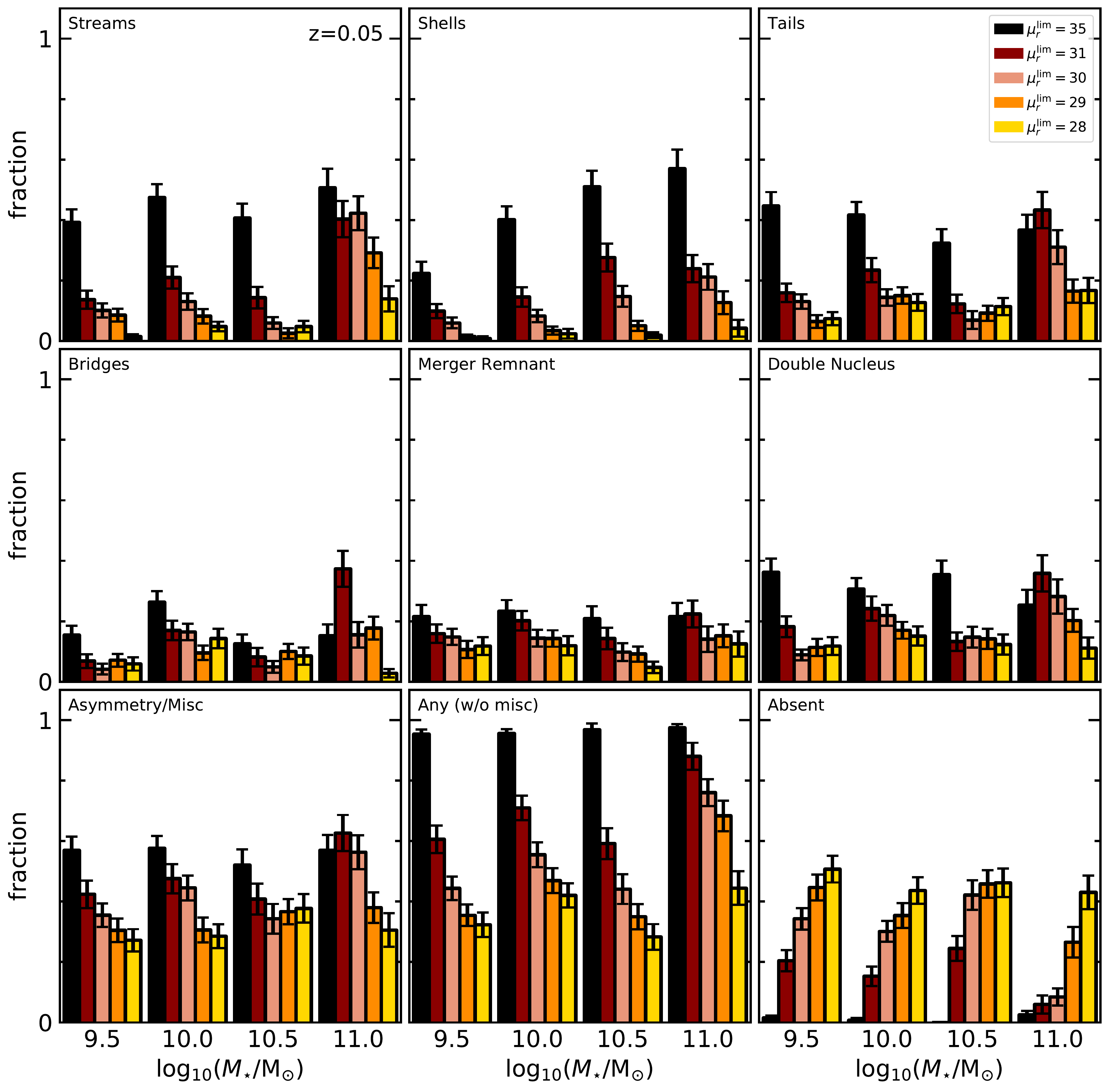}
    \caption{Histograms indicating the prevalence of different classes of tidal feature as a function of stellar mass and limiting surface brightness. Bars indicate the fraction of galaxies that exhibit at least one instance of a given class of tidal feature in each mass bin. The four mass bins have a width of 0.5 dex and run from 9.25~M$_{\odot}$ to 11.25~M$_{\odot}$, and errorbars for each bin are determined by bootstrap. At the highest limiting surface brightness, almost all galaxies exhibit coherent tidal features regardless of their mass. We observe significantly different behaviour in the prevalence of tidal features as a function of limiting surface brightness and stellar mass across different classes.}
    \label{fig:classifications_hist}
\end{figure*}

In this section we consider the frequency at which different classes of tidal feature were identified by our human classifiers. Figure \ref{fig:classifications_hist} shows the fraction of galaxies at $z=0.05$ in which different classes of tidal feature were identified for different stellar masses and limiting surface brightnesses. 

We see similar behaviour in the prevalence of shells as seen in observational studies like \citet[][]{Bilek2020}\footnote{It should be noted that the sample of \citet[][]{Bilek2020} is more strongly dominated by elliptical galaxies and that we have not attempted to replicate their definitions for different tidal feature classes.}, and the decline in the prevalence of shells with stellar mass remains across different limiting surface brightness (with differing normalisation). Interestingly, while we see similar behaviour for tails and streams at lower limiting surface brightnesses, at $\mu_{r}^{\rm lim}(3\sigma,10^{\prime\prime}\times10^{\prime\prime})=35$~mag\,arcsec$^{-2}$, there is little difference in their prevalence across mass bins, indicating that while these features are present at similar levels across the mass range we consider, they are typically fainter and more difficult to detect in lower mass galaxies. 

In the highest mass bin, streams and tails are both detected at similar rates regardless of limiting surface brightness for $\mu_{r}^{\rm lim}(3\sigma,10^{\prime\prime}\times10^{\prime\prime})\geq30$~mag\,arcsec$^{-2}$ indicating that the majority of streams and tidal tails are at least this bright in these more massive galaxies. Meanwhile, merger remnants and double nuclei are quite reliably identified regardless of limiting surface brightness and occur at relatively similar levels across stellar mass.

In all, close to 100 per cent of galaxies exhibit some kind of distinct tidal feature (i.e. not just asymmetries) at $\mu_{r}^{\rm lim}(3\sigma,10^{\prime\prime}\times10^{\prime\prime})=35$~mag\,arcsec$^{-2}$ regardless of mass. However, this number falls fairly significantly as more realistic limiting surface brightnesses are considered. This result is in broad agreement with \citet[][]{VeraCasanova2021}, who show that $\sim 90$ per cent of their Aurgia models show clear LSB features at 31 mag~arcsec$^{-2}$ for a sample of host galaxies with an average stellar mass of $10^{10.8}$~${\rm M_{\odot}}$ (compare with the most massive bin of the `Any' panel of Figure \ref{fig:classifications_hist}). 

\subsection{Visual biases}
\label{sec:visual_z_mu}
\subsubsection{Effect of redshift and limiting surface brightness}

\begin{figure}
    \centering
    \includegraphics[width=0.45\textwidth]{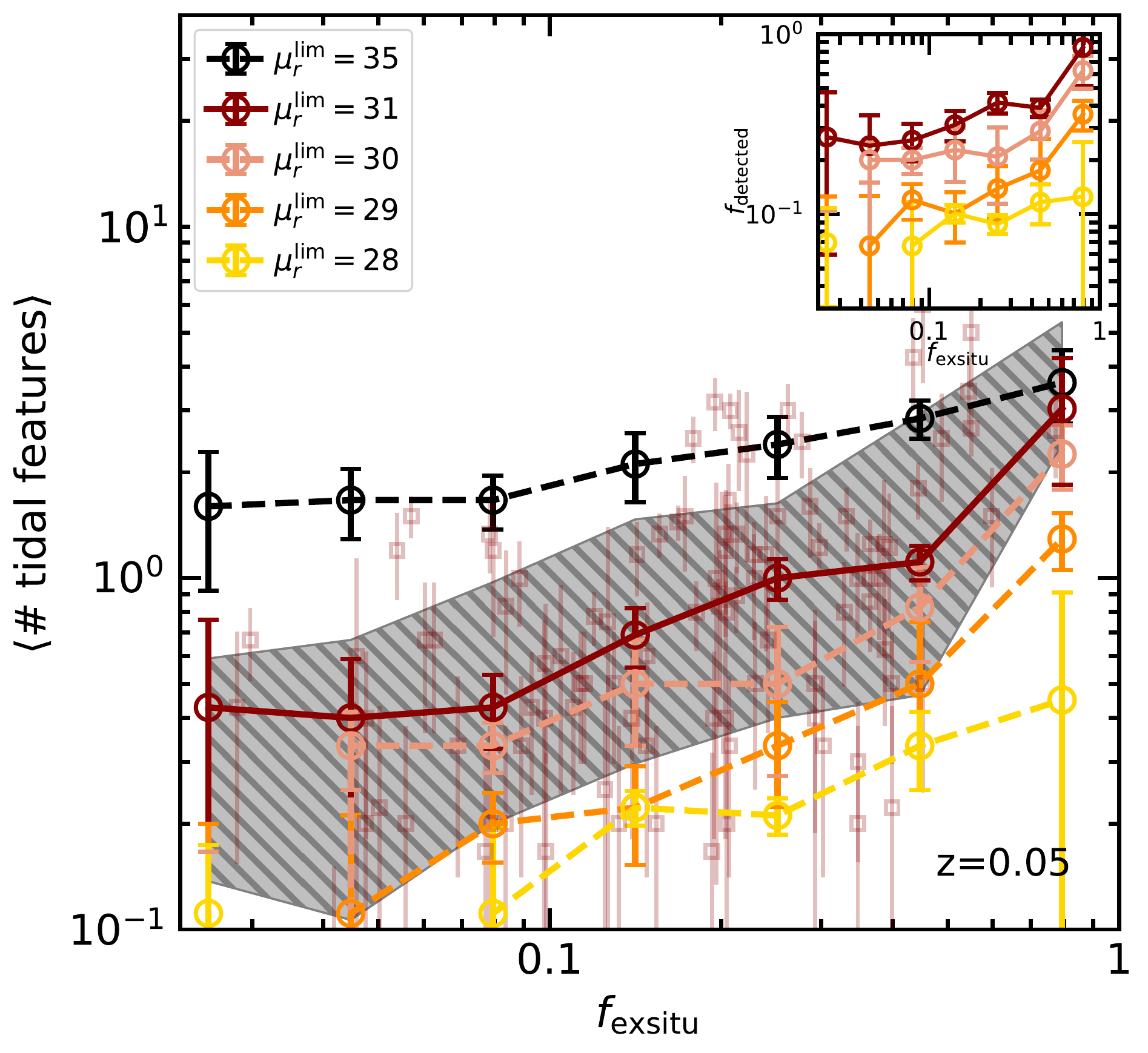}
    \caption{The average number of tidal features identified by classifiers as a function of galaxy ex-situ mass fraction. The solid red line indicates the average number of distinct tidal features identified per galaxy (excluding the merger, double nuclei and asymmetry / misc category) at $z=0.05$ for $\mu_{r}^{\rm lim}(3\sigma,10^{\prime\prime}\times10^{\prime\prime})=31$~mag\,arcsec$^{-2}$ as a function of $f_{\rm exsitu}$, with error bars determined by bootstrap. Coloured dashed lines show the same for different limiting surface brightnesses indicated in the legend. Light red squares with errorbars indicate the number of tidal features counted in individual galaxies for $\mu_{r}^{\rm lim}(3\sigma,10^{\prime\prime}\times10^{\prime\prime})=31$~mag\,arcsec$^{-2}$ with errors determined by the standard deviation across multiple classifiers and different projections. The hatched region indicates the central 68th percentile ($1\sigma$) spread for these points. The inset plot shows instead the number of tidal features recovered as a fraction of the average number of tidal features identified for $\mu_{r}^{\rm lim}(3\sigma,10^{\prime\prime}\times10^{\prime\prime})=35$~mag\,arcsec$^{-2}$ in each $f_{\rm exsitu}$ bin. We observe fairly weak evolution and large scatter in the number of tidal features identified with $f_{\rm exsitu}$, especially at higher limiting surface brightness.}
    \label{fig:visual_fexsitu}
\end{figure}

In this section we study how the number of tidal features identified changes as a function of limiting surface brightness and redshift. Figure \ref{fig:visual_fexsitu} shows the average number of distinct tidal features identified per galaxy (excluding mergers, double nuclei and miscellaneous asymmetries) at $z=0.05$ as a function of galaxy ex-situ mass for overlapping logarithmic bins of width 0.4 dex. Different coloured lines indicate our results based on mock images with different limiting surface brightnesses. We find a similar result to that shown in Figure \ref{fig:fraction_mass_exsitu} -- as $f_{\rm exsitu}$ increases the average number of tidal features identified also tends to increase. Red error bars indicate the number of tidal features detected in individual galaxies (for $\mu_{r}^{\rm lim}(3\sigma,10^{\prime\prime}\times10^{\prime\prime})=31$~mag\,arcsec$^{-2}$), with errors derived from the standard deviation in 3 different projections with each projection classified independently by at least two classifiers. There exists a fairly wide spread, which can be seen in the hatched region indicating the $1\sigma$ dispersion of the points. 

If we use the $\mu_{r}^{\rm lim}(3\sigma,10^{\prime\prime}\times10^{\prime\prime})=35$~mag\,arcsec$^{-2}$ line as a proxy for the true number of tidal features, we see that the average galaxy has at least one identified tidal feature regardless of the ex-situ mass fraction and that the average number of tidal features identified increases only modestly with $f_{\rm exsitu}$, although the trend does appear to strengthen for large $f_{\rm exsitu}$. Considering the large variation over time of the individual galaxy merger histories seen in Figure \ref{fig:merger_history_dy}, this fairly weak correspondence is perhaps not surprising. It is also true that, while there is only a weak dependence in the average number of tidal features with redshift, tidal features in galaxies with low $f_{\rm exsitu}$ are typically weaker so that, at brighter limiting surface brightnesses, the trend strengthens.

The inset plot shows the number of tidal features identified as a fraction of the average number identified at $\mu_{r}^{\rm lim}(3\sigma,10^{\prime\prime}\times10^{\prime\prime})=35$~mag\,arcsec$^{-2}$ in each bin. At all limiting surface brightnesses shown, a greater fraction of tidal features are identified for galaxies with higher ex-situ mass fractions. Considering that high mass or high $f_{\rm exsitu}$ galaxies tend to exhibit stronger tidal features (e.g. Figure \ref{fig:frac_recovered}), we can expect a greater fraction of tidal features to be bright enough to be detected at brighter limiting surface brightnesses. It is worth acknowledging that this reflects a possible observational bias -- that the tidal features present in lower mass haloes have fewer tidal features, but these tidal features are also likely to be weaker -- so that they are more likely to go undetected.

\begin{figure*}
    \centering
    \includegraphics[width=0.95\textwidth]{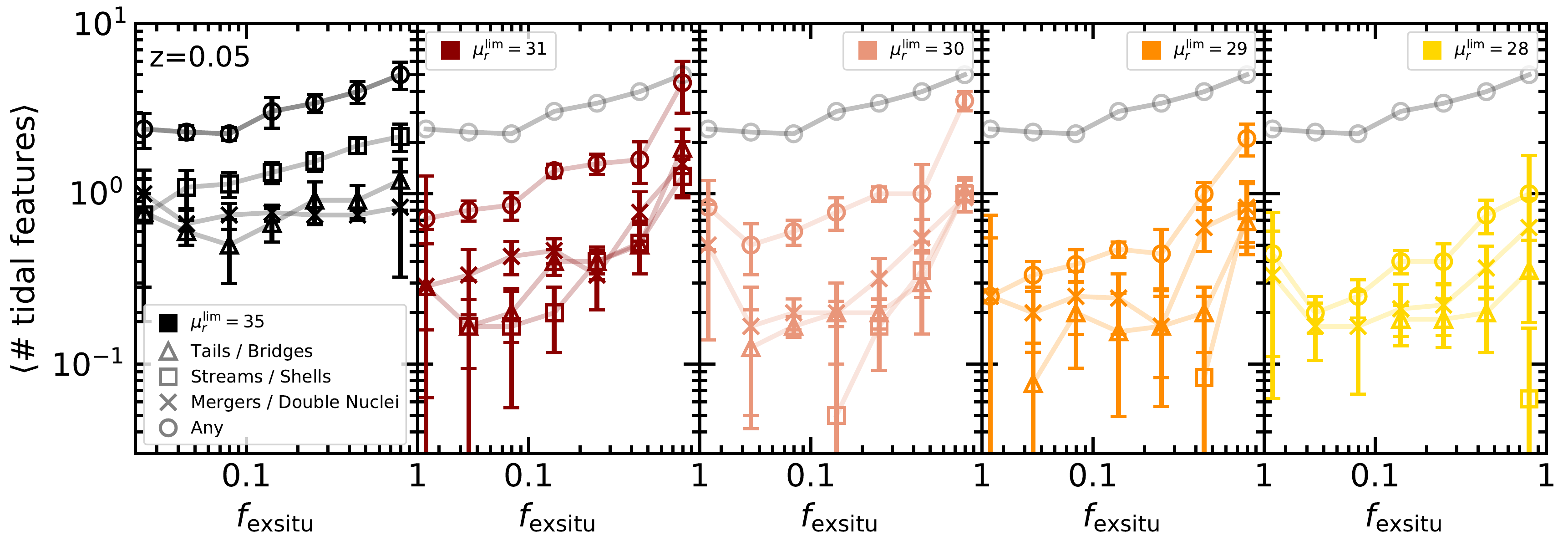}
    \caption{The average number of tidal features identified by classifiers as a function of galaxy ex-situ mass fraction at $z=0.05$ split by limiting surface brightness and tidal feature class. Tidal tails and bridges are indicated by triangular markers, tidal streams and shells by squares, mergers or double nuclei by crosses, and all tidal features (not including the miscellaneous category) by circles. Different panels correspond to different limiting surface brightnesses with a line for the `Any' category at $\mu_{r}^{\rm lim}(3\sigma,10^{\prime\prime}\times10^{\prime\prime})=35$~mag\,arcsec$^{-2}$ plotted in grey for comparison in each panel. Error bars show $1\sigma$ uncertainties obtained from bootstrapping.}
    \label{fig:visual_class}
\end{figure*}

Figure \ref{fig:visual_class} again shows the average number of distinct tidal features identified per galaxy as a function of ex-situ stellar mass. Tidal features are broken down into 3 categories: tidal tails and bridges, streams and shells, and merger remnants or double nuclei denoted by open triangles, squares and crosses respectively. We observe markedly different behaviour in the trends across $f_{\rm exsitu}$ and limiting surface brightness for different classes of tidal feature. Streams and shells are the best tracer of ex-situ mass, with tails, bridges, mergers and double nuclei occurring with roughly constant frequency across the range of ex-situ masses shown.

Being more numerous and longer lasting than other classes of tidal feature \citep[][]{Greco2018}, shells and streams are expected to better sample the average accretion history of the galaxy. However, at the limiting surface brightness achievable by the Rubin Observatory, streams and shells are the least frequently identified class of tidal feature (e.g. Figure \ref{fig:classifications_hist}). Since prevalence of shells and streams appears to decline strongly towards fainter limiting surface brightnesses ($\mu_{r}^{\rm lim}(3\sigma,10^{\prime\prime}\times10^{\prime\prime})\leq31$~mag\,arcsec$^{-2}$), especially compared with other classes of tidal feature such as tidal tails, this indicates a decline in tidal feature strength at smaller ex-situ masses, rather than being a direct tracer of the frequency of accretion events. This is a natural consequence of the fact that features such as streams and shells are typically formed in more unequal mass ratio mergers with lower mass satellites.

Below $\mu_{r}^{\rm lim}(3\sigma,10^{\prime\prime}\times10^{\prime\prime})=$31~mag\,arcsec$^{-2}$, there is almost no evolution in the normalisation of the trend with $f_{\rm exsitu}$ for both the tails/bridges and mergers/double nuclei classes, with a stronger trend emerging at high $f_{\rm exsitu}$. Since both the frequency and strength of tidal features changes with $f_{\rm exsitu}$ it is difficult to disentangle the effect, but the shallower relation seen for the faintest limiting surface brightness suggests that it is driven by these tidal features being generally fainter.

\begin{figure}
    \centering
    \includegraphics[width=0.45\textwidth]{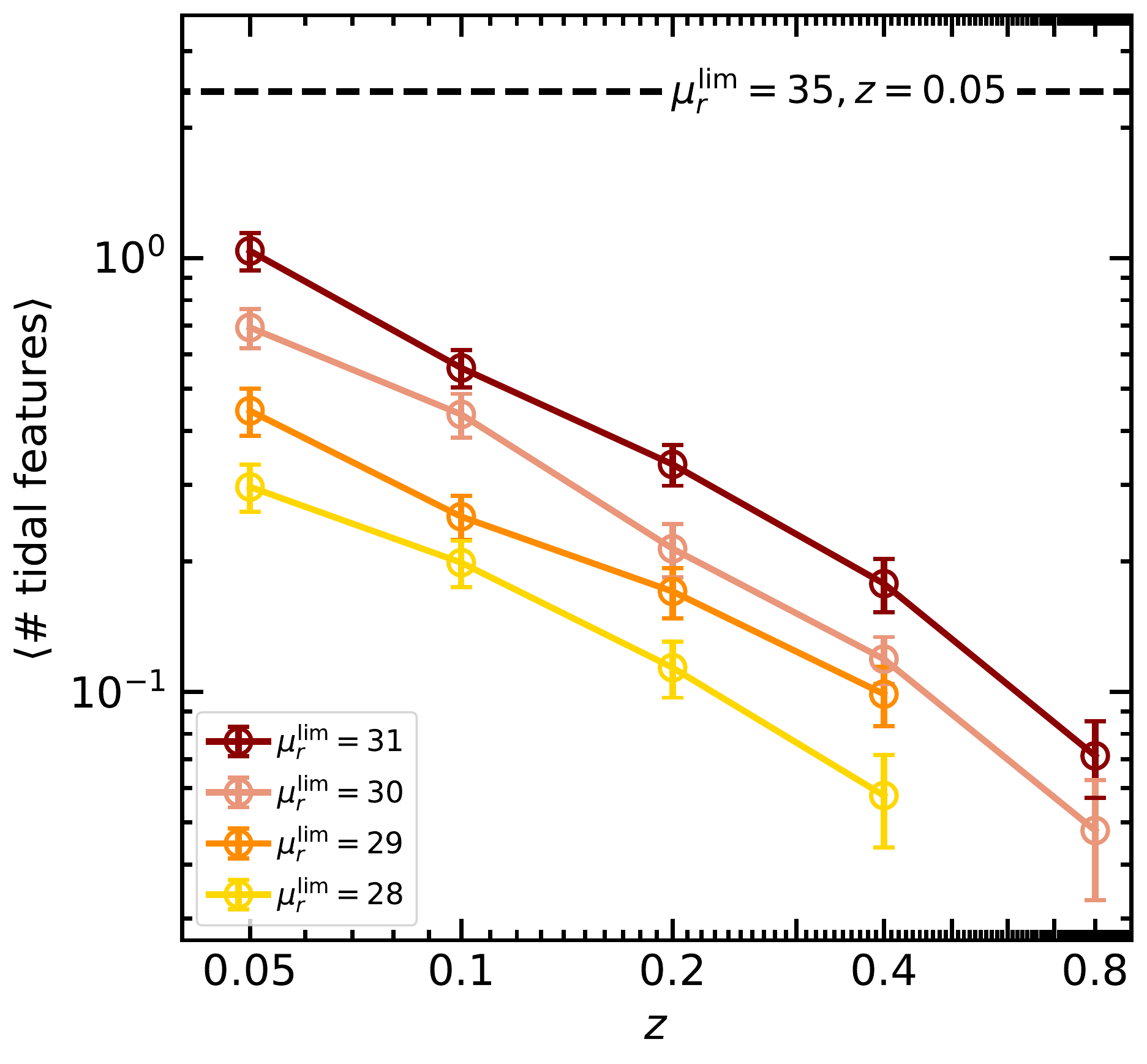}
    \caption{The average number of tidal features identified by classifiers as a function of redshift. Solid coloured lines indicate the average number of distinct tidal features identified per galaxy (excluding the asymmetry / misc category) as a function of redshift and for the different limiting surface brightnesses indicated in the legend. The dashed black line indicates the average number of tidal features identified for $\mu_{r}^{\rm lim}(3\sigma,10^{\prime\prime}\times10^{\prime\prime})=35$~mag\,arcsec$^{-2}$ at $z=0.05$. Lines do not extend to $z=0.8$ for the two brightest limiting surface brightnesses because low signal-to-noise makes visual classification too difficult for a majority of objects.}
    \label{fig:visual_z}
\end{figure}

Figure \ref{fig:visual_z} shows the average number of distinct tidal features identified per galaxy (excluding mergers, double nuclei and miscellaneous asymmetries) as a function of redshift. Different coloured lines correspond to different limiting surface brightnesses and the black dashed line indicates the average number of tidal features identified for $\mu_{r}^{\rm lim}(3\sigma,10^{\prime\prime}\times10^{\prime\prime})=35$~mag\,arcsec$^{-2}$ at $z=0.05$. The average number of tidal features identified per galaxy falls rapidly with redshift so that around ten times fewer tidal features are identified when the same galaxies are viewed at $z=0.8$ compared with $z=0.05$. This decline is principally a consequence of cosmological dimming and the PSF or pixel scale blurring features as apparent size of objects decreases. At fainter limiting surface brightnesses, enough light could be scattered from the central galaxy to its extended tidal features to obscure them, especially as they move further into the core of the PSF towards higher redshifts  While this is likely not a concern at limiting surface brightnesses achievable by the Rubin Observatory, the effect can become more important in deeper imaging (see Appendix \ref{sec:PSF_visibility} for further details).

\subsubsection{Effect of projection}

In this section, we consider the difference in classifications made in 3 different projections 90 degrees apart ($xy$, $xz$, $yz$) comparing the scatter in the number of features identified against the scatter between individual classifiers. We measure the standard deviation of the number of features identified for each galaxy in two ways: In the first case, we measure the standard deviation across classifiers, $\sigma_{\rm classifiers}$, treating each projection of the same object independently. In the second case, we measure the standard deviation across projections, $\sigma_{\rm projection}$, using the mean number of classifications across all classifiers for each projection. Any objects where no features of a given class were identified by any of the classifiers are not considered. We obtain the fractional standard deviation by dividing by the average number of features in each class and then take the RMS of this value over all galaxies.

\begin{figure}
    \centering
    \includegraphics[width=0.45\textwidth]{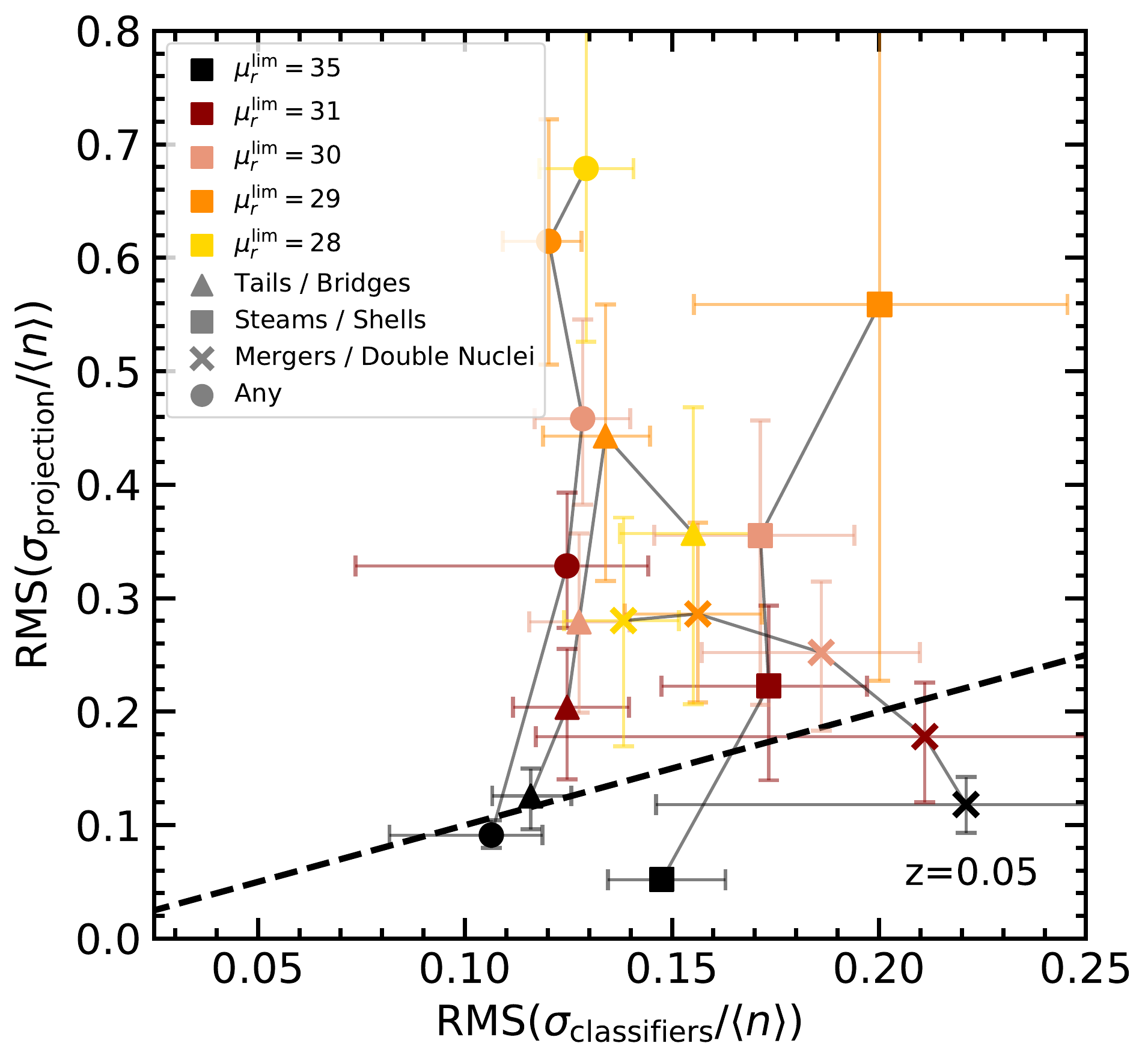}
    \caption{Evolution of the RMS fractional $\sigma_{\rm projection}$ and $\sigma_{\rm classifiers}$ and as a function of limiting surface brightness at $z=0.05$. Tidal tails and bridges are indicated by triangular markers, tidal streams and shells by squares, mergers or double nuclei by crosses and all tidal features (not including the miscellaneous category) by circles. Colours show the limiting surface brightness as indicated in the legend and error bars show $1\sigma$ uncertainties obtained from bootstrapping. No marker is plotted for shells at $\mu_{r}^{\rm lim}(3\sigma,10^{\prime\prime}\times10^{\prime\prime})=28$~mag\,arcsec$^{-2}$ because almost no objects display shells. While fainter limiting surface brightnesses typically improve the accuracy of classifications, the opposite is true in the case of the merger/double nuclei category.}
    \label{fig:tracks_uncert}
\end{figure}

Figure \ref{fig:tracks_uncert} shows the evolution of these quantities as a function of limiting surface brightness and for different classes of tidal feature at $z=0.05$. Tidal tails and bridges are indicated by triangular markers, tidal streams and shells by squares, mergers or double nuclei by crosses, and all tidal features (not including the miscellaneous category) by circles. Colours indicate the limiting surface brightness and error bars show $1\sigma$ uncertainties obtained from bootstrapping. At the brightest limiting surface brightness, the typical scatter in classifications is larger for $\sigma_{\rm classifiers}$, but towards fainter limiting surface brightnesses, $\sigma_{\rm projection}$ quickly becomes larger, while $\sigma_{\rm classifiers}$ does not change very significantly. In other words, at sufficient depth, disagreement between classifiers arising purely from subjective disagreement on the classification of identical images is the dominant source of uncertainty, but uncertainties arising from projection begin to dominate in shallower imaging. At expected LSST surface brightness limits ($30-31$ mag arcsec$^{-2}$), the uncertainty arising from projection is dominant for all but the mergers / double nuclei categories.

Most robust to changing the limiting surface brightness is the mergers and double nuclei category, which only sees a modest increase in uncertainty towards brighter limiting surface brightness. This is likely because features associated with merger remnants and double nuclei are typically bright so are robustly detected regardless of image depth. Interestingly, the variability of $\sigma_{\rm classifiers}$ for mergers and double nuclei is the largest of any category (while still quite modest) with an opposite trend to any of the other classes of tidal feature. This increase in uncertainty with fainter limiting surface brightness is also corroborated in the feedback given by classifiers, who mentioned that very deep imaging made certain features more difficult to classify. The visual appearance of lower surface brightness features, which tend to be significantly more extended, can change significantly through different projections leading to significant variation. In the case of higher surface brightness features, which are already clear at relatively bright limiting surface brightness (and less variable with projection), deeper imaging acts only to increase the prevalence of confounding sources which leads to an increase in the variance between classifiers at fainter limiting surface brightness. 

Based on both qualitative evidence from classifiers themselves and our quantitative analysis, we see that there are inherent uncertainties which cannot be completely removed. While there is typically improvement with deeper imaging, in some circumstances, classifications actually become less robust. Of course, a solution to this problem, which would be effective at least up to the surface brightness of the features themselves, would be to apply brighter surface brightness cuts than the actual limiting surface brightness of the data. However, if we note the difference in normalisation between the $\mu_{r}^{\rm lim}(3\sigma,10^{\prime\prime}\times10^{\prime\prime})=35$~mag\,arcsec$^{-2}$ and $31$~mag\,arcsec$^{-2}$ panels of Figure \ref{fig:visual_class}, it is evident that some merging systems / double nuclei are only revealed at very faint limiting surface brightnesses meaning some fraction of systems would be missed in this case. Also note that we do not model foreground and background objects or other astrophysical contaminants in our mock images, the inclusion of which would likely further reduce the agreement between classifiers.

\section{Summary}

\label{sec:summary}

In this paper we have performed a comprehensive theoretical investigation of the extended diffuse light around galaxies and galaxy groups down to low stellar mass densities and explored the reliability of human classifications under different observational biases. Our sample consists of 37 unique objects from the \textsc{NewHorizon} simulation whose progenitors we select at $z=0.2$, $z=0.4, 0.6$ and $0.8$ giving a total of 148 objects across 4 different redshifts with stellar masses $10^{9.5} < M_{\star}/{\rm M_{\rm \odot}} < 10^{11.5}$. Our main findings based on automated techniques and human visual classification are as follows:

\begin{enumerate}
\item  Distribution of tidal flux:
\begin{enumerate}
\item \textit{A large fraction of tidal flux is expected to be detectable at LSST 10-year depth.} Assuming the LSST pipeline is suitably optimised, 50 per cent of the total flux from substructure identifiable as distinct tidal features is detectable with a limiting surface brightness of $\mu_{r}^{\rm lim}(3\sigma,10^{\prime\prime}\times10^{\prime\prime})=30.5$~mag arcsec$^{-2}$. 90 per cent of the pixels that make up tidal features are brighter than $\mu_{r}^{\rm lim}(3\sigma,10^{\prime\prime}\times10^{\prime\prime})=32$~mag arcsec$^{-2}$ by area. However, almost all of the more diffuse light around galaxies (which makes up around 25 per cent of the total light in tidal features) will remain undetectable at a limiting surface brightness of $\mu_{r}^{\rm lim}(3\sigma,10^{\prime\prime}\times10^{\prime\prime})=30 - 31$~mag arcsec$^{-2}$ outside of very coarse binning.
\\
\item \textit{Much of the tidal flux in galaxies is found at large radii.} While 50 per cent of tidal flux is contained within $7~R_{\rm eff}$ of the galaxy centre on average, close to 100 per cent of flux is only reached by $25~R_{\rm eff}$ or $\sim 0.6~R_{\rm vir}$.
\\
\item \textit{The amount of tidal flux detected is strongly dependent on limiting surface brightness.} At brighter limiting surface brightnesses the normalisation of the relation between mass and tidal flux decreases so that the average $f_{\rm tidal}$ for a MW mass galaxy ($M_{\star}\approx10^{10.5}~{\rm M_{\odot}}$) decreases from $\sim 5$ per cent at $\mu_{r}^{\rm lim}(3\sigma,10^{\prime\prime}\times10^{\prime\prime})=31$~mag arcsec$^{-2}$ to only a fraction of a per cent at $\mu_{r}^{\rm lim}(3\sigma,10^{\prime\prime}\times10^{\prime\prime})=28$~mag arcsec$^{-2}$, while the scatter increases from 0.4~dex at $\mu_{r}^{\rm lim}(3\sigma,10^{\prime\prime}\times10^{\prime\prime})=31$~mag arcsec$^{-2}$ to 1.3~dex at $\mu_{r}^{\rm lim}(3\sigma,10^{\prime\prime}\times10^{\prime\prime})=28$~mag arcsec$^{-2}$.
\\
\item \textit{At predicted LSST limiting 10-year depth, a majority (75 per cent) of tidal flux in MW mass galaxies ($M_{\star} > 10^{10.5}~\rm{M_{\odot}}$) is detectable at $z=0.05$.} At low masses ($M_{\star} < 10^{10}~\rm{M_{\odot}}$), almost no galaxies are expected to exhibit visible tidal features. This is driven by the fact that tidal features are less frequent, but also generally weaker in lower mass galaxies. Even if a shallower final depth of 29.5 mag arcsec$^{-2}$ is assumed, we still expect the Rubin Observatory to detect more than 60 per cent of flux in tidal features for galaxies of MW mass or greater.
\\
\item \textit{Similarly, tidal features become significantly more difficult to detect at higher redshifts so that we would not expect to routinely identify any tidal features around MW mass galaxies beyond $z=0.2$.} While cosmological dimming is the primary driver, smearing of tidal features as they move into the core of the PSF may also play a role, particularly if very faint limiting surface brightnesses are considered. In this case, diffraction-limited, space-based observatories such as Roman \citep[][]{Robertson2019} and Euclid \citep[][]{Borlaff2021} offer an important complement to the Rubin Observatory.  
\\
\item \textit{ex-situ mass fraction correlates with galaxy mass and tidal flux fraction.} Partial correlation coefficients indicate a more favourable correlation with ex-situ mass (at a significance $2.2\sigma$) giving some indication that accretion history drives the tidal flux fraction beyond the simple correlation with mass. We observe a break in the relation between tidal flux and stellar mass at $10^{10.1\pm^{\scriptscriptstyle 0.01}_{\scriptscriptstyle 0.05}}$~${\rm M_{\odot}}$, corresponding to the crossover mass at which mergers are thought to become the dominant process driving galaxy evolution.
\\
\end{enumerate}
\item Reliability of human classification:
\begin{enumerate}
\item \textit{Galaxies in the \textsc{NewHorizon} simulation exhibit a range of analogues to observed tidal features.} While data at sufficient depth are relatively scant and exact comparisons are difficult, the \textsc{NewHorizon} simulation produces tidal features whose frequencies evolve with stellar mass in a way that is comparable to trends seen in available observational data.
\\
\item \textit{At very faint limiting surface brightnesses ($\mu_{r}^{\rm lim}(3\sigma,10^{\prime\prime}\times10^{\prime\prime})=35$~mag arcsec$^{-2}$), expert classifiers were able to identify specific tidal features in close to 100 per cent of galaxies ($M_{\star} > 10^{9.5}~\rm{M_{\odot}}$).} Certain features, like merger remnants were identified at roughly the same frequency regardless of limiting surface brightness, while the detection of shells was found to be much more sensitive to image depth.
\\
\item \textit{A greater number of tidal features were identified in galaxies with high ex-situ mass fractions.} When compared with the number of tidal features identified for a limiting surface brightness of $\mu_{r}^{\rm lim}(3\sigma,10^{\prime\prime}\times10^{\prime\prime})=35$~mag arcsec$^{-2}$, a greater fraction of tidal features are detected in galaxies with higher ex-situ mass fractions due to the fact that they are typically also brighter. This reflects a possible observational bias since the tidal features present in galaxies with a smaller number of tidal features are also likely to be weaker and are, therefore, more likely to go undetected.
\\
\item \textit{Concurrence between classifiers  generally improves with deeper imaging but morphologies can become more complex, introducing uncertainty in precise characterisation.} In particular, classifiers were less likely to concur with each other the presence of a merger remnant and double nuclei when viewing deeper images.
\\
\item \textit{Concurrence between classifiers is quite robust to different limiting surface brightnesses, but brighter limiting surface brightnesses produce much weaker agreement when classifications over different projections of the same object are compared.} Typically different projections of the same object produce a larger scatter in classifications than the scatter between different classifiers viewing the same object in the same orientation. 

\end{enumerate}
\end{enumerate}

Our findings, which are based on realistic Rubin Observatory mock images at the final LSST survey depth ($30-31$ mag arcsec$^{-2}$), indicate that the Rubin Observatory will be well situated to provide high quality observations of the tidal features surrounding galaxies. We expect the Rubin Observatory to open up a new region of discovery space by delivering sufficiently deep imaging down to intermediate redshifts ($z<0.2$) and stellar masses ($M_{\star}>10^{10}\,{\rm M_{\odot}}$) to study these structures in detail.


\section*{Acknowledgements}

GM thanks Peter Yoachim for sharing theoretical results for LSST surface brightness metrics and Yohan Dubois for fruitful discussion. JAB acknowledge financial support from CONICET through PIP 11220170100527CO grant. JLC acknowledges support from National Science Foundation (NSF) grant AST-1816196. CAC acknowledges support from the Science and Technology Research Council under grant ST/S006095/1 and LJMU. FAG  acknowledges financial support from FONDECYT Regular 1211370 and from the Max Planck Society through a Partner Group grant. GG gratefully acknowledges support by the ANID BASAL projects ACE210002 and FB210003. RAJ acknowledges support from the Yonsei University Research Fund (Yonsei Frontier Lab. Young Researcher Supporting Program) of 2021 and from the Korean National Research Foundation (NRF-2020R1A2C3003769). SK acknowledges support from the STFC [ST/S00615X/1] and a Senior Research Fellowship from Worcester College Oxford. JHK acknowledges financial support from the European Union's Horizon 2020 research and innovation programme under Marie Sk\l odowska-Curie grant agreement No 721463 to the SUNDIAL ITN network, from the State Research Agency (AEI-MCINN) of the Spanish Ministry of Science and Innovation under the grant "The structure and evolution of galaxies and their central regions" with reference PID2019-105602GB-I00/10.13039/501100011033, and from IAC project P/300724, financed by the Ministry of Science and Innovation, through the State Budget and by the Canary Islands Department of Economy, Knowledge and Employment, through the Regional Budget of the Autonomous Community. DJP acknowledges funding from an Australian Research Council Discovery Program grant DP190102448. JR acknowledges support from the State Research Agency (AEI-MCINN) of the Spanish Ministry of Science and Innovation under the grant "The structure and evolution of galaxies and their central regions" with reference PID2019-105602GB-I00/10.13039/501100011033. EAS thanks the LSSTC Data Science Fellowship Program, which is funded by LSSTC, NSF Cybertraining Grant \#1829740, the Brinson Foundation, and the Moore Foundation. AEW acknowledges support from the STFC [ST/S00615X/1]. MB acknowledges support from the Science and Technology Facilities Council through grant number ST/N021702/1. FB acknowledges support grants PID2020-116188GA-I00 and PID2019-107427GB-C32 from The Spanish Ministry of Science and Innovation. FD acknowledges support from the STFC [ST/V506709/1]. RD gratefully acknowledges support by the ANID BASAL projects ACE210002 and FB210003.

This work has made use of the Horizon cluster on which the simulation was post-processed, hosted by the Institut d'Astrophysique de Paris. We warmly thank S.~Rouberol for running it smoothly. 

The authors thank the referee, Andrew Cooper, for a detailed and constructive report which helped improve the final paper.

\section*{Data Availability}

The simulation data analysed in this paper were provided by the \textsc{NewHorizon} collaboration. The data will be shared on request to the corresponding author, with the permission of the \textsc{NewHorizon} collaboration or may be requested from \href{https://new.horizon-simulation.org/data.html}{https://new.horizon-simulation.org/data.html}.




\bibliographystyle{mnras}
\bibliography{paper_mnras} 
\vspace{0.1in}
\textbf{AFFILIATIONS}\\

\small{
\noindent $^{1}$Korea Astronomy and Space Science Institute, 776 Daedeokdae-ro, Yuseong-gu, Daejeon 34055, Korea\\
$^{2}$Steward Observatory, University of Arizona, 933 N. Cherry Ave, Tucson, AZ 85719, USA\\
$^{3}$Research Centre for Astronomy, Astrophysics \& Astrophotonics, Macquarie University, Sydney, NSW 2109, Australia\\
$^{4}$Department of Physics \& Astronomy, Macquarie University, Sydney, NSW 2109, Australia\\
$^{5}$Australian Research Council Centre of Excellence for All Sky Astrophysics in 3 Dimensions (ASTRO 3D)\\
$^{6}$INAF-Astronomical Observatory of Capodimonte, Salita Moiariello 16, I80131, Naples, Italy\\
$^{7}$Department of Astronomy, Case Western Reserve University, Cleveland, OH, USA\\
$^{8}$Space Telescope Science Institute, 3700 San Martin Drive, Baltimore, MD 21218, USA\\
$^{9}$Instituto de Astronom\'ia Teórica y Experimental, CONICET-UNC, Laprida. 854, X5000BGR, C\'ordoba, Argentina.\\
$^{10}$Observatorio Astron\'omico de C\'ordoba, Universidad Nacional de C\'ordoba, Laprida 854, X5000BGR,C\'ordoba, Argentina.\\
$^{11}$School of Physics, University of New South Wales, NSW 2052, Australia\\
$^{12}$NSF's NOIRLab/Rubin Observatory Project Office, 950 North Cherry Avenue, Tucson, AZ 85719, USA\\
$^{13}$Astrophysics Research Institute, Liverpool John Moores University, IC2, Liverpool Science Park, Liverpool L3 5RF, UK\\
$^{14}$Universit\'e de Strasbourg, CNRS, Observatoire astronomique de Strasbourg (ObAS), UMR 7550, 67000 Strasbourg, France\\
$^{15}$Instituto de Investigaci\'on Multidisciplinar en Ciencia y Tecnolog\'ia, Universidad de La Serena, Ra\'ul Bitr\'an 1305, La Serena, Chile\\
$^{16}$Departamento de Astronom\'ia, Universidad de La Serena, Av. Juan Cisternas 1200 Norte, La Serena, Chile\\
$^{17}$Instituto de Astrof\'{\i}sica, Pontificia Universidad Cat\'olica de Chile. Vicu\~na Mackenna 4860, Macul, Santiago, Chile\\
$^{18}$Instituto de Astronom\'{i}a. Universidad Nacional Aut\'{o}noma de M\'{e}xico A.P. 70-264, 04510, M\'{e}xico, D.F., México\\
$^{19}$Department of Astronomy,Yonsei University Observatory, Yonsei University, Seoul 03722, Republic of Korea\\
$^{20}$Centre for Astrophysics Research, Department of Physics, Astronomy and Mathematics, University of Hertfordshire, Hatfield, AL10 9AB, UK\\
$^{21}$Instituto de Astrof\'{i}sica de Canarias, V\'{i}a L\'{a}ctea S/N, E-38205 La Laguna, Spain\\
$^{22}$Departamento de Astrof\'{i}sica, Universidad de La Laguna, E-38206 La Laguna, Spain\\
$^{23}$School of Physics \& Astronomy, University of Birmingham, Birmingham, B15 2TT, UK\\
$^{24}$Department of Physics, Lancaster University, Lancaster, LA1 4YB, UK\\
$^{25}$Department of Physics and Astronomy, UCLA, PAB 430 Portola Plaza, Los Angeles, CA 90095-1547, USA\\
$^{26}$Department of Physics and Astronomy, University of California, Davis, One Shields Ave, Davis, CA 95616, USA\\
$^{27}$Center for Interdisciplinary Exploration and Research in Astrophysics (CIERA) and Department of Physics and Astronomy, Northwestern University, 1800 Sherman Ave, Evanston, IL 60201, USA\\
$^{28}$Institut d'Astrophysique de Paris, Sorbonne Universit\'es, UMPC Univ Paris 06 et CNRS, UMP 7095, 98 bis bd Arago, 75014 Paris, France\\
$^{29}$School of Physics, Korea Institute for Advanced Study (KIAS), 85 Hoegiro, Dongdaemun-gu, Seoul, 02455, Republic of Korea\\
$^{30}$Institute for Astronomy, University of Edinburgh, Royal Observatory, Edinburgh EH9 3HJ, UK\\
$^{31}$IPAC, California Institute of Technology, 1200 E. California Blvd., Pasadena, CA 91125\\
$^{32}$Departamento de F\'{i}sica Te\'{o}rica, At\'{o}mica y \'{O}ptica, Universidad de Valladolid, 47011 Valladolid, Spain\\
$^{33}$Instituto de Astrof\'{\i}sica e Ci\^{e}ncias do Espa\c{c}o, Universidade de Lisboa, OAL, Tapada da Ajuda, PT1349-018 Lisbon, Portugal\\
$^{34}$INAF - Catania Astrophysical Observatory, Via S. Sofia, 78, 95123 - Catania - Italy\\
$^{35}$Departamento de Astronom\'ia, Facultad de Ciencias F\'isicas y Matem\'aticas,
Universidad de Concepci\'on, Concepci\'on, Chile\\
$^{36}$Universidad Internacional de Valencia. Carrer del Pintor Sorolla 21, 46002, Valencia, Spain.\\
$^{37}$Centre for Astrophysics and Supercomputing, Swinburne University of Technology, Hawthorn, VIC 3122, Australia\\
$^{38}$Department of Physics and Astronomy, University of Louisville, Natural Science Building 102, 40292 KY, Louisville, USA\\
$^{39}$Astrophysics Research Centre, University of KwaZulu-Natal, Westville Campus, Durban 4041, South Africa\\
$^{40}$Dept. of Physics and Astronomy, The University of Alabama, Tuscaloosa, AL 35487\\
$^{41}$Minnesota State University, Mankato; Trafton North 141 Mankato MN 56001 USA\\
$^{42}$Inter-University Centre for Astronomy and Astrophysics, Post Bag 4, Ganeshkhind, Pune 411007, India\\
$^{43}$OzGrav-Swinburne, Centre for Astrophysics and Supercomputing, Swinburne University of Technology, Hawthorn, VIC 3122, Australia\\
$^{44}$Departamento de F\'isica, Facultad de Ciencias, Universidad Nacional Aut\'onoma de M\'exico, Ciudad Universitaria, CDMX 04510, M\'exico\\
$^{45}$National Radio Astronomy Observatory, 520 Edgemont Road, Charlottesville, VA 22033\\
}
\appendix

\section{Effective resolution limit for detecting shells}
\label{sec:res_limit}

The stellar particle mass resolution of a cosmological simulation puts limits on its ability to resolve structures. For instance, an image of a structure with a total stellar mass close to the stellar mass resolution of a simulation may not have sufficient contrast against the background of the host galaxy to be detectable, even if the particles that make up the structure are kinematically distinct from the galaxy. Given the stellar mass of a structure in a simulation, one can estimate the number of stellar particles in the structure and use this number to predict whether it is resolved (detectable) in a simulation with known stellar particle mass. In this section, we present the lower limits on detectable shells based on the \textsc{NewHorizon} stellar particle mass resolution ($1.3\times10^4~{\rm M_{\odot}}$).

Following Bazkiaei et al. (in preparation), we combine analytical profiles of shells based on \citet{Sanderson2010} and \citet{Sanderson2013} with \citet{Sersic1968} models in order to estimate the numerical limits of the simulation to resolve tidal features.

Table \ref{tab:galcat} specifies four host galaxy S\'{e}rsic models which we select to bracket a realistic range of parameter space and to be roughly representative of galaxies found in the four mass bins shown in Figure \ref{fig:frac_recovered}. Shell models are generated according to Equations (1) and (19) of \citet{Sanderson2013}. Following the same notation used by \citet{Sanderson2013}, we generate models for a range of characteristic widths, $\delta_{r}$, galactocentric radii $r_{s}$ and opening angles, $\alpha$, as well as for a range of stellar masses.

\begin{table}
	\centering
	\caption{Summary of the parameters used to generate each model galaxy: $a$, model name; $b$, total stellar mass; $c$, S\'{e}rsic index; $d$, effective radius.}
	\label{tab:galcat}
	\begin{tabular}{@{}llllll@{}} 
		\toprule
		Model$\,^{a}$ & $M_{*}/{\rm M_{\odot}}\,^{b}$ & $n\,^{c}$ & $r_{\rm eff}\ [{\rm kpc}]\,^{d}$ \\
		\midrule
		Galaxy 1 & 10$^{9.5}$ & 0.5 & 2\\
		Galaxy 2 & 10$^{10.0}$ & 1.0 & 4\\
		Galaxy 3 & 10$^{10.5}$  & 2.5 &  6\\
		Galaxy 4 & 10$^{11.0}$ & 4.0 &  8\\
		\bottomrule
	\end{tabular}
\end{table}

To find the stellar mass within a shell, we produce mass maps for every combination of galaxy and shell model matching the Rubin Observatory 0.2$^{\prime \prime}$ pixel scale (corresponding to a physical size of 0.19, 0.37, 0.66, 1.08, 1.51 kpc per pixel for the lowest to highest redshift of this work). The maps are then re-binned to 1$^{\prime \prime}$ to mimic the procedure we use to produce mock images from \textsc{NewHorizon} galaxies (see Section \ref{sec:making_mocks_class}). Two types of re-binned 1$^{\prime \prime}$ image created:
\begin{enumerate}
    \item \textit{1$^{\prime \prime}$ mass maps} -- we perform a simple re-binning of the original 0.2$^{\prime \prime}$ maps to 1$^{\prime \prime}$ -- used to determine the stellar mass of each shell model and the region of each galaxy overlapped by the shell model.
    \item \textit{PSF convolved surface brightness maps} -- the 0.2$^{\prime \prime}$ mass maps are used to calculate the $r$-band surface brightnesses of the tidal feature models following the same procedure as detailed in Section \ref{sec:raw_mock_images} and assuming all stars making up the tidal features are born at $z=2$ and have a metallicity of $Z=0.1$. They are then convolved with the Hyper Suprime-Cam PSF \citep[][]{Montes2021} and re-binned to 1$^{\prime \prime}$.

\end{enumerate}

Using the first set of mass maps, we calculate the signal-to-noise-ratio (SNR$_{\rm shell}$) for each shell as follows:
\begin{equation}
    \label{eqn:shellsnrsa}
    {\rm SNR}_{\rm shell} = \frac{n_{\rm shell}}{\sigma_{\rm shell}} = \frac{n_{\rm shell}}{\sqrt{2\,n_{\rm galaxy} + n_{\rm shell}}},
\end{equation}

\noindent where $n_{\rm shell}$ and $n_{\rm galaxy}$ are the total number of star particles that comprise the arc of the shell (i.e. the brightest part) and galaxy respectively within region described by the arc of the shell model given the stellar mass resolution of the simulation. Shells are considered to be detected if ${\rm SNR}_{\rm shell}>5$.

Then, using the PSF convolved surface brightness maps, we calculate the average surface brightness across the arc of the shell in order to find the faintest detectable shell from among all the detected shell models.

\begin{figure}
	\centering
	\includegraphics[width=0.45\textwidth]{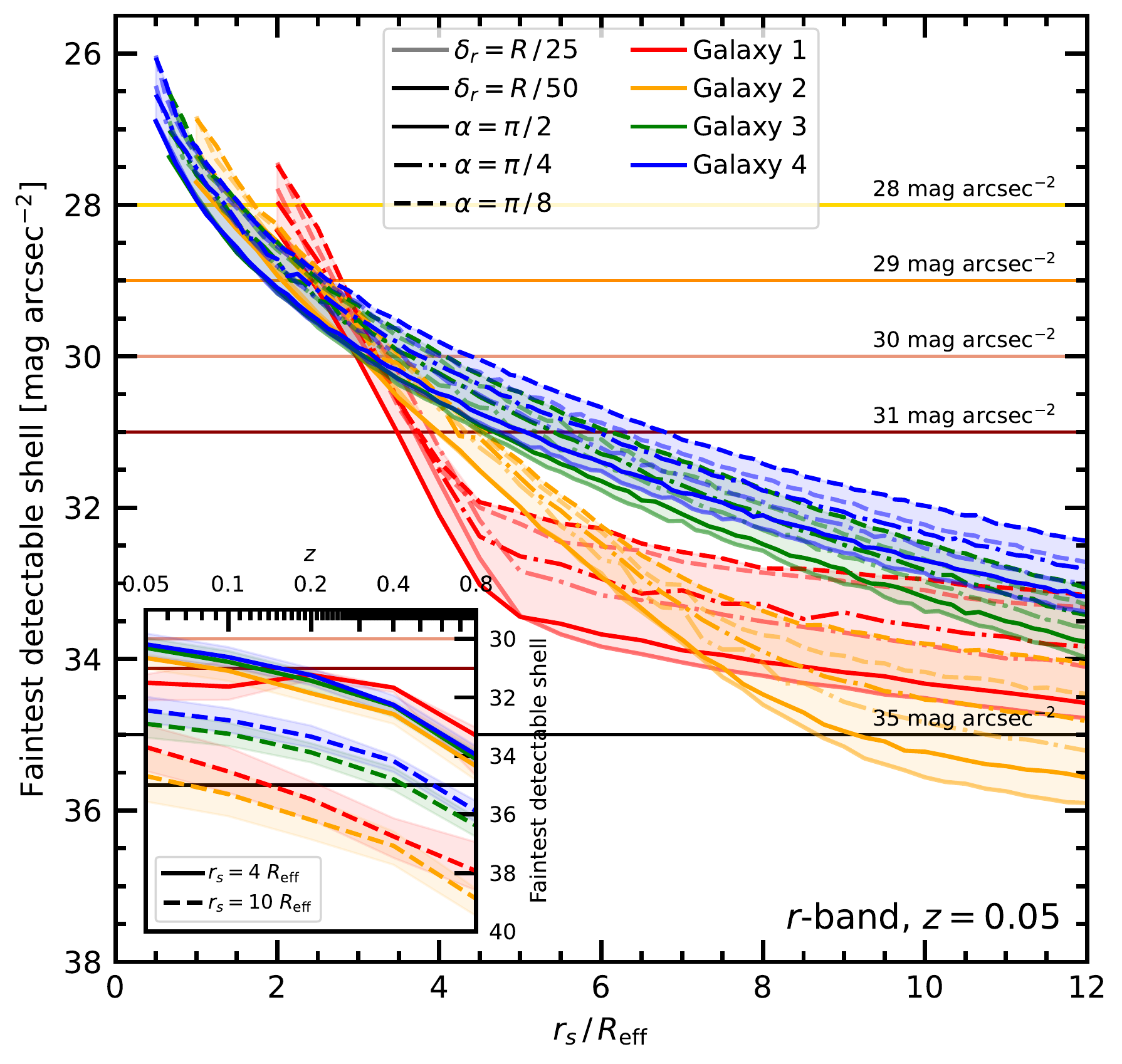}
    \caption{The main plot shows numerical lower limits for the average surface brightness of the faintest detectable shell as a function of galactocentric radius for different galaxy and shell models. Different coloured lines indicate different galaxy S\'{e}rsic models, while solid, dot-dashed and dashed lines indicate different opening angles. Darker or lighter colours indicate different characteristic widths with all values shown in the legend. We expect any shells (with a given set of parameters and at a given radius) with surface brightnesses fainter than these lines to be undetectable in \textsc{NewHorizon} mock images, regardless of the image depth. For reference we also show coloured horizontal lines, which highlight the surface brightness limits of mock images used throughout this work. The inset plot shows the surface brightness of the faintest detectable shell averaged over the different parameter values of $\delta_{r}$ and $\alpha$ and measured at 4~$R_{\rm eff}$ (solid lines) and 10~$R_{\rm eff}$ (dashed lines) as a function of redshift. Line colours again indicate the S\'{e}rsic models while the shaded region indicate the range of values measured across the different shell parameters.}
    \label{fig:faintest_detectable_shells}
\end{figure}

The surface brightness of faintest detectable shells, based on the \textsc{NewHorizon} stellar particle mass resolution are presented in Figure \ref{fig:faintest_detectable_shells}. This Figure shows the numerical lower limits for the faintest $r$-band surface brightness at which shells around each of the model galaxies are detected as a function of galactocentric radius. Different galaxy models and shell model parameters are represented by different colours and line styles respectively.

Even with its relatively high stellar mass resolution, we expect \textsc{NewHorizon} to struggle to resolve shells with sufficient contrast close to the central parts of galaxies. Around our most massive model galaxy in particular, signal-to-noise is not sufficient to detect any of our model shells with surface brightnesses fainter than 31 mag arcsec$^{-2}$ within galactocentric radii smaller than 4.5~$R_{\rm eff}$. For less massive galaxy models, it is possible to detect faint shells at significantly smaller radii, however.

Overall, \textsc{NewHorizon} is expected to resolve most of the shells (and likely other types of tidal feature) which which would be detectable at limiting surface brightnesses realistically achievable by the main LSST survey ($\mu_{r}^{\rm lim}(3\sigma,10^{\prime\prime}\times10^{\prime\prime})\leq31$~mag\,arcsec$^{-2}$) outside of the central few effective radii. More care needs to be taken in interpreting predictions at higher limiting surface brightnesses as observationally detectable features may not reach sufficient signal-to-noise to be resolved in the simulation out to relatively large radii ($r_{s}\gg10~R_{\rm eff}$).

Note that in our treatment we do not account for possible differences in the stellar populations of the galaxy and tidal features which result in differing mass-to-light ratios between the galaxy and shell. Tidal features may also be easier to detect in false colour images in cases where their colours differ enough from host galaxy, even if there is insufficient signal-to-noise in any single band. We cannot quantify how much this would improve our results, except to say that we do see this effect in at least some mock images. Without knowing the underlying spatial distribution of shells, it is difficult quantify the significance of the impact that an inability to detect shells close to the central galaxy has, however. On the other hand, depending on viewing angle, shells may be expected to be brighter at their maximum than other types of tidal feature since stars accumulate at the shell apocenter. However, since we consider an average signal-to-noise across the whole arc of the shell, we expect this effect to be lessened.

\section{Effect of the PSF on the visibility of tidal features}
\label{sec:PSF_visibility}

\begin{figure}
    \centering
    \includegraphics[width=0.45\textwidth]{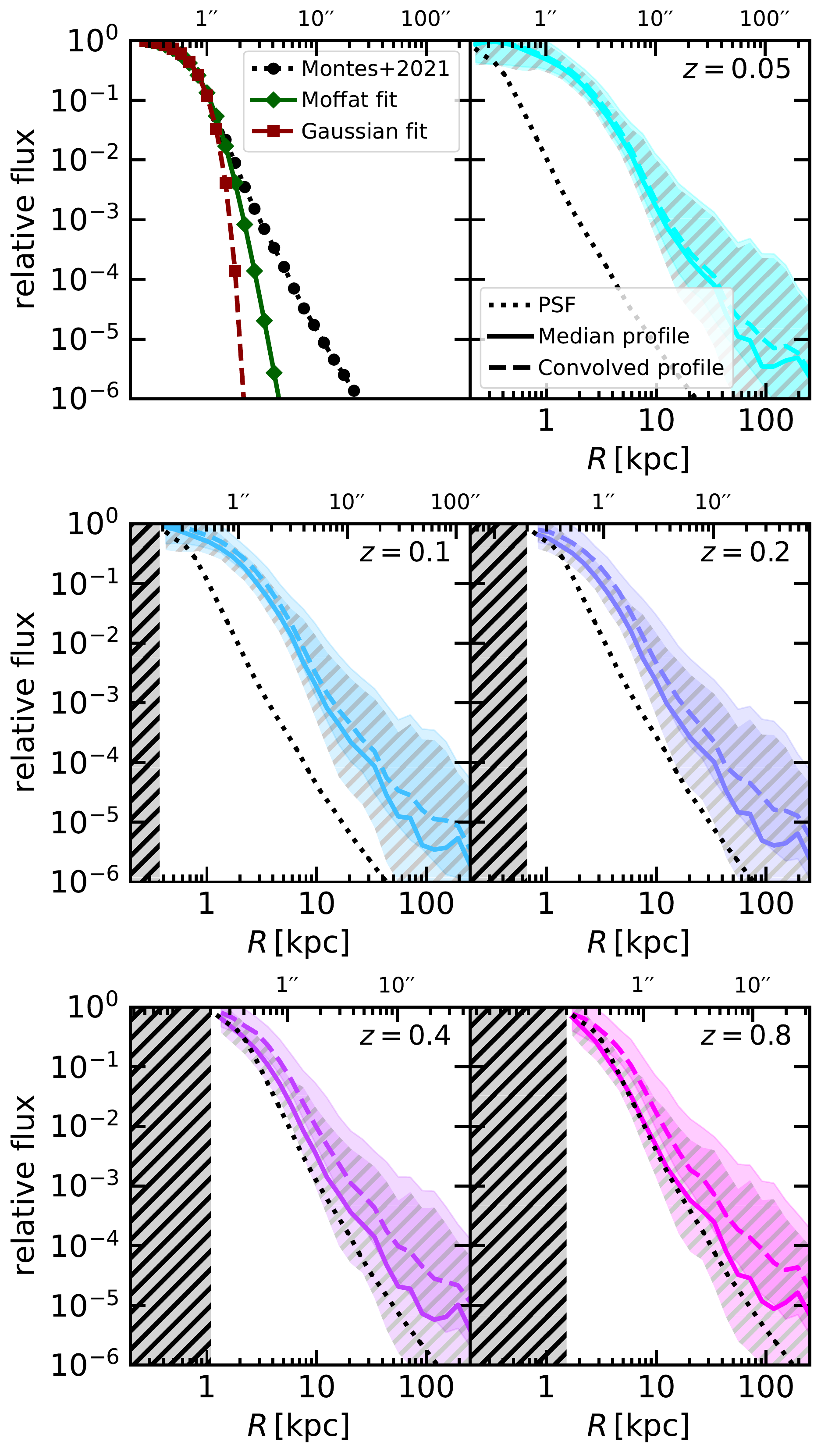}
    \caption{\textbf{Top left}: Comparison of the measured 1D PSF from \citet[][]{Montes2021} with a the best fit \citet[][]{Moffat1969} and Gaussian distributions. \textbf{Other panels}: surface brightness profiles compared to the PSF for different redshifts (indicated in the top right corner of each panel). Solid lines show the median relative surface brightness profile of the raw mock images, dashed lines show the median relative surface brightness profiles after they have been convolved with the PSF, and the dotted black line shows the \citet[][]{Montes2021} PSF. Coloured filled regions indicate the central 68th percentile ($1\sigma$) of profiles where the hatched region indicates the unconvolved profile. The $x$-axis limits are fixed in physical units between 0.2~kpc and 250~kpc with the equivalent scale in arcseconds for each redshift shown on the top $x$-axis of each panel. The grey hatched region indicates radii smaller than the $0.2^{\prime\prime}$ Rubin Observatory pixel size.}
    \label{fig:PSF}
\end{figure}

Figure \ref{fig:PSF} illustrates the effect of the PSF on the median surface brightness profile of our galaxies. In the top right panel we show a comparison of the measured PSF from \citet[][]{Montes2021} and best fitting \citet[][]{Moffat1969} and Gaussian distributions. The wings of the PSF are not well described by either Gaussian or Moffat distributions beyond a few arcseconds and this becomes increasingly severe towards large radii. We therefore find that neither are appropriate choices for modelling the effect of the PSF in the faint outskirts of the galaxy. The remaining panels show median surface brightness profiles compared to the PSF for different redshifts. The solid lines show the median relative surface brightness profile of the raw mock images, dashed lines show the median relative surface brightness profiles after they have been convolved with the PSF and the dotted black line indicates the PSF. The profiles and PSF in each panel are scaled to a fixed physical scale, with the angular scale indicated separately on the upper $x$-axis. Towards higher redshift, as the increasing angular scale means that fainter regions of the galaxy surface brightness profile move further into the core of the PSF, the convolved profiles start to depart significantly from the original profiles. As we neglect any possible contribution of scattered light from stars, this effect is entirely the result of smearing and the scattering of light from the bright core of the galaxy into the fainter outskirts. We therefore expect that, for very deep imaging, the PSF will have some impact on how well we are able to detect tidal features. For a noise level of $\mu_{r}^{\rm lim}(3\sigma,10^{\prime\prime}\times10^{\prime\prime})=31$~mag\,arcsec$^{-2}$, we find that the visibility of tidal features is noticeably impacted by the PSF, becoming especially apparent after $z=0.4$. However, we do not find that this effect is strong enough that previously visible tidal features routinely become invisible, particularly at typical limiting surface brightnesses accessible to the Rubin Observatory. As instruments improve further and it becomes possible to probe even deeper into the outskirts of galaxies, we can expect that this effect will become more important.

\section{Selecting the ex-situ mass time interval}
\label{sec:exsitu_interval}

We try to choose a time interval that gives the tightest relation between ex-situ mass and halo mass, however it is not obvious what this interval should be. While longer timescales probe more of the accretion history galaxy, they may not reflect the current state of the galaxy (for example, if the galaxy recently underwent a merger) and so at some point may begin to correlate poorly with the halo mass. Equally a timescale that is too small will be effected more strongly by the stochasticity inherent in galaxy accretion histories.

By changing the value of $t_{\rm min}$ in Equation \eqref{eqn:exsitu}, we vary the time over which we measure the ex-situ mass fraction. Here we consider the affect of adopting different time intervals $\Delta t$ such that $t_{\rm min} = t_{\rm max} - \Delta t$. We then measure the distance correlation coefficient \citep[][]{Szekely2007} between $f_{\rm exsitu}$ and $M_{h}$.

\begin{figure}
    \centering
    \includegraphics[width=0.45\textwidth]{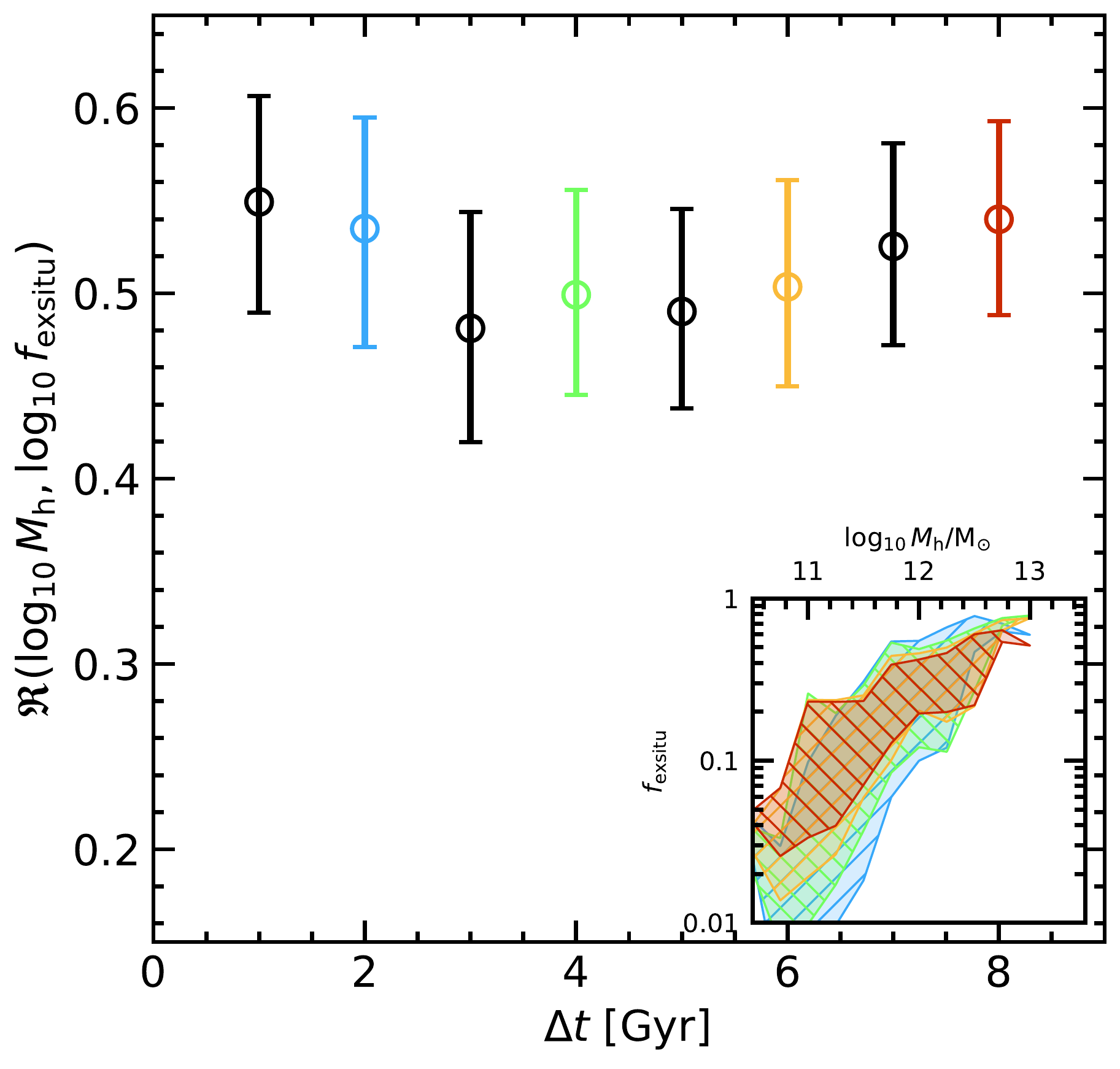}
    \caption{Correlation coefficient for different time intervals with $1\sigma$ uncertainties indicated by error bars. The inset plot shows a region enclosing the $1\sigma$ scatter for the $M_{h}$--$f_{\rm exsitu}$ relation for $\Delta t = [2,4,6,8]$ where the colour of the region corresponds to the value $\Delta t$ indicated by the coloured error bars in the main plot.}
    \label{fig:dt_test}
\end{figure}

\begin{figure}
    \centering
    \includegraphics[width=0.45\textwidth]{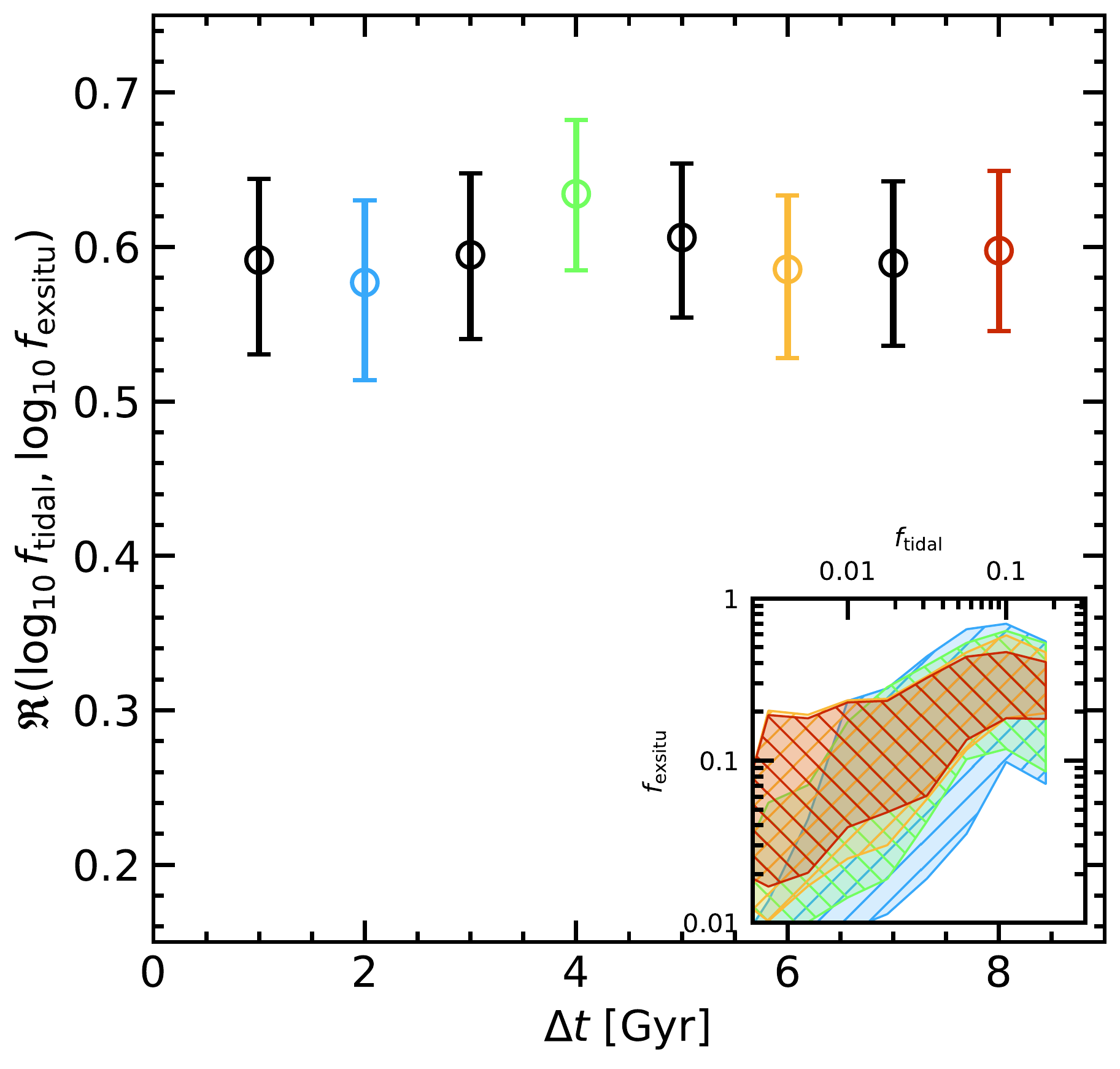}
    \caption{Correlation coefficient for different time intervals with $1\sigma$ uncertainties indicated by error bars. The inset plot shows a region enclosing the $1\sigma$ scatter for the $M_{h}$--$f_{\rm exsitu}$ relation for $\Delta t = [2,4,6,8]$ where the colour of the region corresponds to the value $\Delta t$ indicated by the coloured error bars in the main plot.}
    \label{fig:dt_test_exsitu}
\end{figure}

Figures \ref{fig:dt_test} and \ref{fig:dt_test_exsitu} show how the correlation coefficient behaves for different values of $\Delta t$. Figure \ref{fig:dt_test} shows the correlation between the correlation between halo mass and $f_{\rm exsitu}$ and Figure \ref{fig:dt_test_exsitu} shows the correlation between $f_{\rm tidal}$ and $f_{\rm exsitu}$. Open circles with error bars indicate the value of the correlation coefficient and associated $1\sigma$ uncertainty as a function of $\Delta t$. In the inset panel we plot filled and hatched regions which enclose the $1\sigma$ scatter of the $M_{h}$--$f_{\rm exsitu}$ relation for multiple values of $\Delta t$ ($\Delta t = [2,4,6,8]$). The colour of each region corresponds to the value $\Delta t$ and is indicated by the 4 coloured error bars in the main plot.

We find that increasing the timescale does not have a significant influence on the level of correlation in either case. We therefore adopt the maximum possible value of $\Delta t$ for each galaxy (i.e. over the whole lifetime of the galaxy) as this better reflects the overall accretion history of the galaxy.

\section{Redshift evolution}
\label{sec:z_evo}

We calculate the change in $f_{\rm exsitu}$ and $f_{\rm tidal}$ for our sample of 37 galaxies in the time interval between highest and lowest redshifts that we consider ($z=0.8$ and $z=0.2$) and define the growth rate of $f_{\rm exsitu}$, $\Gamma_{\rm exsitu}$; Equation \eqref{eqn:gamma_exsitu}, and the tidal mass fraction, $\Gamma_{\rm tidal}$; Equation \eqref{eqn:gamma_tidal}, as follows:
\begin{equation}
    \label{eqn:gamma_exsitu}
    \Gamma_{\rm exsitu} = \frac{f_{\rm exsitu}(z=0.2, z = \infty) - f_{\rm exsitu}(z=0.8, z = \infty)}{f_{\rm exsitu}(z=0.2, z = \infty)\, \Delta t},
\end{equation}

\noindent where $f_{\rm exsitu}$ is defined in Equation \ref{eqn:exsitu} and $\Delta t$ is the time between $z=0.8$ and $z=0.2$ ($\sim 4$~Gyr).
\begin{equation}
    \label{eqn:gamma_tidal}
    \Gamma_{\rm tidal} = \frac{f_{\rm tidal}(z=0.2) - f_{\rm tidal}(z=0.8)}{f_{\rm tidal}(z=0.2)\, \Delta t}
\end{equation}

\noindent where $f_{\rm tidal}$ is defined in Equation \ref{eqn:ftidal}.
\begin{figure}
    \centering
    \includegraphics[width=0.45\textwidth]{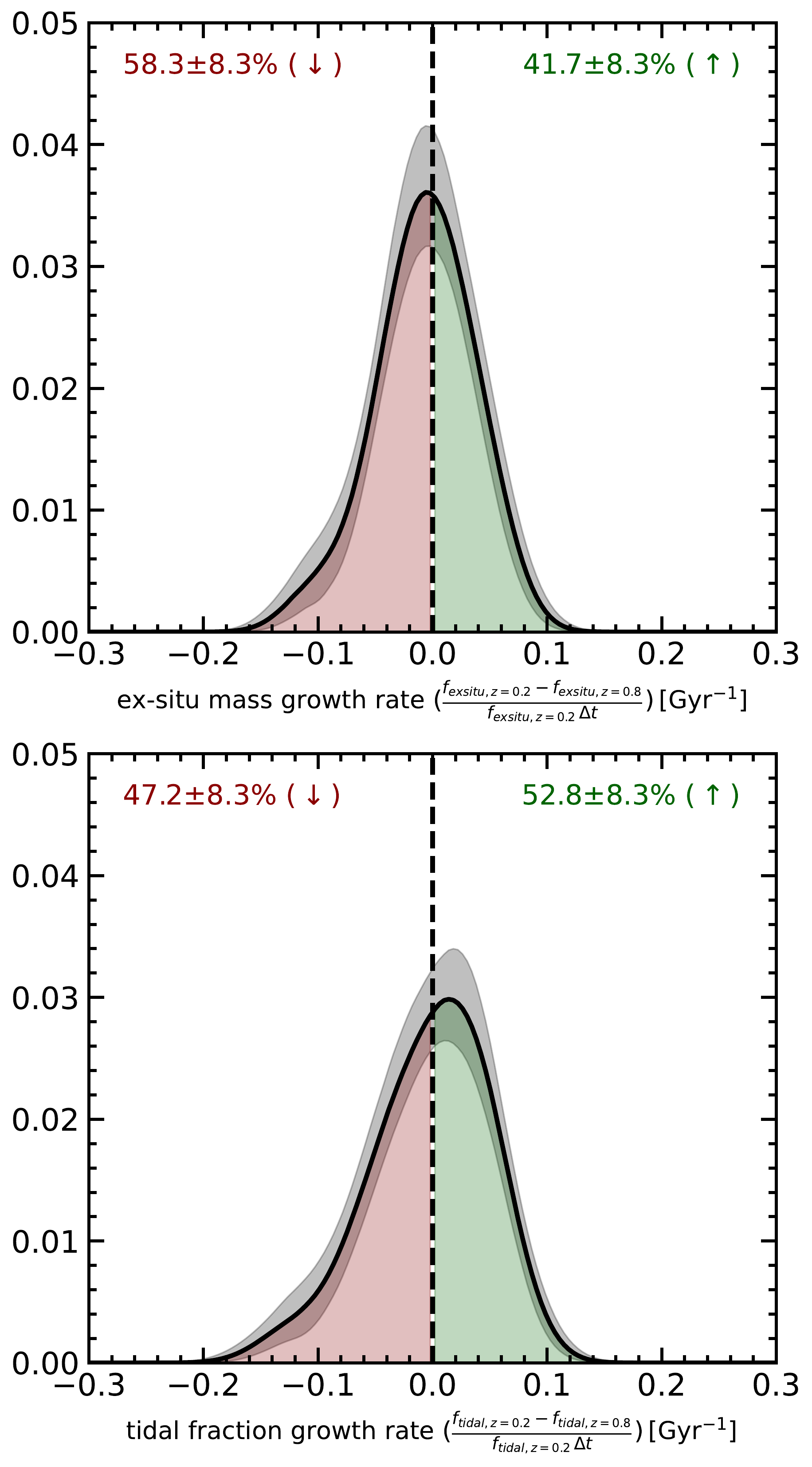}
    \caption{\textbf{Top}: the distribution of the $f_{\rm exsitu}$ growth rate ($\Gamma_{\rm exsitu}$) from Gaussian kernel density estimates from 10,000 bootstraps. \textbf{Bottom}: the distribution of the $f_{\rm tidal}$ growth rate ($\Gamma_{\rm tidal}$) from Gaussian kernel density estimates from 10,000 bootstraps. Both growth rates are calculated between $z=0.8$ and $z=0.2$. The grey region in both panels shows the $1\sigma$ uncertainty in the kernel density estimate.}
    \label{fig:fraction_flux_dy}
\end{figure}

Figure \ref{fig:fraction_flux_dy} shows the distribution $\Gamma_{\rm exsitu}$ (top panel) and $\Gamma_{\rm tidal}$ (bottom panel) both obtained from Gaussian kernel density estimates using 10,000 bootstraps. The numbers in red and green at the top of both panels indicate the percentage of galaxies whose growth rate is either negative or positive respectively with their associated $1\sigma$ errors. The standard deviation of the distribution of $\Gamma_{\rm exsitu}$ and $\Gamma_{\rm tidal}$ is 0.15~Gyr$^{-1}$ and 0.18~Gyr$^{-1}$ respectively, indicating relatively large swings in the growth rate (amounting to a greater than $\sim 50$ per cent change over the 4 Gyrs between $z=0.8$ and $z=0.2$ in 40--50 per cent of the population). In both cases the growth rate in the redshift range between $z=0.8$ and $z=0.2$ is consistent with an equal number of galaxies having negative and positive growth rates.

Additionally, the median $\Gamma_{\rm exsitu}$ and the median $\Gamma_{\rm tidal}$ are both consistent with no average change ($-0.0186\pm 0.0261$~Gyr$^{-1}$ for $\Gamma_{\rm exsitu}$ and $-0.0230\pm 0.0652$~Gyr$^{-1}$ for the $\Gamma_{\rm tidal}$). This is also true if we consider the overall fractional change in the net $f_{\rm exsitu}$ and $f_{\rm tidal}$, which we calculate by taking the mean value weighted by the host galaxy masses ($-0.0292\pm 0.0295$~Gyr$^{-1}$ for $\Gamma_{\rm exsitu}$ and $-0.0187\pm 0.0246$~Gyr$^{-1}$ for the $\Gamma_{\rm tidal}$).

\begin{figure}
    \centering
    \includegraphics[width=0.45\textwidth]{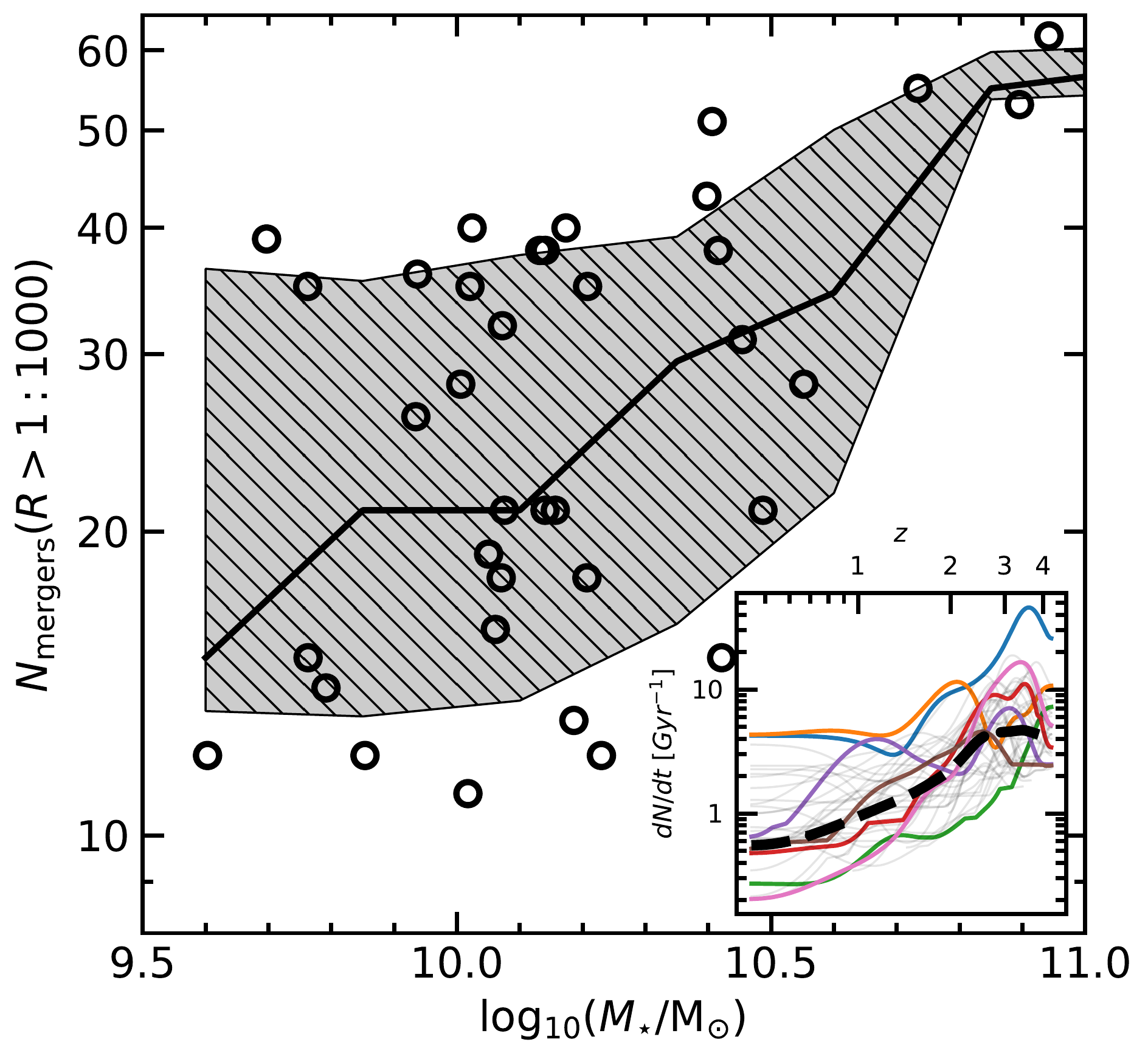}
    \caption{Number of mergers with mass ratio R < 1:1000 undergone per galaxy since $z=5$ as a function of galaxy stellar mass. The hatched region encloses the $1\sigma$ scatter and the black line indicates the median. The inset plot shows the merger rate (R < 1:1000) per Gyr of individual galaxies as a function of redshift. The thick dashed line shows the evolution of the median merger rate of all the galaxies in the sample, thin grey lines show the merger rate history for each galaxy individually, with a small random subset highlighted with thicker coloured lines for clarity.}
    \label{fig:merger_history_dy}
\end{figure}

Finally, Figure \ref{fig:merger_history_dy} shows the total number of mergers undergone by the galaxies in our sample, which we adopt as a rough proxy for the number of discrete units of mass entering the galaxy halo over time (i.e. objects that could be disrupted in the galaxy halo to form tidal features). The main plot shows the total number of mergers with mass ratio $R > 1:1000$ that each galaxy has undergone as a function of their stellar mass and the inset plot shows individual tracks indicating the change in the merger rate history of each galaxy as a function of redshift for the same sample of galaxies. The merger rate history for a small sub-sample of galaxies is highlighted with thicker coloured lines, while the remaining galaxy merger histories are shown as thin grey lines. While we observe a clear average evolution in the merger rate, there is a very significant spread in both the total number of mergers that galaxies have undergone at fixed stellar mass as well as in the overall shape and normalisation of galaxy merger histories. In the range of redshifts that we consider in this paper ($z=0.2$ to $z=0.8$), we see galaxy merger rates decreasing on average, but the variation in the merger histories between $z=0.2$ and $z=0.8$ is very large, with merger rates increasing significantly (fractional increase greater than 0.1) in $\sim 15$ per cent of cases or remaining roughly flat (fractional change of less than 0.1) for a further $\sim 30$ per cent of cases. 

Together, these results indicate that, although there is a clear average evolution in galaxy accretion histories over cosmic time, the merger histories of individual galaxies are sufficiently stochastic that we do not expect to observe this trend in individual galaxies over the timescale that we consider in this study.

\section{Tidal feature detection method}
\label{sec:binary_fill}

\begin{figure}
    \centering
    \includegraphics[width=0.45\textwidth]{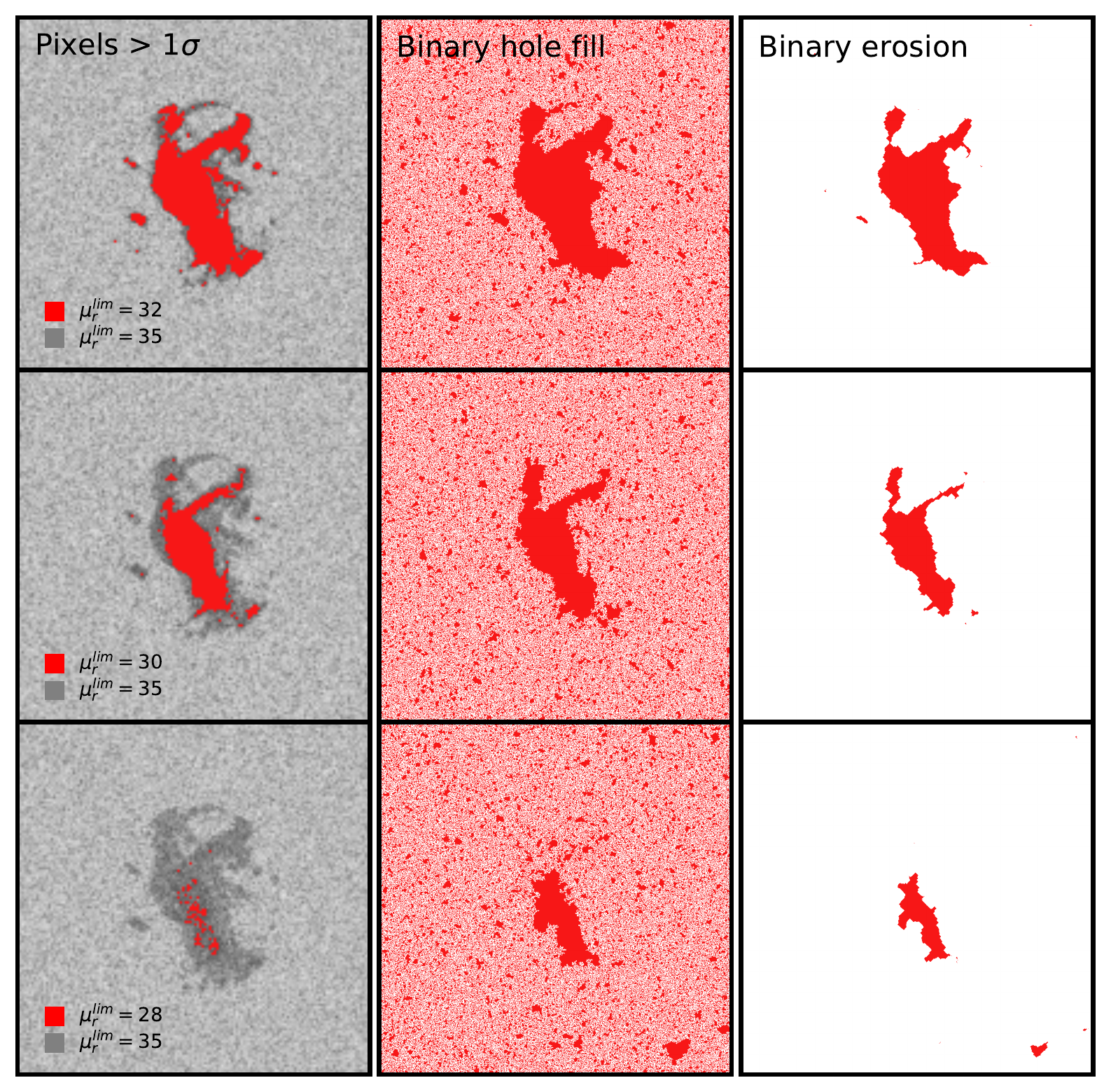}
    \caption{Example images showing the process of locating \textit{detected structures} for limiting surface brightnesses of $\mu_{r}^{\rm lim}(3\sigma,10^{\prime\prime}\times10^{\prime\prime})=32$~mag\,arcsec$^{-2}$, $30$~mag\,arcsec$^{-2}$ and $28$~mag\,arcsec$^{-2}$. Each panel of the first column shows a map of pixels that are at least $1\sigma$ brighter than the noise level for each given limiting surface brightness level (red) plotted over the same map for a limiting surface brightness of $35$~mag\,arcsec$^{-2}$ (gray). For this column only, the images are interpolated for illustrative purposes in order to average out the effect of noise, which otherwise makes it difficult to distinguish between the red and grey maps. The second column shows the binary mask resulting from a $3\times3$ element binary hole fill and the third column shows the final mask after applying a binary erosion.}
    \label{fig:binary_fill}
\end{figure}

Here, we describe our method for determining detected pixels which is used in the definition of the detection fraction used in Section \ref{sec:recovered}. We take into account the fact that it is generally possible to detect contiguous structures by eye, even if they are made up of pixels that are mostly fainter than the surface brightness limit. This is because only relatively few detected pixels grouped close together are required for a contiguous structure to be recognised even if these detected pixels make up a small fraction of the total area of the visible tidal feature. Our aim is to identify these structures in a way that produces similar results to the human eye while rejecting regions of noise.

 We adopt a definition for \textit{detected structures} based on the connections between pixels that are $1\sigma$ above the noise level in images produced from particles that are part of dense tidal features only (e.g. Figure \ref{fig:residual_image}, panel \textbf{c}). We proceed as follows:
\begin{enumerate}
    \item We first use the \texttt{binary\_fill\_holes} function implemented in SciPy \citep[][]{2020SciPy} using a cross-shaped $3\times3$ structuring element, which allows us to construct a mask consisting of every pixel lying within the boundary of a connected region (i.e. we fill any undetected pixels that are surrounded by detected pixels).
    \item  In order to remove small isolated structures which arise from spurious detections in the noise, we then perform a binary erosion on the mask with enough iterations that structures no longer appear in isolated regions of the image.
    \item Any flux found in pixels that are within the mask is considered to be detected.
\end{enumerate}

Figure \ref{fig:binary_fill} shows our method performed on an example galaxy. Left-hand panels show detection maps consisting of pixels that are at least $1\sigma$ brighter the noise level for each given limiting surface brightness level (red) plotted over the same map for a limiting surface brightness of $35$~mag\,arcsec$^{-2}$ (gray). These images are interpolated (smoothed) in order to average out the effect of noise, which otherwise makes it difficult to distinguish between the red and grey maps (for illustrative purposes only). The second column shows the binary mask resulting from a $3\times3$ element binary hole fill and the third column shows the final mask after applying a binary erosion. Although the interpolated detection maps and final masks appear similar at fainter limiting surface brightnesses, the utility of this method becomes more apparent at limiting surface brightnesses closer to that of the tidal features. In this case the area of visible tidal features is significantly larger than the area of pixels that make up the detection map.







\bsp	
\label{lastpage}
\end{document}